\documentclass[aps,preprint,showpacs,preprintnumbers,floatfix,eqsecnum,prb]{revtex4-1}
\usepackage{amsfonts,amsmath,amssymb}
\usepackage{hyperref}

\usepackage[pctex32]{graphicx}
\usepackage{dcolumn}
\usepackage{bm}
\usepackage{xfrac}

\begin{document}

\newcommand{\Area}{\ensuremath{\mathcal{A}}}
\newcommand{\Dbox}{\ensuremath{\Delta_{\text{Box}} }}
\newcommand{\delh}{\ensuremath{\delta h }}
\newcommand{\epsf}{\epsilon_{\textrm{F}}}
\newcommand{\frth}{\ensuremath{\frac{1}{4}}}
\newcommand{\hef}{$^{4}$\textrm{He }}
\newcommand{\het}{$^{3}$\textrm{He }}
\newcommand{\hetf}{$^{3}$\textrm{He--}$^{4}$\textrm{He }}
\newcommand{\hfz}{\ensuremath{h_{4}^{0}}}
\newcommand{\hz}{\ensuremath{h_{0}}}
\newcommand{\hlf}{\ensuremath{\frac{1}{2}}}
\newcommand{\htm}{\ensuremath{\frac{\hbar^{2}}{2 \mstar}}}
\newcommand{\htp}{\ensuremath{\hbar^{2} 2 \pi}}
\newcommand{\kB}{\ensuremath{k_{\textrm{B}}}}
\newcommand{\kT} {\ensuremath{\kappa_{\Tee}}}
\newcommand{\lthree} {\ensuremath{\ell_{3}}}
\newcommand{\ltsq}{\ensuremath{\lambda_{T}^{2}}}
\newcommand{\mstar}{m^{*}}
\newcommand{\muh}{\ensuremath{\mu_{m} \mathcal{H}_{0}}}
\newcommand{\mum}{\ensuremath{\mu_{m}}}
\newcommand{\ntz}{\ensuremath{n_{3}^{0} }}
\newcommand{\nfz}{\ensuremath{n_{4}^{0} }}
\newcommand{\nell}{\ensuremath{n_{\ell}}}
\newcommand{\non}{\ensuremath{n_{\text{onset}}}}
\newcommand{\Numb}{\ensuremath{\mathcal{N}}}
\newcommand{\Tee}{\ensuremath{\mathcal{T}}}
\newcommand{\Pee}{\ensuremath{\mathcal{P}}}
\newcommand{\ua}{\ensuremath{\uparrow}}
\newcommand{\da}{\ensuremath{\downarrow}}

\newcommand{\vf}{\ensuremath{v_{\text{F}}}}
\newcommand{\vfs}{\ensuremath{v_{\text{F}}^{\sigma}}}
\newcommand{\vfsmp}{\frac{v_{\text{F}}^{-\sigma}}{v_{\text{F}}^{\sigma}}}
\newcommand{\vfsp}{\ensuremath{v_{\text{F}}^{\sigma'}}}
\newcommand{\Ns}{\ensuremath{N_{0}^{\sigma}}}
\newcommand{\Nsp}{\ensuremath{N_{0}^{\sigma'}}}
\newcommand{\Ntilde}{\ensuremath{\tilde{N}_{0}}}
\newcommand{\Fzs}{\ensuremath{F^{s}_{0}}}
\newcommand{\Fos}{\ensuremath{F^{s}_{1}}}
\newcommand{\Fza}{\ensuremath{F^{a}_{0}}}
\newcommand{\Foa}{\ensuremath{F^{a}_{1}}}
\newcommand{\gs}{\ensuremath{g_{\sigma}}}
\newcommand{\gu}{\ensuremath{g_{\uparrow}}}
\newcommand{\gd}{\ensuremath{g_{\downarrow}}}
\newcommand{\hu}{\ensuremath{h_{\uparrow}}}
\newcommand{\hd}{\ensuremath{h_{\downarrow}}}
\newcommand{\xis}{\ensuremath{\xi_{\sigma}}}
\newcommand{\xu}{\ensuremath{\xi_{\uparrow}}}
\newcommand{\xd}{\ensuremath{\xi_{\downarrow}}}
\newcommand{\Sigs}{\ensuremath{\Sigma_{\sigma}}}
\newcommand{\Su}{\ensuremath{\Sigma_{\uparrow}}}
\newcommand{\Sd}{\ensuremath{\Sigma_{\downarrow}}}
\newcommand{\nusp}[1]{\ensuremath{\nu^{\sigma}_{#1}}}
\newcommand{\nusm}[1]{\ensuremath{\nu^{-\sigma}_{#1}}}
\newcommand{\mpsm}{\ensuremath{\frac{m^{*}_{\sigma}}{m}}}
\newcommand{\mmsm}{\ensuremath{\frac{m^{*}_{-\sigma}}{m}}}
\newcommand{\nbar}{\ensuremath{\overline{n}}}
\newcommand{\te}{\ensuremath{\tilde{\epsilon}}}
\newcommand{\tn}{\ensuremath{\tilde{n}}}
\newcommand{\okq}{\ensuremath{\omega_{k q}}}
\newcommand{\kfs}{\ensuremath{k_{\text{F} \sigma}}}
\newcommand{\kfss}{\ensuremath{k^{2}_{\text{F} \sigma}}}
\newcommand{\pol}{\mathcal{P}}

\newcommand{\kvec}{\mathbf k}
\newcommand{\kvep}{\mathbf k^{\prime}}
\newcommand{\qvec}{\mathbf q}
\newcommand{\pvec}{\mathbf p}
\newcommand{\pvep}{\mathbf p^{\prime}}
\newcommand{\ppq}{\ensuremath{{\mathbf p} + \frac{\mathbf q}{2}}}
\newcommand{\mpq}{\ensuremath{-{\mathbf p} + \frac{\mathbf q}{2}}}
\newcommand{\pppq}{\ensuremath{{\mathbf p^{\prime}} + \frac{\mathbf q}{2}}}
\newcommand{\mppq}{\ensuremath{-{\mathbf p^{\prime}} + \frac{\mathbf q}{2}}}
\newcommand{\mvec}{\mathbf m}
\newcommand{\rvec}{\mathbf r}
\newcommand{\rvep}{{\mathbf r}^{\prime}}
\newcommand{\uvec}{\mathbf u}
\newcommand{\vvec}{\mathbf v}
\newcommand{\vu}{\ensuremath{v_{\sigone}}}
\newcommand{\vd}{\ensuremath{v_{-\sigone}}}
\newcommand{\pu}{\ensuremath{p_{\sigone}}}
\newcommand{\pd}{\ensuremath{p_{-\sigone}}}
\newcommand{\sigone}{\ensuremath{\sigma_{1}}}

\newcommand{\ssqttwo}{\sin^{2}\left(\frac{\theta_{kk^{\prime}}}{2}\right)}
\newcommand{\tsqttwo}{\tan^{2}\left(\frac{\theta_{kk^{\prime}}}{2}\right)}

\newcommand{\tkkp}{\theta_{k k^{\prime}}}
\newcommand{\ctkkp}{\cos{\tkkp}}
\newcommand{\ctppp}{\cos{\theta_{p p^{\prime}}}}
\newcommand{\cpq}{\cos{\theta_{p q}}}
\newcommand{\stkkp}{\sin{\tkkp}}
\newcommand{\spq}{\sin{\theta_{p q}}}
\newcommand{\ttkkp}{\tan{\tkkp}}
\newcommand{\ctkkptwo}{\ensuremath{ \cos{(\frac{\tkkp}{2}} })}
\newcommand{\stkkptwo}{\ensuremath{ \sin{(\frac{\tkkp}{2}} })}
\newcommand{\ttkkptwo}{\ensuremath{ \tan{(\frac{\tkkp}{2}} })}
\newcommand{\kup}{k_{\uparrow}}
\newcommand{\kdn}{k_{\downarrow}}
\newcommand{\kmkp}{\vert \kvec - \kvep \vert}
\newcommand{\kpkp}{\vert \kvec + \kvep \vert}
\newcommand{\tpq}{\left(\frac{2p}{q}\right)}

\newcommand{\sgn}[1]{\ensuremath{\text{sgn}(#1)}}

%
%

\mathchardef\LL="024C
\def\bra#1{\left\langle#1\right|}
\def\ket#1{\left|#1\right\rangle}
\def\duone{\delta u_1(\rvec)}
\def\duonp{\delta u_1(\rvep)}
\def\duonei#1{\delta u_1(\rvec_#1)}
\def\dutwo#1#2{\delta u_2(\rvec_#1,\rvec_#2)}
\def\ronpi#1{\rho_1(\rvep_#1)}
\def\drone{\delta\rho_1(\rvec)}
\def\dronp{\delta\rho_1(\rvep)}
\def\dronei#1{\delta\rho_1(\rvec_#1)}
\def\drtwo#1#2{\delta\rho_2(\rvec_#1,\rvec_#2)}
\def\dgtwo#1#2{\delta g(\rvec_#1,\rvec_#2)}
\def\difr#1#2{|\rvec_#1-\rvec_#2|}
\def\vph#1#2{V_{\rm p-h}(\rvec_#1,\rvec_#2)}
\def\hm#1{\frac{\hbar^2}{#1m}}
\def\he#1{$^{#1}$He}
\def\half{\frac{1}{2}}
\def\uone{u_1(\rvec)}
\def\uonp{u_1(\rvep)}
\def\uonei#1{u_1(\rvec_#1)}
\def\uonpi#1{u_1(\rvep_#1)}
\def\utwo#1#2{u_2(\rvec_#1,\rvec_#2)}
\def\rone{\rho_4(\rvec)}
\def\ronp{\rho_4(\rvep)}
\def\ronei#1{\rho_4(\rvec_#1)}
\def\rtwo#1#2{\rho_2(\rvec_#1,\rvec_#2)}
\def\gtwo#1#2{g(\rvec_#1,\rvec_#2)}
\def\htwo#1#2{h(\rvec_#1,\rvec_#2)}ARTICLE
\def\comment#1{\bigskip\hrule\smallskip#1\smallskip\hrule\bigskip}
\def\I{{\rm i}}

\title{Transport in thin polarized Fermi-liquid films}
\author{David Z. Li}
\email{zhaozhe.li@email.wsu.edu}
\altaffiliation[Present address: ]{Centre for Quantum and Optical Science,
Swinburne University of Technology,
Melbourne, VIC 3122
Australia.}
\author{R. H. Anderson}
\email{rha@spu.edu}
\author{M. D. Miller}
\email{mdm@wsu.edu}
\affiliation{Department of
Physics and Astronomy, Washington State University, Pullman, WA
99164-2814, USA} 

\date{\today}

\begin{abstract}
We calculate expressions for the state-dependent quasiparticle lifetime,  the thermal conductivity $\kappa$, the shear viscosity $\eta$, and discuss the spin diffusion coefficient $D$ for Fermi-liquid films in two dimensions.  The expressions are valid for low temperatures and arbitrary polarization. In two dimensions, as in three dimensions, the integrals over the transition rates factor into energy and angular parts.  However, the angular integrations contain a weak divergence. This problem is addressed using the method of Miyake and Mullin.  The low-temperature expressions for the transport coefficients are essentially exact. We find that $\kappa^{-1} \sim T \ln{T}$, and $\eta^{-1} \sim T^{2}$ for arbitrary polarizations $0 \le \pol \le 1$.   These results are in agreement with earlier zero-polarization results of Fu and Ebner, but are in contrast with the discontinuous change in temperature dependence from $T^{2} \ln{T}$ at $\pol =0$ to $T^{2}$ at $0 < \pol < 1$ that was found by Miyake and Mullin for $D$.  We note that the shear viscosity requires a unique analysis.    We utilize previously determined values for the density and polarization dependent Landau parameters to calculate the transition probabilities in the lowest order  ``$\ell = 0$ approximation", and thus we obtain  predictions for the density, temperature and polarization dependence of  the thermal conductivity, shear viscosity, and spin diffusion coefficient for thin \he3 films. Results are shown for second layer \he3 films on graphite, and thin \he3-\he4 superfluid mixtures. The density dependence is discussed in detail. For $\kappa$ and $\eta$ we find roughly an order of magnitude increase in magnitude from zero to full polarization.  For $D$ a simialr large increase is predicted from zero polarization to the polarization where $D$ is a maximum ($\sim 0.74$). We discuss the applicability of \he3 thin films to the question of the existence of a universal lower bound for the ratio of the shear viscosity to the entropy density. 
\end{abstract}

\maketitle

\section{ \label{sec:Intro} Introduction}

 Fermi-liquid theory, developed by Landau~\cite{Landau56,Landau57}  in the mid-1950's,  showed how low-temperature collective excitations and thermodynamic properties of strongly-interacting normal many-fermion systems could be encoded in a few parameters, the Landau parameters, and that these parameters were related to a certain limiting value of the microscopic scattering function.~\cite{Landau58}   In Ref.~\onlinecite{Landau57}, Landau also introduced a kinetic equation to describe the nonequilibrium properties of a Fermi liquid. The kinetic equation is of the same form as the classical Boltzmann equation with the local quasiparticle energy $\tilde{\epsilon}_{\pvec \sigma}(\rvec, t)$ playing the role of a Hamiltonian. The application of the linearized Landau kinetic equation to the calculation of transport coefficients for bulk \he3 has been very successful. In this manuscript we shall apply this approach to a strongly interacting many-fermion system in two dimensions.   Reviews of the bulk calculations at zero polarization can be found in the works of Abrikosov and Khalatnikov,~\cite{AK1958} Pines and Nozi\`{e}res,~\cite{PinesNoz1966} and Baym and Pethick.~\cite{BP1991}  The calculation of transport coefficients for Fermi liquids in three dimensions with arbitrary polarization can be found in Anderson, Pethick and Quader~\cite{APQ1987}, and Meyerovich.~\cite{Meyerovich1983} The former set of authors used a slick general notation that emphasized the similarities in the calculations of the various coefficients. There exist some measurements of transport coefficients as a function of polarization for bulk \he3. Buu, Forbes, Puech, and Wolf~\cite{Buu...1999}, and also  Akimoto,  Xia,  Adams, Candela,  Mullin,  and Sullivan~\cite{Akimoto...2002} studied the shear viscosity.  Sawkey, Puech, and Wolf~\cite{Sawkey...2006} studied the thermal conductivity.  

Abrikosov and Khalatnikov (AK) in particular  showed that the integrals involved in the collision integral factor neatly into a product of  integrals involving angular variables and those involving energy variables. The resulting expression for the kinetic equation could then be brought into the form of a linear integral eigenvalue problem for essentially the non-equilibrium part of the fermion distribution function. The exact solutions of these integral eigenvalue problems are  derived, and discussed in detail by Sykes and Brooker~\cite{SykesBrooker1970}  and also Jensen, Smith, and Wilkins.~\cite{JSW1969}

In recent work, we have utilized the kinetic equation approach to study the transition between collective excitations in the ballistic regime (zero sound) and collective excitations in the hydrodynamic regime  (first sound) in thin, arbitrarily polarized Fermi-liquid films.~\cite{LAM_PRB2013}  For sound, the kinetic equation is usually solved by rewriting the integral equation as an (infinite) set of algebraic equations by using a Fourier expansion, and then taking moments with respect to the angular functions.  This procedure is not unique, and we have compared and discussed in detail the predicted propagation speeds and attenuation for two different approaches.~\cite{LAMC2014}  In the above cited works we have utilized previously calculated~\cite{LAM_PRB2012,LAM_PRB2013} density and polarization dependent Landau parameters in order to obtain numerical predictions for thermodynamic and collective excitations for the specific case of \he3 films. In this paper we shall use these same Landau parameters to calculate predicted values for the density and polarization dependent transport coefficients in thin \he3 films. 

The calculation of transport coefficients for thin \he3 films has been considered by Fu and Ebner,~\cite{FuEbner1974} and also by Miyake and Mullin.~\cite{MiyakeMullin1983,*MiyakeMullin1984} Fu and Ebner applied the variational approach that was developed by Baym and Ebner~\cite{BaymEbner1970} in order to calculate transport coefficients for \he3 in superfluid \he4 bulk solutions.  The variational approach of Fu and Ebner as applied in two dimensions does not lend itself to analytic solution, nevertheless, they were able to extract the lowest order temperature dependencies together with numerically determined coefficients for the thermal conductivity $\kappa$, the first (or shear) viscosity $\eta$, and the spin diffusion coefficient $D$ all at zero polarization.   Fu and Ebner obtained $\ln{T}$ behavior for two of the coefficients, and pointed out that the source was a weak divergence in the momentum space integrals. 

Miyake and Mullin (MM) derived an exact expression for the spin diffusion coefficient for two-dimensional fermions with arbitrary polarization.  They indicated that in two dimensions one obtains a logarithmic divergence at \textit{finite} temperature in one of the angular integrals if one proceeds by strictly following the three-dimensional approach developed by AK.  They identified the source of the divergence at finite temperature as an artifice of using zero-temperature values for the Fermi momenta in an integrand of one of the angular integrals in the kinetic equation.   In a very clever analysis, by generalizing the analysis to low but finite temperature they were able to extract an expression that yielded a logarithmic divergence only in the zero-temperature limit.  In Sec.~\ref{sec:tau} we shall derive this fundamental result in detail. 

In Sec.~\ref{sec:tau} we apply the MM method to calculate the state-dependent quasiparticle lifetime at arbitrary polarization.  This calculation is similar to that of the transport coefficients but is simpler.  This allows us to utilize the MM approach in a clear context.  We shall compare the present result for the quasiparticle lifetime to a previous one~\cite{LAM_PRB2013} that was obtained using a method  that is completely independent of MM.  In Secs. \ref{sec:therm} and  \ref{sec:visco} we calculate the thermal conductivity, and  the shear viscosity, respectively. In Sec.~\ref{sec:diff} we include only a brief summary of the calculation of the spin diffusion coefficient since that transport coefficient was  analyzed in detail by MM.    We note that as in three dimensions the calculations of the thermal conductivity and the spin diffusion calculation are very similar. However, unlike three dimension, for two dimensions we find that the analysis for the shear viscosity needs significant modification. As in the case addressed by MM, the problem in the shear viscosity calculation is identified as being due to the incorrect use of the zero-temperature limit in the integrands of the angular integrals.   In Sec.~\ref{sec:he3} we utilize Landau parameters that were previously determined for second layer \he3 films on a graphite substrate, and also for thin film \he3-\he4 mixtures to compute density, temperature and polarization dependencies for the transport coefficients. Our results for the shear viscosity are used to calculate the ratio of the shear viscosity to the entropy density.    Sec.~\ref{sec:Conclusion} is the conclusion.

\section{\label{sec:tau} Quasiparticle lifetime}
We examine a system of $N = N_{\ua} + N_{\da}$, spin-$\frac{1}{2}$ fermions in a box of area $L^{2}$.  The particles have bare mass $m$, and interact with two-body potential $V(r)$ that is assumed to depend only on the scalar distance between the particles. The particles fill two Fermi seas up to Fermi momenta $\kup$ and  $\kdn$, and we introduce the convention that the spin-down Fermi sea will always be the minority Fermi sea in the case of nonzero polarization. The term \textit{polarization} denotes the magnetization per particle which will be denoted by $\pol$, thus $\pol \equiv M/N = \left( N_{\ua} - N_{\da}\right) / N$.  The terms coverage and areal density $\left( N/L^{2} \right)$ are used interchangeably. The system is assumed to be at some finite but low temperature $T$ in the sense that $T << T_{\text{F} \da}$.

The quasiparticle lifetime due to quasiparticle-quasiparticle interactions in two-dimensional Fermi-liquids was calculated in Ref.~\onlinecite{LAM_PRB2013}.  The method used in that reference was borrowed from two-dimensional electron theory, and took advantage of the similarity in structure between the collision integral and the free fermion dynamic structure function. The fluctuation-dissipation theorem together with Stern's analytic expression~\cite{SternPRL1967} for the two-dimensional susceptibility yielded an analytic expression for the low-temperature lifetime. In this section we shall repeat this calculation using the Miyake-Mullin method.  This is convenient because the lifetime calculation is similar but simpler than that for the transport coefficients. The results from this section are in agreement with our previous results, and will be used in the following sections for the transport coefficients.  

\begin{figure}[t]
\includegraphics[trim = 0cm 23cm 10cm 0cm, clip]{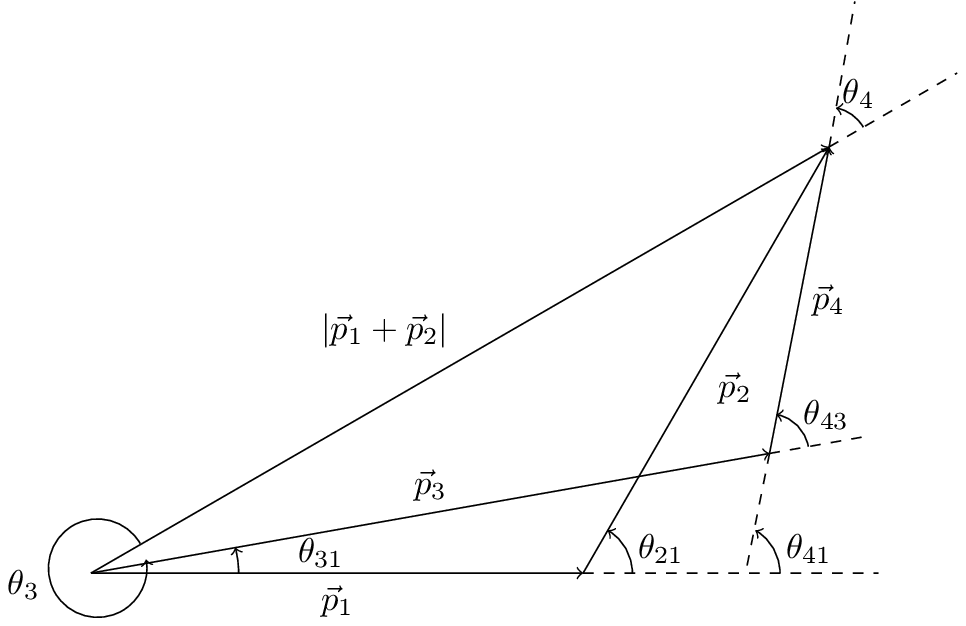}
\caption{\label{fig:forward} The momentum space diagram for the forward scattering process, $\pi < \theta_{3} \leq  2 \pi$. From momentum conservation $\pvec_{1} + \pvec_{2} = \pvec_{3} + \pvec_{4}$. The angle $\theta$ is the angle of $\pvec_{2}$ as measured from the direction of $\pvec_{1}$;  $\theta_{3}$ is the angle of $\pvec_{3}$ as measured from the direction of   $\pvec_{1} + \pvec_{2}$.  We shall also need:  the angles  $\Phi_{3} \equiv \theta_{31}$ and $\Phi_{4} \equiv \theta_{41}$ which are the angles of $\pvec_{3}$ and $\pvec_{4}$ as measured from the direction of $\pvec_{1}$, respectively; $\alpha \equiv \theta_{43}$ is the angle of $\pvec_{4}$ measured relative to $\pvec_{3}$.}
\vspace{1truein}
\end{figure}

\begin{figure}[t]
\includegraphics[trim = 0cm 24cm 12cm 0cm, clip]{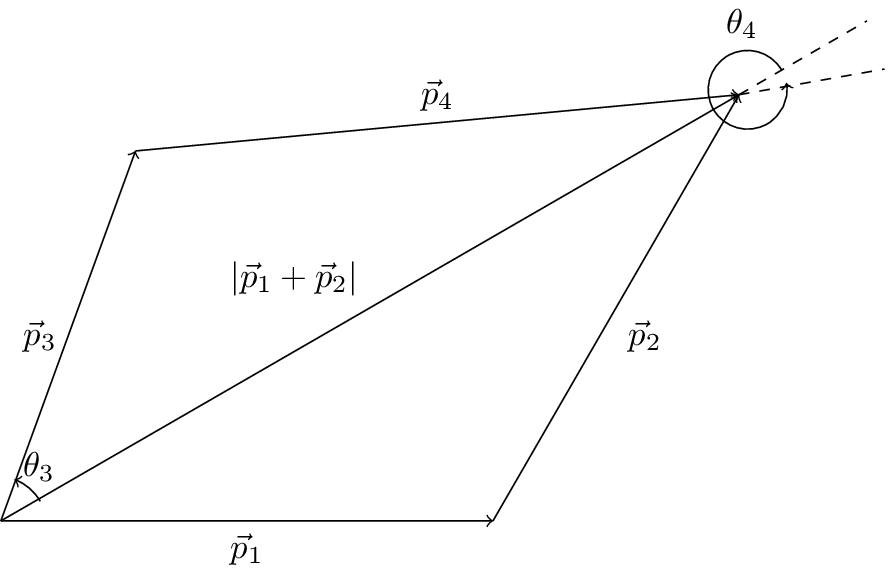}
\caption{\label{fig:backward} The momentum space diagram for the backwards scattering process, $0 < \theta_{3} \leq \pi$. We note that this figure can be obtained from Fig.~\ref{fig:forward} by exchanging $\pvec_{3}$ and $\pvec_{4}$.  From momentum conservation $\pvec_{1} + \pvec_{2} = \pvec_{3} + \pvec_{4}$. }
\vspace{1truein}
\end{figure}

After some simplification,~\cite{BP1991} the quasiparticle collision frequency  is given by:
\begin{equation} \label{eq:tausig}
\frac{1}{\tau_{\sigma_{1}} (\pvec_{1})} = \sum_{ \pvec_{2}, \sigma_{2}} \sum_{ \pvec_{3}, \sigma_{3}} \sum_{ \pvec_{4}, \sigma_{4}}   W(1, 2;3,4) \delta(\epsilon_{1} + \epsilon_{2}  - \epsilon_{3} - \epsilon_{4}) \delta_{\pvec_{1} + \pvec_{2},  \pvec_{3} + \pvec_{4}} \delta_{\sigma_{1} + \sigma_{2},  \sigma_{3} +\sigma_{4}}  n_{2} \nbar_{3}  \nbar_{4}  \,,
\end{equation} 
where $n_{\pvec \sigma} \equiv 1 / [\exp(\beta (\epsilon_{\pvec \sigma} -\pvec \cdot \uvec - \mu_{\sigma})) + 1 ]$ is the Fermi distribution function, $\beta \equiv 1 / k_{B} T$,  $\mu_{\sigma}$ is the chemical potential for the $\sigma\text{th}$ Fermi sea, and $\uvec$ is the fluid velocity. In this section we can set $\uvec =0$.  The $W$'s are transition rates, and we have defined
\begin{equation} \label{eq:nbar}
\nbar_{\pvec \sigma} \equiv 1 - n_{\pvec \sigma} = \frac{1}{1 + e^{-\beta (\epsilon_{\pvec \sigma} - \pvec \cdot \uvec - \mu_{\sigma}) }} \,.
\end{equation} The standard treatment in three-dimensions follows  Abrikosov and Khalatnikov,~\cite{Khalatnikov1958,*AK1959} and 
introduces new integration variables in terms of energies and angles. These integrations are independent of one another, and in lowest order in temperature one can find a closed form expression for $1 / \tau$ in terms of an angular average of the transition rates. Label the incoming quasiparticles as $\pvec_{1}, \pvec_{2}$ and the outgoing quasiparticles $\pvec_{3}, \pvec_{4}$.  The standard angular variables, $\theta$ and $\varphi$, are defined as follows: $\theta$ is the angle of $\pvec_{2}$ measured relative to the direction of $\pvec_{1}$, and $\varphi$ is the angle between the planes formed by the pairs of vectors $\{ \pvec_{1}, \pvec_{2} \}$ and $\{ \pvec_{3},\pvec_{4} \}$. As discussed by MM, in two dimensions $\varphi$ can only take on two values: $0, \pi$. We illustrate these two possibilities in Figs.~\ref{fig:forward} and \ref{fig:backward}. Along with MM we shall refer to these two processes as  forward and backward scattering, respectively.  We note that the forward and backward scattering processes have a direct and exchange relationship since Fig.~\ref{fig:backward} is obtained from Fig.~\ref{fig:forward} by exchanging $\pvec_{3}$ and $\pvec_{4}$. 

Figs.~\ref{fig:forward} and \ref{fig:backward} define the conventions that we shall use throughout this paper to label the angles associated with the quasiparticle momenta. All angles are measured counter-clockwise. The angles $\theta_{i}$ with $i = 1,2,3,4$ are the angles of $\pvec_{i}$ as measured from $\pvec_{1} + \pvec_{2}$.  The angle $\theta_{ij}$ is the angle of $\pvec_{i}$ as measured from the direction of $\pvec_{j}$.  In the discussion below, we shall find it convenient to use the following definitions: $\Phi_{i} \equiv \theta_{i1}$, $\alpha \equiv \theta_{43}$, and as noted above $\theta \equiv \Phi_{2}$.

Since the interaction is assumed spin-independent, total spin is conserved in the collisions as indicated by the Kronecker delta in (\ref{eq:tausig}).  Thus, we must have $\sigma_{3} = \sigma_{1}$ and $\sigma_{4} = \sigma_{2}$.  In the spin parallel case $\sigma_{2} = \sigma_{1}$ the exchange diagram is identical to the direct diagram, and therefore they must be counted only once in the phase space integrations. In the spin anti-parallel case $\sigma_{1} = - \sigma_{2}$ the direct and exchange diagrams give different contributions to the total transition probability. Performing the spins sums, replacing the momentum sums by integrations, and performing an integration over $\pvec_{4}$ yields:
\begin{align} 
\frac{1}{\tau_{\sigma_{1}} (\pvec_{1})} &= \dfrac{1}{h^{4}} \int d\pvec_{2} d\pvec_{3}  \left( \hlf W^{\sigma_{1} \, \sigma_{1}} + W^{\sigma_{1} \, -\sigma_{1}} \right) \delta(\epsilon_{1} + \epsilon_{2}  - \epsilon_{3} - \epsilon_{4}) n_{2} \nbar_{3}  \nbar_{4}  \,, \nonumber \\
&\equiv \dfrac{1}{\tau_{\sigma_{1} \sigma_{1}}} + \dfrac{1}{\tau_{\sigma_{1} -\sigma_{1}}} \label{eq:taugen}\,.
\end{align} 
where for later use  we have introduced spin parallel and spin anti-parallel collision frequencies.  In this expression we have set $A = 1$.  Thus, the units of the $W^{\sigma \sigma'}$'s are energy-time$^{-1}$-area$^{2}$. As usual, the factor of one-half appearing with the spin-parallel transition probability prevents over counting as discussed above.~\cite{BP1991}   

We first consider the spin-parallel lifetime, and separate out the angular integrals:
\begin{align} \label{eq:tauang}
\frac{1}{\tau_{\sigma_{1} \sigma_{1}} } &= \dfrac{1}{h^{4}} \int_{0}^{\infty} d \pvec_{2}  \int_{0}^{\infty} d \pvec_{3}    \left( \hlf W^{\sigma_{1} \, \sigma_{1}}(\theta) \right) \delta(\epsilon_{1} + \epsilon_{2}  - \epsilon_{3} - \epsilon_{4}) n_{2} \nbar_{3}  \nbar_{4}  \,, \nonumber \\
&= \dfrac{2}{h^{4}} \int_{0}^{\infty} p_{2} d p_{2}  \int_{0}^{\infty} p_{3} d p_{3} \int_{0}^{\pi} d\theta \,  \int_{\pi}^{2 \pi} d\theta_{3} \, W_{f}^{\sigma_{1}  \, \sigma_{1}}(\theta)  \delta(\epsilon_{1} + \epsilon_{2}  - \epsilon_{3} - \epsilon_{4}) n_{2} \nbar_{3}  \nbar_{4}  \,, \nonumber \\
\end{align} 
where we have taken advantage of the symmetry in $\theta$ about $\pi$, $\theta_{3}$ is defined as the angle of $\pvec_{3}$ measured with respect to $\pvec_{1} + \pvec_{2}$, see Fig.~\ref{fig:forward}, and the subscript $f$ or $b$ on $W$ identifies the transition probability as that for forward scattering  ($\pi < \theta_{3} \leq 2 \pi$) or backward scattering ($0 < \theta_{3} \leq \pi$), respectively.  We now rewrite the angular variable $\theta_{3}$ in a more useful form.   With an eye on Fig.~\ref{fig:forward} the law of cosines yields:
\begin{align}
p_{4}^{2} &= p_{3}^{2} + \ell^{2} - 2 p_{3}\ell \cos{\theta_{3}} \,, \label{eq:p4sq} \\
d\theta_{3} &= \dfrac{dp_{4}^{2}}{2 p_{3}\ell \sin{\theta_{3}}} \,, \label{eq:dt3}
\end{align}
where following MM the useful variable $\ell$ has been defined:
\begin{equation} \label{eq:ell}
\ell \equiv |\pvec_{1} + \pvec_{2}| \,.
\end{equation}

It is now convenient to introduce the angle $\alpha \equiv \theta_{43}$ as shown in Fig.~\ref{fig:forward}:
\begin{equation} \label{eq:sina}
- p_{3}\ell \sin{\theta_{3}} = p_{3}p_{4} \sin{\alpha}  \,.
\end{equation}
From the law of cosines again:
\begin{align}
\ell^{2} &= p_{3}^{2} + p_{4}^{2} - 2 p_{3} p_{4} \cos{(\pi - \alpha)} \,, \\
p_{1}^{2} + p_{2}^{2} +2 p_{1} p_{2}\cos{\theta} &=  p_{3}^{2} + p_{4}^{2} + 2 p_{3} p_{4} \cos{\alpha} \,. \label{eq:loc2}
\end{align}
Eq.~(\ref{eq:loc2}) can be simplified using energy conservation:
\begin{equation}
\varepsilon_{1} +\varepsilon_{2} =\varepsilon_{3} + \varepsilon_{4} \,,
\end{equation}
where we have defined $\varepsilon_{i} \equiv  p_{i}^{2} / 2 m_{i}^{\ast}$.
The quasiparticle label on the effective masses is needed since at finite polarization the effective masses are state dependent.  We find:
\begin{equation} \label{eq:cosa}
p_{3} p_{4} \cos{\alpha} = \left(m_{2}^{\ast} - m_{1}^{\ast}\right) \xi_{3}  + p_{1} p_{2} \cos{\theta} \,,
\end{equation}
where we have defined the important quantity $\xi_{3}$:
\begin{equation} \label{eq:xi3}
\xi_{3} \equiv \dfrac{p_{3}^{2} - p_{1}^{2}}{2 m_{1}^{\ast}} \,,
\end{equation}
where $\xi_{3} \sim O(k_{B}T)$.  We use (\ref{eq:cosa}) to eliminate $\alpha$ in (\ref{eq:sina}):
\begin{equation} \label{eq:sinsqa}
p_{3}^{2} p_{4}^{2}  \sin^{2}{\alpha}  = p_{1}^{2} p_{2}^{2}  \sin^{2}{\theta} + \left(m_{1}^{\ast} p_{2}^{2} - m_{2}^{\ast} p_{1}^{2} - p_{1} p_{2} ( m_{2}^{\ast} - m_{1}^{\ast}) \cos{\theta} \right) \xi_{3} - (m_{1}^{\ast} + m_{2}^{\ast})^{2} \xi_{3}^{2} \,.
\end{equation}
Finally, we combine Eqs.~(\ref{eq:dt3}), (\ref{eq:sina}), and (\ref{eq:sinsqa}) to yield:
\begin{equation} \label{eq:dt3exact}
d \theta_{3} = - \dfrac{dp_{4}^{2}}{2 p_{1} p_{2} \sqrt{\sin^{2}{\theta} - \dfrac{1}{\epsilon_{1 2}} \xi_{3} -\dfrac{(m_{1}^{\ast} + m_{2}^{\ast})^{2}}{p_{1}^{2} p_{2}^{2}} \xi_{3}^{2} }} \qquad (\pi < \theta_{3} \leq 2\pi) \,,
\end{equation}
where
\begin{align} 
\dfrac{1}{\epsilon_{1 2}} &\equiv  \left( 1 + \frac{p_{2}}{p_{1}} \cos{\theta} \right)  \frac{1}{ \epsilon_{2}} -  \left( 1 + \frac{p_{1}}{p_{2}} \cos{\theta} \right)  \frac{1}{ \epsilon_{1}} \,, \nonumber \\
&\approx \left( 1 + \frac{p_{F 2}}{p_{F 1}} \cos{\theta} \right)  \frac{1}{ \epsilon_{F 2}} -  \left( 1 + \frac{p_{F 1}}{p_{F 2}} \cos{\theta} \right)  \frac{1}{ \epsilon_{F 1}} \,. \label{eq:eps12}
\end{align}
The $F$ subscripts on the Fermi energies and Fermi momenta indicate that we only need the zero-temperature limit for $\epsilon_{1 2}$ since $\xi_{3}$ itself is $O(k_{B}T)$.   

We note that Eq.(\ref{eq:dt3exact}) is exact. For the spin parallel case the linear term in $\xi_{3}$ vanishes making the thermal correction term $\sim O((k_{B}T)^{2})$. The thermal correction changes from linear order  in $\xi_{3}$ for anti-parallel spin scattering to quadratic order  in $\xi_{3}$ for parallel spin scattering. With Eq.~(\ref{eq:dt3exact}), Eq.~(\ref{eq:tauang}) becomes: 
\begin{align} 
\frac{1}{\tau_{\sigma_{1} \sigma_{1}} }  
&= \dfrac{1}{h^{4}} \int_{0}^{\infty}  d p_{2} \,   d p_{3} \, d p_{4}^{2} \int_{0}^{\pi} \frac{d \theta}{\sqrt{\sin^{2}{\theta}  - \left(\frac{\xi_{3}}{\epsilon_{1}} \right)^{2}}} \, W_{f}^{\sigma_{1} \, \sigma_{1}} \delta(\epsilon_{1} + \epsilon_{2}  - \epsilon_{3} - \epsilon_{4}) n_{2} \nbar_{3}  \nbar_{4}  \,.
\end{align}

The integrals are brought to their final form by introducing dimensionless variables $x_{i} \equiv \beta (\epsilon_{i} - \mu)$:
\begin{align} \label{eq:taup0}
\frac{1}{\tau_{\sigma_{1} \sigma_{1}} }   &= \dfrac{2 (m^{\ast}_{\sigma_{1}})^{3} (k_{B}T)^{2}}{h^{4}  p_{\sigma_{1}}^{2}} \int_{-\infty}^{\infty} d x_{2} d x_{3} d x_{4} 
\delta(x_{1} + x_{2}  - x_{3} - x_{4}) n_{2} \nbar_{3}  \nbar_{4} \int_{0}^{\pi} d\theta \dfrac{W_{f}^{\sigma_{1} \, \sigma_{1}}}{\sqrt{\sin^{2}{\theta} -\left( \frac{x_{3}}{\beta \mu }\right)^{2} }}     \,.
\end{align}
It is convenient to split the $\theta$ integral into pieces:
\begin{equation}
\int_{0}^{\pi} d\theta  = \int_{\Delta}^{\pi - \Delta} d \theta + \left( \int_{0}^{\Delta} d \theta + \int_{\pi -\Delta}^{\pi} d \theta   \right) \,,
\end{equation}
where following MM we have defined $\Delta \equiv 1/(\beta \mu)$.  At low temperature $\Delta \ll 1$ and so the $\theta$  integral can be approximated as
\begin{equation}
\int_{\Delta}^{\pi - \Delta}  \dfrac{d \theta}{\sin{\theta}}+ \left( \int_{0}^{\Delta} + \int_{\pi -\Delta}^{\pi}    \right) \dfrac{d \theta}{\sqrt{\theta^{2} -\left( \frac{x_{3}}{\beta \mu }\right)^{2} }}\,.
\end{equation}
The integrand in the second term is not singular, and so the second term can be neglected relative to the first since it does not contribute to lowest order with a logarithmic temperature dependence.  Thus we obtain the MM form for the angular part of the two-dimensional collision integral:
\begin{align} \label{eq:tauMM}
\frac{1}{\tau_{\sigma_{1} \sigma_{1}} }  \approx \dfrac{2 (m^{\ast}_{\sigma_{1}})^{3} (k_{B}T)^{2}}{h^{4}  p_{\sigma_{1}}^{2}}\int_{-\infty}^{\infty} d x_{2} d x_{3} d x_{4}  \delta(x_{1} + x_{2}  - x_{3} - x_{4}) n_{2} \nbar_{3}  \nbar_{4}  \int_{\Delta}^{\pi - \Delta} d \theta \dfrac{W_{f}^{\sigma_{1} \, \sigma_{1}} }{\sin{\theta}}  \,.
\end{align}
The energy integrals are evaluated by Morel and Nozi\`{e}res~\cite{MorelNozieres1962} and the final result is $\pi^{2}/4$. 

The transition rates can be Fourier analyzed as usual~\cite{LAM_PRB2013} yielding: 
\begin{equation} \label{eq:WFourier}
W_{f, b}^{\sigma \sigma'}(\theta) = \sum_{\ell=0}^{\infty}\alpha_{\ell}  T_{\ell}(\cos{\theta}) W^{\sigma \sigma'}_{\ell \, f,b}\,,
\end{equation}
where the $T_{\ell}(\cos{\theta})= \cos{(\ell \theta)}$ are Chebyshev polynomials of the first kind,\cite{A&S} and the parameters $\alpha_{0} = 1 \text{ and } \alpha_{\ell} = 2 \text{ for } \ell \geq 1$. We can now introduce the lowest order ``$\ell = 0$'' approximation by replacing the full transition rate by its $\ell = 0 $ value $W_{f}^{\sigma_{1} \, \sigma_{1}} \approx W_{f, 0}^{\sigma_{1} \, \sigma_{1}}$. The remaining angular integral can now be evaluated:
\begin{equation} \label{eq:I1}
\int_{\Delta}^{\pi - \Delta} d \theta \dfrac{1 }{\sin{\theta}} \approx 2 \ln{\left( \frac{2 \epsilon_{1}}{k_{B} T}\right)} \,.
\end{equation}
The final result is:
\begin{align} \label{eq:taup}
\frac{1}{\tau_{\sigma_{1} \sigma_{1}} }  = \dfrac{\pi^{2}}{2} \dfrac{ (m^{\ast}_{\sigma_{1}})^{2}}{h^{4}} W_{f, 0}^{\sigma_{1} \, \sigma_{1}} \dfrac{(k_{B}T)^{2}}{\epsilon_{1}} \ln{\left( \frac{2 \epsilon_{1}}{k_{B} T}\right)} \,.
\end{align}

The calculation for anti-parallel spins proceeds analogously. From Eq.~(\ref{eq:taugen}):
\begin{align} 
\frac{1}{\tau_{\sigma_{1} -\sigma_{1}} } &= \dfrac{1}{h^{4}} \int_{0}^{\infty} d \pvec_{2}  \int_{0}^{\infty} d \pvec_{3} \int_{0}^{\pi}   W^{\sigma_{1} \, -\sigma_{1}} \delta(\epsilon_{1} + \epsilon_{2}  - \epsilon_{3} - \epsilon_{4}) n_{2} \nbar_{3}  \nbar_{4}  \,, \nonumber \\
&= \dfrac{2 m^{\ast}_{\sigma_{1}} m^{\ast}_{-\sigma_{1}}}{h^{4}} \int_{0}^{\infty}  d \epsilon_{2} \, d \epsilon_{3} \int_{0}^{\pi} d\theta \, \Bigl( \int_{0}^{\pi} d\theta_{3} \, W_{f}^{\sigma_{1} \, -\sigma_{1}} + \int_{\pi}^{2 \pi} d\theta_{3} \, W_{b}^{\sigma_{1} \, -\sigma_{1}} \Big) \nonumber \\ &\phantom{\dfrac{2 m^{\ast}_{\sigma_{1}} m^{\ast}_{-\sigma_{1}}}{h^{4}} } \times \delta(\epsilon_{1} + \epsilon_{2}  - \epsilon_{3} - \epsilon_{4}) n_{2} \nbar_{3}  \nbar_{4}  \,. 
\end{align} 
We now use (\ref{eq:dt3exact}) for anti-parallel spins:
\begin{equation} \label{eq:dt3ap}
d \theta_{3} = \dfrac{\pm d p_{4}^{2}}{2 p_{1} p_{2} \sqrt{\sin^{2}{\theta} - \frac{1}{\epsilon_{1 2}}\xi_{3} }}  \,,
\end{equation}
where $-$ is for forward scattering $\pi < \theta_{3} \leq 2 \pi$, and $+$ is for backward scattering $ 0< \theta_{3} \leq \pi$.
\begin{align} \label{eq:tauanti}
 \dfrac{1}{\tau_{\sigma_{1} -\sigma_{1}}} = &\dfrac{2  m^{\ast}_{\sigma_{1}} ( m^{\ast}_{-\sigma_{1}})^{2}}{h^{4}  p_{\sigma_{1}} p_{-\sigma_{1}}} \int_{0}^{\infty} d \epsilon_{2} d \epsilon_{3} d \epsilon_{4} \delta(\epsilon_{1} + \epsilon_{2}  - \epsilon_{3} - \epsilon_{4}) n_{2} \nbar_{3}  \nbar_{4} \nonumber \\ 
& \times \int_{0}^{\pi} \left[W_{f}^{\sigma_{1} \, -\sigma_{1}} + W_{b}^{\sigma_{1} \, -\sigma_{1}} \right] \frac{d\theta}{\sqrt{ \sin^{2}{\theta} - \frac{1}{\epsilon_{1 2}} \xi_{3}  }}     \,.
\end{align}
Performing the energy integrals yields:
\begin{align} \label{eq:tauapint}
 \dfrac{1}{\tau_{\sigma_{1} -\sigma_{1}}} = &\dfrac{\pi^{2}  m^{\ast}_{\sigma_{1}} ( m^{\ast}_{-\sigma_{1}})^{2}}{2 h^{4}  p_{\sigma_{1}} p_{-\sigma_{1}}} (k_{B}T)^{2} \int_{\Delta}^{\pi - \Delta} \left[W_{f}^{\sigma_{1} -\sigma_{1}} + W_{b}^{\sigma_{1} -\sigma_{1}} \right] \frac{d\theta}{\sin{\theta}}     \,,
\end{align}
where in this case  $\Delta = \sqrt{k_{B}T / |\epsilon_{1 2}|}$.  Utilizing the $\ell = 0$ approximation we obtain:
\begin{align} \label{eq:tauap}
 \dfrac{1}{\tau_{\sigma_{1} -\sigma_{1}}} = & \dfrac{\pi^{2}}{2 h^{4}}\dfrac{m^{\ast}_{\sigma_{1}} ( m^{\ast}_{-\sigma_{1}})^{2}}{ p_{\sigma_{1}} p_{-\sigma_{1}}} \left( W_{f, 0}^{\sigma_{1} \, -\sigma_{1}} + W_{b, 0}^{\sigma_{1} \, -\sigma_{1}} \right) (k_{B}T)^{2} \ln{\left(\frac{4|\epsilon_{1 2}|}{k_{B}T}\right)}   \,.
\end{align}
By combining (\ref{eq:taup}) and (\ref{eq:tauap}) we obtain the total quasiparticle-quasiparticle collision frequency at finite polarization:
\begin{align} \label{eq:tauP}
\frac{1}{\tau_{\sigma_{1}}} = \dfrac{\pi^{2}}{2 h^{4}} \left[ (m^{\ast}_{\sigma_{1}})^{2} W_{f, 0}^{\sigma_{1} \sigma_{1}} \ln{\left( \frac{2 \epsilon_{1}}{k_{B} T}\right)} +   ( m^{\ast}_{-\sigma_{1}})^{2} \frac{p_{\sigma_{1}}}{p_{-\sigma_{1}}} \hlf \left( W_{f, 0}^{\sigma_{1} \, -\sigma_{1}} + W_{b, 0}^{\sigma_{1} \, -\sigma_{1}} \right)  \ln{\left(\frac{4|\epsilon_{1 2}|}{k_{B}T} \right)} \right]   \dfrac{(k_{B}T)^{2}}{\epsilon_{1}} \,.
\end{align}
In this expression the momenta and energies are zero-temperature Fermi momenta and Fermi energies (see also Eq.~(\ref{eq:eps12})).  

\subsubsection{\label{sssec:tauzp} Zero polarization}
At zero polarization, Eq.~(\ref{eq:taup}) for the spin-parallel collision frequency is still valid.  However for the spin anti-parallel collision frequency $1 / \epsilon_{1 2} = 0$ in the zero polarization limit, and thus the leading order correction in the denominator of $d \theta_{3}$ is quadratic. For zero polarization Eq.~(\ref{eq:tauap}) becomes: 
\begin{align}
\dfrac{1}{\tau_{\sigma -\sigma}} = &\dfrac{\pi^{2}  ( m^{\ast})^{2}}{2 h^{4} } \left( W_{f, 0}^{\sigma \, -\sigma} + W_{b, 0}^{\sigma \, -\sigma} \right) \frac{(k_{B}T)^{2}}{\epsilon_{F}} \ln{\left(\frac{2 \epsilon_{F}}{k_{B}T}\right)}   \,.
\end{align}
By combining this result with (\ref{eq:taup}) we obtain the total zero-polarization collision frequency:
\begin{align} \label{eq:tauP0}
\dfrac{1}{\tau_{0}} \equiv \dfrac{1}{\tau_{\sigma}}(\pol = 0) = &\dfrac{\pi^{2}  ( m^{\ast})^{2}}{2 h^{4} } \left[W_{f, 0}^{\sigma \, \sigma} + W_{f, 0}^{\sigma \, -\sigma} + W_{b, 0}^{\sigma \, -\sigma} \right] \frac{(k_{B}T)^{2}}{\epsilon_{F}} \ln{\left(\frac{2 \epsilon_{F}}{k_{B}T}\right)}   \,,
\end{align}
where the $\tau_{0}$ notation will be used below.  The same quantity in Ref.~\onlinecite{LAM_PRB2013} Eq.~(3.24) differs by the appearance of a factor of $3/8$ in the leading coefficient instead of $1/2$. This is not a problem since in that derivation the coefficient of the log term is uncertain with regards factors of $O(1)$  because of the vagaries of the low-temperature limiting process.
We note that one cannot obtain (\ref{eq:tauP0}) by taking the zero-polarization limit of (\ref{eq:tauP}).

\subsubsection{\label{sssec:fullP} Full polarization}
In the full polarization limit one simply omits the contribution from the anti-parallel spins in (\ref{eq:tauP}):
\begin{align} \label{eq:tauP1}
\dfrac{1}{\tau_{1}} \equiv \dfrac{1}{\tau_{\ua}}(\pol = 1) = &\dfrac{\pi^{2}  ( m_{\ua}^{\ast})^{2}}{2 h^{4} } W_{f, 0}^{\ua \ua}  \frac{(k_{B}T)^{2}}{\epsilon_{F \ua}} \ln{\left(\frac{2 \epsilon_{F \ua}}{k_{B}T}\right)}   \,.
\end{align}

\section{\label{sec:transport} Transport}
The derivation of the transport coefficients in a two-dimensional Fermi liquid proceeds in a very similar way to that in three dimensions. Thus, this and the following sections on transport coefficients will necessarily be brief. For the details we refer the reader to Baym and Pethick~\cite{BP1991} for example. We shall concentrate on those aspects  that are specific to finite polarization and two dimensions. 
The general transport equation can be written:
\begin{equation} \label{eq:trans}
\dfrac{\partial \tn_{\pvec \sigma}}{\partial t}  + \bm{\nabla} \tn_{\pvec \sigma} \cdot \bm{\nabla}_{\pvec} {\te}_{\pvec \sigma} - \bm{\nabla}_{\pvec} \tn_{\pvec \sigma} \cdot \bm{\nabla} {\te}_{\pvec \sigma} = I_{\pvec \sigma} \,,
\end{equation}
where $\tn_{\pvec \sigma}(\rvec)$ is the local quasiparticle distribution function defined with the local quasiparticle energies ${\te}_{\pvec \sigma}(\rvec)$ . As usual the local quasiparticle distribution function is expanded around \textit{local equilibrium}:~\cite{BP1991,PinesNoz1966} 
\begin{equation} \label{eq:ntilde}
\tn_{\pvec \sigma}(\rvec) = n_{\pvec \sigma}\left[{\te}_{\pvec \sigma}(\rvec)\right] + \delta n_{\pvec \sigma}(\rvec) \,,
\end{equation}
where $n_{\pvec \sigma}$ is the Fermi distribution function. Equivalently, we can expand the local quasiparticle energies around a set of local equilibrium  energies $\epsilon_{\pvec \sigma}(\rvec)$:
\begin{equation}
{\te}_{\pvec \sigma}(\rvec) = \epsilon_{\pvec \sigma} + \sum_{\pvep} f^{\sigma \sigma'}_{\pvec \pvep} \delta n_{\pvep \sigma'}(\rvec) \,.
\end{equation}

The collision integral on the  right hand side of the transport equation can be written:
\begin{equation} \label{eq:Icoll}
\begin{split}
I(n_{1})  = -\sum_{\pvec_{2}, \pvec_{3}, \pvec{4}} &W(1,2;3,4) \delta_{\pvec_{1} + \pvec_{2}, \pvec_{3}+\pvec_{4}}\delta_{\sigma_{1}+\sigma_{2}, \sigma_{3}+\sigma_{4}} \delta(\te_{1} + \te_{2}  - \te_{3} - \te_{4})  \\ 
&\times \left[ \tn_{1} \tn_{2}(1 -\tn_{3}) (1 - \tn_{4}) - (1 - \tn_{1}) (1 - \tn_{2}) \tn_{3} \tn_{4}  \right]  \,.
\end{split}
\end{equation}
As discussed in Sec.~\ref{sec:tau}, the sums over $\pvec_{3}$ and $\pvec_{4}$ include only distinguishable final states.  We now expand the collision integral (\ref{eq:Icoll}) to first order in the $\delta n_{\pvec \sigma}(\rvec)$.  Consider the products of distribution functions in the square brackets of (\ref{eq:Icoll}), and  substitute (\ref{eq:ntilde}).  This yields:
\begin{equation}
\left[ \ldots  \right] = - \beta \left( \zeta_{1} + \zeta_{2} - \zeta_{3} - \zeta_{4}  \right) n_{1} n_{2} \nbar_{3} \nbar_{4} \,,
\end{equation}
where we have defined 
\begin{equation} \label{eq:zeta}
\delta n_{i} \equiv \frac{\partial n_{i}}{\partial \epsilon_{i}}  \zeta_{i} = -\beta n_{i} \nbar_{i} \zeta_{i} \,,
\end{equation}
and we have made use of the identity:
\begin{equation}
\left[ n_{1} n_{2} \nbar_{3} \nbar_{4} - \nbar_{1} \nbar_{2} n_{3} n_{4}  \right] \delta(\te_{1} + \te_{2}  - \te_{3} - \te_{4}) = 0 \,.
\end{equation}
Performing the spin sums the collision integral becomes:
\begin{equation} \label{eq:IW}
\begin{split}
I(n_{1})  = \beta & \sum_{\pvec_{2}, \pvec_{3},\pvec_{4}} \left( \hlf W^{\sigone \sigone}(\theta)  + W^{\sigone -\sigone}(\theta)   \right)  \delta_{\pvec_{1} + \pvec_{2}, \pvec_{3}+\pvec_{4}}
\delta(\epsilon_{1} + \epsilon_{2}  - \epsilon_{3} - \epsilon_{4})  \\ 
&\times \left( \zeta_{1} + \zeta_{2} - \zeta_{3} - \zeta_{4}  \right) n_{1} n_{2} \nbar_{3} \nbar_{4}  \,,
\end{split}
\end{equation}
where to lowest order we have now replaced all of the remaining local energies $\te$ by local equilibrium energies $\epsilon$.  As discussed in Sec.~\ref{sec:tau} the value of $\theta_{3}$ determines whether the transition rate is given by the forward or backward scattering diagrams in Figs.~\ref{fig:forward} and \ref{fig:backward}, respectively.  Eq.~(\ref{eq:IW}) is the starting point for the collision integral for each of the transport coefficients discussed below.

\subsection{\label{sec:therm} Thermal conductivity}

The thermal current is given by
\begin{equation} \label{eq:jth}
\bm{j_{th}} = \sum_{\sigma} \int \frac{d \kvec}{h^{2}} \, \delta n_{\kvec \sigma} (\epsilon_{\kvec \sigma} - \mu_{\sigma}) \vvec_{\kvec \sigma} \,,
\end{equation}
where
\begin{equation} \label{eq:deltankT}
\delta n_{\kvec \sigma} = \frac{\partial n_{\kvec \sigma}}{\partial \epsilon_{\kvec \sigma}} \vvec_{\kvec \sigma} \cdot \bm{\nabla} (k_{B}T)  q_{\sigma} (\epsilon_{\kvec \sigma}) \,,
\end{equation}
and Eq.~(\ref{eq:deltankT}) defines the function $q_{\sigma}$.  With no loss of generality, we can assume that $ \bm{\nabla} (k_{B} T)$ is in the $x$-direction. The y-component of $\bm{j_{th}}$ vanishes, and (\ref{eq:jth}) becomes:
\begin{align}
(j_{th})_{x} &= \frac{1}{h^{2}} \int_{0}^{\infty} \, d k \, k \sum_{\sigma}\frac{\partial n_{\kvec \sigma}}{\partial \epsilon_{\kvec \sigma}} (\epsilon_{\kvec \sigma} - \mu_{\sigma}) |\bm{\nabla} (k_{B}T)|  q_{\sigma} (\epsilon_{\kvec \sigma})  v_{F \sigma}^{2} \int_{0}^{2 \pi} \,d\phi \cos^{2}{\phi}  \,, \nonumber \\ 
                  &= \frac{\pi k_{B}^{2}}{h^{2}} T |\bm{\nabla} (T)|  \sum_{\sigma} m_{\sigma}^{*} v_{F \sigma}^{2} \int_{-\infty}^{\infty} \,dx  \frac{\partial n}{\partial x}  \, q_{\sigma}(x)\, x \,.
\end{align}
By definition the thermal conductivity is given by $\bm{j_{th}} = - \kappa \bm{\nabla}T$, thus we obtain:
\begin{equation} \label{eq:KapDef}
\kappa = -\frac{\pi k_{B}^{2}}{h^{2}} T   \sum_{\sigma} m_{\sigma}^{*} v_{F \sigma}^{2} \int_{-\infty}^{\infty} \,dx  \frac{\partial n}{\partial x}  \, q_{\sigma}(x)\, x  \,.
\end{equation}

The integral over $q_{\sigma}$ can be evaluated exactly from the kinetic equation by using the results of Sykes and Brooker.~\cite{SykesBrooker1970}  In Eq.~(\ref{eq:IW}) we have expanded the collision integral (\ref{eq:Icoll}) to linear order in the nonequilibrium distribution functions $\delta n_{\pvec \sigma}(\rvec)$. Now we need to proceed to do the same for the left hand side of (\ref{eq:trans}). Since the system is assumed to be in steady state we can immediately set $\dfrac{\partial \tn_{\pvec \sigma}}{\partial t}  = 0$.  We systematically ignore the gradients of $\delta n_{\pvec \sigma}(\rvec)$. Then with $\uvec = 0$ and no polarization gradient we have:
\begin{align}
\bm{\nabla} \tn_{\pvec \sigma}  &\simeq \dfrac{\partial n_{\pvec \sigma} }{\partial \epsilon_{\pvec \sigma}} \bm{\nabla} {\te}_{\pvec \sigma} + \dfrac{\partial n_{\pvec \sigma} }{\partial T}  \bm{\nabla} T(\rvec)  \,, \\
 \bm{\nabla}_{\pvec} \tn_{\pvec \sigma}  &\simeq \dfrac{\partial n_{\pvec \sigma} }{\partial \epsilon_{\pvec \sigma}}  \bm{\nabla}_{\pvec} \te_{\pvec \sigma} \,,
\end{align}
and thus the left hand side of (\ref{eq:IW}) becomes: $\dfrac{\partial n_{\pvec \sigma} }{\partial T}  \bm{\nabla} T(\rvec) \cdot \vvec_{\pvec \sigma} = \beta \left( \epsilon_{1} - \mu_{1}  \right) \dfrac{n_{1} \nbar_{1}}{T} (\bm{\nabla} T) \cdot \vvec_{1}$.  To lowest order $\bm{\nabla}_{\pvec} \te_{\pvec \sigma} \approx \bm{\nabla}_{\pvec} \epsilon_{\pvec \sigma} =  \vvec_{\pvec \sigma}$ is the Fermi velocity. 

We note that we have omitted a $\nabla \mu$ contribution in the expansion. This term, proportional to the entropy, was shown by Sykes and Brooker to contribute only to the even part of $q$, and that this is of higher order than the odd part. Below we shall point out that our $q_{\sigma}(x)$'s have odd symmetry.

Using (\ref{eq:zeta}) and (\ref{eq:deltankT}) we find:
\begin{equation} \label{eq:qdef}
\zeta_{i} = \vvec_{i} \cdot \bm{\nabla}(k_{B}T) q_{\sigma}(\epsilon_{i}) \,.
\end{equation}
The kinetic equation can then be written:
\begin{equation} \label{eq:Iq}
\begin{split}
 \beta &\left( \epsilon_{1} - \mu_{1}  \right) n_{1} \nbar_{1} (\vvec_{1} \cdot \bm{\nabla} T )  =  \sum_{\pvec_{2}, \pvec_{3},\pvec_{4}, \sigone'}  \delta_{\pvec_{1} + \pvec_{2}, \pvec_{3}+\pvec_{4}} 
\delta(\epsilon_{1} + \epsilon_{2}  - \epsilon_{3} - \epsilon_{4}) n_{1} n_{2} \nbar_{3} \nbar_{4} \\
\times &W^{\sigone \sigone'}(\theta)  
\left[  (\vvec_{1} \cdot \bm{\nabla} T) q_{\sigone}(\epsilon_{1}) +  (\vvec_{2} \cdot \bm{\nabla} T) q_{\sigone'}(\epsilon_{2})  
      -  (\vvec_{3} \cdot \bm{\nabla} T) q_{\sigone}(\epsilon_{3}) -  (\vvec_{4} \cdot \bm{\nabla} T) q_{\sigone'}(\epsilon_{4}) \right] \,,
\end{split}
\end{equation}
where it is understood that in the case of spin parallel scattering the range of the $\theta_{3}$ integration is restricted from $0 \text{ to } \pi$.  It is straightforward to show that the angle between $\vvec_{1}$ and $\bm{\nabla} T$ is arbitrary.  Define $\gamma$ such that $\vvec_{1} \cdot \bm{\nabla} T = v_{1} |\bm{\nabla} T| \cos{\gamma}$. Then similarly for the other quasiparticle velocities: $\vvec_{i} \cdot \bm{\nabla} T = v_{i} |\bm{\nabla} T| \cos{\theta_{i}} = v_{i} |\bm{\nabla} T| \cos{(\gamma + \theta_{i1})} =  v_{i} |\bm{\nabla} T|\left( \cos{\gamma} \cos{\theta_{i1}}   - \sin{\gamma} \sin{\theta_{i1}} \right) $, where $i = 2,3,4$. As a reminder, from Figs.~\ref{fig:forward} and \ref{fig:backward} we have $\theta_{21} \equiv \Phi_{2} = \theta$, $\theta_{31} \equiv \Phi_{3}$, and $\theta_{41} \equiv \Phi_{4}$.  We note for further reference that we will need the following expressions in our analysis of backward scattering at $0 < \pol <1$:
\begin{equation} \label{eq:cossin}
\begin{aligned}
\cos{\Phi_{3}} &= 1 - \frac{2 p_{2}^{2} \sin^{2}{\theta}}{\ell^{2}} \,,  & \qquad 
\sin{\Phi_{3}} &=   \sin{\theta} \frac{2 p_{2} (p_{1} + p_{2}\cos{\theta}) }{\ell^{2}}  \,, \\ 
\cos{\Phi_{4}} &= \cos{\theta} + \frac{2 p_{1} p_{2} \sin^{2}{\theta}}{\ell^{2}} \,, & \qquad 
\sin{\Phi_{4}} &=  \sin{\theta} \frac{ (p_{2}^{2} - p_{1}^{2}) }{\ell^{2}} \,,
\end{aligned}
\end{equation}
where $\bm{\ell} \equiv \pvec_{1} + \pvec_{2}$ as defined in (\ref{eq:ell}).  The integral (\ref{eq:Iq}) is symmetric in $\theta $ with respect to the intervals $[0, \pi]$ and $[\pi, 2\pi]$.  Thus the terms containing $\sin{}\theta_{i1}$ must cancel out. From (\ref{eq:cossin}) it is clear that this is true for any value of $\theta_{3}$. 
The term $|\bm{\nabla} T| \cos{\gamma} $ cancels out from both sides. We then divide through by $v_{1}$ to obtain:
\begin{equation} \label{eq:KEq}
\begin{split}
 \beta \left( \epsilon_{1} - \mu_{1}  \right) n_{1} \nbar_{1}  &=  \sum_{\pvec_{2}, \pvec_{3},\pvec_{4}, \sigone'}  \delta_{\pvec_{1} + \pvec_{2}, \pvec_{3}+\pvec_{4}} 
\delta(\epsilon_{1} + \epsilon_{2}  - \epsilon_{3} - \epsilon_{4}) n_{1} n_{2} \nbar_{3} \nbar_{4} \\
\times &W^{\sigone \sigone'}(\theta)  
\left[   q_{\sigone}(\epsilon_{1}) +   \frac{v_{2}}{v_{1}}  \cos{\theta_{21}} q_{\sigone'}(\epsilon_{2}) \right. \\
      &\left.-    \cos{\theta_{31}} q_{\sigone}(\epsilon_{3}) -   \frac{v_{4}}{v_{1}}   \cos{\theta_{41}} q_{\sigone'}(\epsilon_{4}) \right] \,.
\end{split}
\end{equation}

 We now transform the momentum integrations into energy and angular integrals as was done for the collision time in the previous section:
\begin{multline} \label{eq:KEqeps}
 \beta \left( \epsilon_{1} - \mu_{1}  \right) n_{1} \nbar_{1}  =  \frac{2}{h^{4}} \int_{0}^{\infty} d\epsilon_{2} d\epsilon_{3} d\epsilon_{4}  \,
\delta(\epsilon_{1} + \epsilon_{2}  - \epsilon_{3} - \epsilon_{4})   n_{1} n_{2} \nbar_{3} \nbar_{4}   \\
\times \Big\{\frac{(\mstar_{\sigone})^{3}}{p_{\sigone}^{2}}  \int_{0}^{\pi} d\theta \frac{W^{\sigone \, \sigone}_{f}(\theta)  }{\sqrt{\sin^{2}{\theta} - (\frac{\xi_{3}}{\epsilon_{F \sigone}})^{2}}} 
\Big[   q_{\sigone}(\epsilon_{1}) +   q_{\sigone}(\epsilon_{2}) \cos{\theta}  
      -   q_{\sigone}(\epsilon_{3}) -  q_{\sigone}(\epsilon_{4}) \cos{\theta} \Big]     \\
+ \frac{\mstar _{\sigone}(\mstar_{-\sigone})^{2}}{p_{\sigone} p_{-\sigone}}  \int_{0}^{\pi} d\theta \frac{W^{\sigone \, -\sigone}_{f}(\theta)}{\sqrt{\sin^{2}{\theta} - (\frac{\xi_{3}}{\epsilon_{1 2}})}}  
\Big[   q_{\sigone}(\epsilon_{1}) +  \frac{v_{-\sigone}}{v_{\sigone}} q_{-\sigone}(\epsilon_{2}) \cos{\theta}  
      -   q_{\sigone}(\epsilon_{3}) - \frac{v_{-\sigone}}{v_{\sigone}} q_{-\sigone}(\epsilon_{4}) \cos{\theta} \Big]\Big\}  \\   
+ \frac{\mstar _{\sigone}(\mstar_{-\sigone})^{2}}{p_{\sigone} p_{-\sigone}}  \int_{0}^{\pi} d\theta  \frac{W^{\sigone \, -\sigone}_{b}(\theta)   }{\sqrt{\sin^{2}{\theta} - (\frac{\xi_{3}}{\epsilon_{1 2}})}}  
\Big[   q_{\sigone}(\epsilon_{1}) +  \frac{v_{-\sigone}}{v_{\sigone}} q_{-\sigone}(\epsilon_{2}) \cos{\theta}  
      -   q_{\sigone}(\epsilon_{3}) \cos{\Phi_{3}} - \frac{v_{-\sigone}}{v_{\sigone}} q_{-\sigone}(\epsilon_{4}) \cos{\Phi_{4}} \Big] \Big\}    \,,
\end{multline}
where we have used the symmetry in the integrals about $\theta = \pi$, and $\xi_{3}$ and $\epsilon_{12}$ are defined in Eqs.~(\ref{eq:xi3}) and (\ref{eq:eps12}), respectively. At this point we can show that to lowest order in temperature the terms in (\ref{eq:KEqeps}) that are proportional to $\cos{\theta}$ can be omitted. For either parallel or anti-parallel spins we have: 
\begin{align}
 \int_{0}^{\pi} d\theta \frac{W^{\sigma \sigma'}_{f,b}(\theta) \cos{\theta} }{\sqrt{\sin^{2}{\theta} - (\frac{\xi_{3}}{\epsilon_{F \sigone}})^{2}}} &\approx W_{f,b, \, 0}^{\sigma \sigma'} \int_{\Delta}^{\pi-\Delta} d\theta \frac{ \cos{\theta} }{\sin{\theta}} = 0  \,.
\end{align}
For the  spin parallel case: $\Delta = k_{B}T/\epsilon_{F \sigma} \ll 1$, and for the spin anti-parallel case:  $\Delta = \sqrt{k_{B}T/\epsilon_{12}} \ll 1$.  We note that from (\ref{eq:eps12}) the quantity $\epsilon_{12}$ depends explicitly on $\theta$:
\begin{equation}
\frac{1}{\epsilon_{1 2}} \equiv \frac{1}{\epsilon_{2}} - \frac{1}{\epsilon_{1}} + \frac{2 (\mstar_{2} - \mstar_{1})}{p_{1}p_{2}} \cos{\theta} \,.
\end{equation}
This expression would seem to introduce a problem at $\pol = 1$.  However at full polarization these terms are removed by the minority-spin Fermi velocities in (\ref{eq:KEqeps}), and make no contribution to the physics in that limit (see, for example Eq.~(\ref{eq:tauP1})). At finite polarizations $0 < \pol < 1$ the $\theta$ dependence in $\epsilon_{1 2}$ introduces different values of $\Delta$ at the lower and upper limits.  Nevertheless since these differences are of $O(1)$ or less we still have $\Delta = \sqrt{k_{B}T/\epsilon_{12}} \ll 1$ at both limits. 

Following the same line of argument as in three dimensions, we introduce dimensionless variables: $x_{i} \equiv \beta (\epsilon_{i} - \mu_{i})$. If we then let $x_{i} \rightarrow -x_{i}$, we now see that $q_{\sigma}(x_{i})$ is an \textit{odd} function of its argument: $q_{\sigma}(x_{i}) = - q_{\sigma}(-x_{i})$, in lowest order of temperature for the thermal conductivity.  If we let $x_{3} \rightarrow -x_{3}$ and $x_{4} \rightarrow -x_{4}$, we obtain:
\begin{multline} \label{eq:KEqx234}
 x_{1} n_{1} \nbar_{1}  =  \frac{2}{h^{4}} (k_{B}T)^{2} \int_{-\infty}^{\infty} dx_{2} dx_{3} dx_{4}  \,
\delta(x_{1} + x_{2}  + x_{3} + x_{4})   n_{1} n_{2} n_{3} n_{4}   \\
\times \Big\{ \Big[ \frac{(\mstar_{\sigone})^{3}}{p_{\sigone}^{2}}  \int_{0}^{\pi} d\theta \frac{W^{\sigone \sigone}_{f}(\theta)  }{\sqrt{\sin^{2}{\theta} - (\frac{\xi_{3}}{\epsilon_{F \sigone}})^{2}}} + \frac{\mstar _{\sigone}(\mstar_{-\sigone})^{2}}{p_{\sigone} p_{-\sigone}}  \int_{0}^{\pi} d\theta \frac{W^{\sigone -\sigone}_{f}(\theta)}{\sqrt{\sin^{2}{\theta} - (\frac{\xi_{3}}{\epsilon_{1 2}}) }}   \\ 
+ \frac{\mstar _{\sigone}(\mstar_{-\sigone})^{2}}{p_{\sigone} p_{-\sigone}}  \int_{0}^{\pi} d\theta  \frac{W^{\sigone -\sigone}_{b}(\theta)   }{\sqrt{\sin^{2}{\theta} - (\frac{\xi_{3}}{\epsilon_{1 2}}) }}  \Big]
\Big[   q_{\sigone}(x_{1}) +  q_{\sigone}(x_{3})  \Big]     \\
+ \frac{\mstar _{\sigone}(\mstar_{-\sigone})^{2}}{p_{\sigone} p_{-\sigone}}  \int_{0}^{\pi} d\theta  \frac{W^{\sigone -\sigone}_{b}(\theta)   }{\sqrt{\sin^{2}{\theta} - (\frac{\xi_{3}}{\epsilon_{1 2}}) }}  
\Big[   - \frac{2 p_{-\sigone}^{2} \sin^{2}{\theta}}{\ell^{2}} q_{\sigone}(x_{3}) + \frac{v_{-\sigone}}{v_{\sigone}} \frac{2 p_{-\sigone} p_{\sigone} \sin^{2}{\theta}}{\ell^{2}} q_{-\sigone}(x_{4})  \Big] \Big\}    \,.
\end{multline}
At this point we can relabel the variables $x_{3}, x_{4} \rightarrow x_{2}$ due to symmetry under the integral. Further, we note from Eq.~(\ref{eq:tauP}) that the first set of angular integrals is just the collision time:
\begin{align} \label{eq:tausigdef}
\frac{4}{\pi^{2}} \frac{1}{\tau_{\sigma}} =   \frac{2}{h^{4}} (k_{B}T)^{2} \Big[ &\frac{(\mstar_{\sigma})^{3}}{p_{\sigma}^{2}}  \int_{0}^{\pi} d\theta \frac{W^{\sigma \sigma}_{f}(\theta)  }{\sqrt{\sin^{2}{\theta} - (\frac{\xi_{3}}{\epsilon_{F \sigma}})^{2}}} + \frac{\mstar _{\sigma}(\mstar_{-\sigma})^{2}}{p_{\sigma} p_{-\sigma}}  \int_{0}^{\pi} d\theta \frac{W^{\sigma -\sigma}_{f}(\theta)}{\sqrt{\sin^{2}{\theta} - (\frac{\xi_{3}}{\epsilon_{1 2}}) }} \nonumber  \\ 
+ &\frac{\mstar _{\sigma}(\mstar_{-\sigma})^{2}}{p_{\sigma} p_{-\sigma}}  \int_{0}^{\pi} d\theta  \frac{W^{\sigma -\sigma}_{b}(\theta)   }{\sqrt{\sin^{2}{\theta} - (\frac{\xi_{3}}{\epsilon_{1 2}}) }} \Big]  \,.
\end{align}
We follow MM (see Sec.~{\ref{sec:diff}} below) and introduce generalized frequencies:
\begin{subequations} \label{eq:nuKappa}
\begin{align}
\frac{4}{\pi^{2}} \nu_{\sigma} &\equiv  \frac{2}{h^{4}} (k_{B}T)^{2} \frac{\mstar _{\sigma}(\mstar_{-\sigma})^{2}}{p_{\sigma} p_{-\sigma}}  \int_{0}^{\pi} d\theta  \frac{W^{\sigma -\sigma}_{b}(\theta)   }{\sqrt{\sin^{2}{\theta} - (\frac{\xi_{3}}{\epsilon_{1 2}}) }}  
\frac{2 p_{-\sigma}^{2} \sin^{2}{\theta}}{\ell^{2}}  \,, \\
\frac{4}{\pi^{2}} \left( \frac{p_{-\sigma}}{p_\sigma} \right)^{2} \nu_{-\sigma}  &\equiv \frac{2}{h^{4}} (k_{B}T)^{2} \frac{\mstar _{\sigma}(\mstar_{-\sigma})^{2}}{p_{\sigma} p_{-\sigma}}  \int_{0}^{\pi} d\theta  \frac{W^{\sigma -\sigma}_{b}(\theta)   }{\sqrt{\sin^{2}{\theta} - (\frac{\xi_{3}}{\epsilon_{1 2}}) }}  
\Big[ \frac{v_{-\sigma}}{v_{\sigma}} \frac{2 p_{\sigma}p_{-\sigma} \sin^{2}{\theta}}{\ell^{2}} \Big] \,.
\end{align}
\end{subequations}
For simplicity in notation we have switched from $\sigone$ to $\sigma$. Then the kinetic equation becomes:
\begin{align} 
 x_{1} n_{1} \nbar_{1}  =  &\int_{-\infty}^{\infty} dx_{2} dx_{3} dx_{4} \delta(x_{1} + x_{2}  + x_{3} + x_{4})   n_{1} n_{2} n_{3} n_{4}   \nonumber \\
&\Big\{\frac{4}{\pi^{2}} \frac{1}{\tau_{\sigma}} \Big[   q_{\sigma}(x_{1}) +  q_{\sigma}(x_{2})  \Big] - \frac{4}{\pi^{2}} \nu_{\sigma}q_{\sigma}(x_{2})
+ \frac{4}{\pi^{2}} \bigg( \frac{\nbar_{-\sigma}}{\nbar_{\sigma}} \bigg) \nu_{-\sigma} q_{-\sigma}(x_{2}) \Big\} \,.
\end{align}
In the third term in curly brackets we have introduced the notation $\nbar_{\sigma}$ for the \textit{areal density} of the $\sigma^{th}$ Fermi sea, and this should not be confused with the similar looking Fermi distribution function (\ref{eq:nbar}) whose subscript is a momentum label and not a Fermi sea label, thus:  
\begin{equation} \label{eq:ndens}
\nbar_{\sigma} \equiv \frac{N_{\sigma}}{A} \,.
\end{equation}
The energy integrals can be found in Appendix A of Baym and Pethick:~\cite{BP1991}
\begin{align} {\label{eq:BP}}
\int_{-\infty}^{\infty} \,dx_{3} \,dx_{4} \,\delta(x_{1} + x_{2}  + x_{3} + x_{4}) \, n_{3} n_{4}  &= \frac{x_{1}+x_{2}}{1 - e^{-(x_{1}+x_{2})}} \,, \\
\int_{-\infty}^{\infty} \,dx_{2} \,dx_{3} \,dx_{4} \,\delta(x_{1} + x_{2}  + x_{3} + x_{4}) \, n_{2} n_{3} n_{4} &= \hlf \frac{x_{1}^{2}+\pi^{2}}{1 + e^{-x_{1}}} \,.
\end{align}

Thus, the kinetic equation can be brought into non-diagonal Sykes-Brooker form:
\begin{equation}  \label{eq:KEThmixed}
\begin{split}
\left( \frac{\pi^{2}}{4} \right) \tau_{\sigma} x_{1}
=  \int_{-\infty}^{\infty} dx_{2} \, K(x_{1}, x_{2}) \Big[ q_{\sigma}(x_{1}) -  \left( 1 - \nu_{\sigma} \tau_{\sigma} \right) q_{\sigma}(x_{2})  - \bigg( \frac{\nbar_{-\sigma}}{\nbar_{\sigma}} \nu_{-\sigma} \tau_{\sigma} \bigg)  q_{-\sigma}(x_{2}) \Big]    \,,
\end{split}
\end{equation}
where the Sykes-Brooker kernel is defined as:
\begin{equation} 
K(x_{1}, x_{2})  = \dfrac{(1+e^{-x_{1}})(x_{2}-x_{1})}{(1+e^{-x_{2}})(e^{x_{2}-x_{1}}-1)} \label{eq:Kernel}\,.
\end{equation}

This kinetic equation mixes the two components of $q_{\sigma}$. It is in very similar form as the kinetic equation for spin diffusion  as derived by MM.  In Sec.~\ref{sec:diff} we shall briefly write down the relevant expressions for the sake of comparison.  We rewrite (\ref{eq:KEThmixed}) with a matrix representation of the coefficients of  $q_{\sigma}(x_{2})$:
\begin{equation}  \label{eq:KEThMat}
\begin{split}
\left( \frac{\pi^{2}}{4} \right) \tau_{\sigma} x_{1} 
=  \int_{-\infty}^{\infty} dx_{2} \, K(x_{1}, x_{2}) \Big[ q_{\sigma}(x_{1}) -  \sum_{\sigma'} \lambda_{\sigma \sigma'} q_{\sigma'}(x_{2})   \Big]    \,,
\end{split}
\end{equation}
 The coefficient matrix is given by: 
\begin{equation}  \label{eq:lamkapmat}
\lambda = \begin{pmatrix}
1 - \nu_{\ua}\tau_{\ua} & \frac{\nbar_{\da}}{\nbar_{\ua}} \nu_{\da}\tau_{\ua} \\
\frac{\nbar_{\ua}}{\nbar_{\da}} \nu_{\ua}\tau_{\da}     & 1 - \nu_{\da}\tau_{\da}
\end{pmatrix} \,.
\end{equation}
The matrix diagonalization proceeds by using the general method described in Anderson, Pethick, and Quader.~\cite{APQ1987} We note that in this case $\lambda$ is not symmetric. The eigenvalues of $\lambda$ are:
\begin{equation} \label{eq:lambdaKappa}
\lambda_{+} = 1\,, \qquad \lambda_{-} = 1 - \left(  \nu_{\ua}\tau_{\ua} +  \nu_{\da}\tau_{\da} \right) \,.
\end{equation}
The $\pm$ subscripts on the $\lambda$'s refer to the plus and minus roots of the quadratic equation generated by diagonalizing (\ref{eq:lamkapmat}). In a spinor sense, plus and minus label the top and bottom rotated spin state, respectively.

We introduce transformed variables $\tilde{\tau} = S \tau$ and $\tilde{q} = Sq$ where the transformation matrix $S$ and its inverse are given by:
\begin{equation}
\begin{aligned}
&S = \frac{1}{\nu_{\ua}\tau_{\ua} +  \nu_{\da}\tau_{\da}} \begin{pmatrix} \frac{\tau_{\da}}{\nbar_{\da}} &\frac{\tau_{\ua}}{\nbar_{\ua}} \\ \frac{\nu_{\ua}}{\nbar_{\da}} & -\frac{\nu_{\da}}{\nbar_{\ua}} \end{pmatrix} \,, \\
&S^{-1} = \begin{pmatrix} \nbar_{\da}\nu_{\da} &\nbar_{\da}\tau_{\ua} \\ \nbar_{\ua}\nu_{\ua} &-\nbar_{\ua}\tau_{\da} \end{pmatrix} \,.
\end{aligned}
\end{equation}
We find:
\begin{align}
&\tilde{\tau} =  \frac{1}{\nu_{\ua}\tau_{\ua} +  \nu_{\da}\tau_{\da}}  \begin{pmatrix} \tau_{\ua}\tau_{\da} (\frac{1}{\nbar_{\da}} + \frac{1}{\nbar_{\ua}}) \\ \frac{\tau_{\ua}\nu_{\ua}}{\nbar_{\da}}  -\frac{\tau_{\da}\nu_{\da}}{\nbar_{\ua}} \end{pmatrix} \,, \\
&\tilde{q} = \frac{1}{\nu_{\ua}\tau_{\ua} +  \nu_{\da}\tau_{\da}}  \begin{pmatrix} \frac{\tau_{\da}q_{+}}{\nbar_{\da}} + \frac{\tau_{\ua}q_{-}}{\nbar_{\ua}} \\ \frac{\nu_{\ua}q_{+}}{\nbar_{\da}}  - \frac{\nu_{\da}q_{-}}{\nbar_{\ua}} \end{pmatrix} \label{eq:KEThqtil} \,.
\end{align}
In terms of these variables the diagonalized pair of kinetic equations are:
\begin{align} 
\frac{\pi^{2}}{4} \tilde{\tau}_{\sigma} x_{1} &= \int_{-\infty}^{\infty} dx_{2} \, K(x_{1}, x_{2}) \Big[ \tilde{q}_{\sigma}(x_{1}) - \lambda_{\sigma}\tilde{q}_{\sigma}(x_{2})   \Big] \,.
\end{align}

As shown in Eq.~(\ref{eq:KapDef}), the important quantity is not $\tilde{q}_{\sigma}$ itself but rather the integrated quantities:
\begin{equation} \label{eq:KEThQtil}
\tilde{Q}_{\sigma} \equiv -\int_{-\infty}^{\infty} dx \frac{\partial n}{\partial x} \tilde{q}_{\sigma}(x) x\,.
\end{equation}
From Sykes and Brooker Eq.~(60) the solutions to the diagonalized problem can be written:
\begin{align}
&\tilde{Q}_{\sigma} = \dfrac{\pi^{2}\tilde{\tau_{\sigma}}}{2(3-\lambda_{\sigma})} H(\lambda_{\sigma}) \,,
\end{align}
where $H(\lambda)$ is an infinite series involving the eigenvalues $\lambda_{\pm}$ that will be explicitly written below. We substitute (\ref{eq:KEThqtil}) into (\ref{eq:KEThQtil}) and then inverse-transform using:
\begin{equation} \label{eq:QsigKappa}
\begin{pmatrix} Q_{\ua} \\Q_{\da} \end{pmatrix} =
\begin{pmatrix} \nbar_{\da}\nu_{\da} \tilde{Q}_{+} + \nbar_{\da}\tau_{\ua} \tilde{Q}_{-} \\ \nbar_{\ua}\nu_{\ua} \tilde{Q}_{+} - \nbar_{\ua} \tau_{\da} \tilde{Q}_{-} \end{pmatrix} \,.
\end{equation}
The exact solution for the low-temperature thermal conductivity in two dimensions can be written:
\begin{align} \label{eq:kappafinal}
\kappa = \frac{\pi k_{B}^{2}}{h^{2}} T   \sum_{\sigma} m_{\sigma}^{*} v_{F \sigma}^{2} Q_{\sigma}  \,, 
\end{align}
where
\begin{subequations} \label{eq:KThQpm}
\begin{align}
Q_{\ua} &= \frac{1}{\nu_{\ua}\tau_{\ua} +  \nu_{\da}\tau_{\da}} \frac{\pi^{2}}{2} 
\Big[ \nu_{\da}\tau_{\da}(1 + (\nbar_{\da}/\nbar_{\ua}))\dfrac{H(\lambda_{+})}{3 - \lambda_{+}}   + (\nu_{\ua}\tau_{\ua} - (\nbar_{\da}/\nbar_{\ua})\nu_{\da}\tau_{\da})\dfrac{H(\lambda_{-})}{3 - \lambda_{-}} \Big] \tau_{\ua} \,, \\
Q_{\da} &= \frac{1}{\nu_{\ua}\tau_{\ua} +  \nu_{\da}\tau_{\da}} \frac{\pi^{2}}{2} 
\Big[\nu_{\ua}\tau_{\ua}(1 + (\nbar_{\ua}/\nbar_{\da}))\dfrac{H(\lambda_{+})}{3 - \lambda_{+}}   + (\nu_{\da}\tau_{\da} - (\nbar_{\ua}/\nbar_{\da})\nu_{\ua}\tau_{\ua})\dfrac{H(\lambda_{-}) }{3 - \lambda_{-}}\Big] \tau_{\da} \,,
\end{align}
\end{subequations}
together with the explicit expression for $H(\lambda)$:
\begin{equation} 
H(\lambda) = \frac{3 - \lambda}{4} \sum_{n=0}^{\infty} \dfrac{4n + 5}{(n+1)(2n+3)[(n+1)(2n+3) - \lambda]}  \,.
\end{equation}
According to Eqs.~(\ref{eq:tauP}) and (\ref{eq:nuKappa}) we find $\nu_{\sigma} \tau_{\sigma} \sim 1/\ln{T}$. Thus, at very low temperatures we can set $\lambda_{-} \approx \lambda_{+} = 1$. In this limit then Eqs.~(\ref{eq:KThQpm}) simplify to $Q_{\sigma} = \frac{\pi^{2}}{4} H(1) \tau_{\sigma} \,,$
and the thermal conductivity becomes:
\begin{equation} \label{eq:KappaLowT}
\kappa = \frac{\pi^{3} }{2 h^{2}} k_{B}^{2} T H(1)  \sum_{\sigma} \epsilon_{F \sigma} \tau_{\sigma} \,.
\end{equation}

\subsubsection{Zero polarization}
At zero polarization the eigenvalues are $\lambda_{+} = 1 \text{ and } \lambda_{-} = 1 - 2\nu \tau_{0}$. By inspection of (\ref{eq:KThQpm}) we have  $Q_{\ua} = Q_{\da} = (\pi^{2}/4)  H(1) \tau_{0}$ and therefore:
\begin{equation} \label{eq:kappaP0}
\kappa(\pol = 0)= \frac{\pi^{3}}{h^{2}}  k_{B}^{2}T  \epsilon_{F}  H(1) \tau_{0}  \,,
\end{equation}
in agreement with (\ref{eq:KappaLowT}).  In the $\ell = 0$ approximation this becomes:
\begin{equation}
\kappa(\pol =0) = \pi h^{2} k_{B} \frac{v_{F}^{2}}{m^{*}} \frac{H(1)}{ \left[W_{f, 0}^{\ua \, \ua} + W_{f, 0}^{\ua \, \da} + W_{b, 0}^{\ua \, \da} \right]} \frac{(\epsilon_{F}/k_{B} T)}{\ln{\left(2 \epsilon_{F} /k_{B} T \right) }}
\end{equation}

\subsubsection{Full polarization}
At full polarization all quasiparticles are in the $\ua$ Fermi sea, and thus the terms with the spin anti-parallel contributions $\nu_{\ua} \text{ and } \nu_{\da}$ do not appear. By inspection of the kinetic equation (\ref{eq:KEThMat}) the eigenvalue $\lambda = 1 $.  From (\ref{eq:KThQpm}) we have  $Q_{\ua} = (\pi^{2} H(\lambda)/2(3 - \lambda)) \tau_{1}$ and therefore:
\begin{align} \label{eq:kappaP1}
\kappa(\pol = 1) =  \frac{\pi^{3}}{2 h^{2}}  k_{B}^{2}T  \epsilon_{F \ua}  H(1)  \tau_{1}  \,,
\end{align}
in agreement with (\ref{eq:KappaLowT}). 

\subsubsection*{Summary}
We find that the temperature dependence for the thermal conductivity at arbitrary polarization $0 \leq \Pee \leq 1$ is $\kappa^{-1} \sim T \ln{T}$.  This is in agreement with the zero-polarization results of Fu and Ebner.~\cite{FuEbner1974}

\subsection{\label{sec:visco} Shear viscosity}

We consider a Fermi-liquid film flowing with speed $u_{x}$ in the $x$-direction. The flow is not uniform. There exists a small non-zero $y$-gradient of the velocity  $\partial u_{x} / \partial y$ that will drive a transverse momentum flux $\sigma_{x y}$. The coefficient of proportionality $\eta$ is the first or shear viscosity:
\begin{equation} \label{eq:eta}
\sigma_{x y} = \eta \frac{\partial u_{x}}{\partial y} \,.
\end{equation}
The stress tensor $\sigma_{x y}$ can be written in terms of the non-equilibrium part of the distribution function:~\cite{BP1991}
\begin{equation} \label{eq:sigxy}
\sigma_{x y} = - \sum_{\sigma} \int \frac{d \pvec}{h^{2}} \, p_{x} \left( \frac{\partial \epsilon_{\pvec \sigma}}{\partial p_{y}}  \right) \delta n_{\pvec \sigma} \,,
\end{equation}
we note that $v_{\pvec \sigma, y} = \partial \epsilon_{\pvec \sigma} / \partial p_{y}$.  Then using (\ref{eq:zeta}) we can write $\delta n_{\pvec \sigma}$ in terms of the driving field:
\begin{equation} \label{eq:etadn}
\delta n_{i} \equiv \frac{\partial n_{i}}{\partial \epsilon_{i}} \zeta_{i} = \frac{\partial n_{i}}{\partial \epsilon_{i}} \hlf \left(p_{ix} v_{iy}  + p_{iy} v_{ix} \right) \frac{\partial u_{x}}{\partial y} q_{\sigma} \, (\epsilon_{i})\,,
\end{equation}
where from symmetry: $p_{ix} v_{iy}  = p_{iy} v_{ix} = m^{*}_{i} v_{ix} v_{iy} $.  For the shear viscosity we will need the first two terms of (\ref{eq:etadn}) in powers of $(\epsilon_{\pvec \sigma} - \mu)$.  Following Sykes and Brooker we obtain:
\begin{equation}
\sigma_{x y} = - \Big[  \sum_{\sigma} \int \frac{d \pvec}{h^{2}} \, \left[ p_{x}   v_{\pvec \sigma, y} \right] \frac{\partial n_{\pvec \sigma}}{\partial \epsilon_{\pvec \sigma}} (  1 + \frac{x_{\sigma}}{\beta \epsilon_{F \sigma}} ) m^{*}_{\sigma} v_{\sigma, x} v_{\sigma, y} q_{\sigma}(\epsilon_{\pvec \sigma}) \Big]  \frac{\partial u_{x}}{\partial y} \,.
\end{equation}
where $x_{\sigma} = \beta (\epsilon_{\pvec \sigma} - \mu)$ was introduced before Eq.~(\ref{eq:taup0}).  The shear viscosity is then:
\begin{equation}
\eta = - \sum_{\sigma} (p_{F \sigma} v_{F \sigma})^{2}  \int \frac{d \pvec}{h^{2}} (  1 + \frac{x_{\sigma}}{\beta \epsilon_{F \sigma}} )  \cos^{2}{\theta}  \sin^{2}{\theta}   \frac{\partial n_{\pvec \sigma}}{\partial \epsilon_{\pvec \sigma}} q_{\sigma}(\epsilon_{\pvec \sigma}) \,. 
\end{equation}
We now perform the angular integration, and change integration variable from $p$ to $x$:
\begin{align} 
\eta = - \frac{\pi}{4 h^{2}} \sum_{\sigma} \frac{(p_{F \sigma})^{4}}{m_{\sigma}^{*}} \int_{-\infty}^{\infty} dx \, (  1 + \frac{x_{\sigma}}{\beta \epsilon_{F \sigma}} )  \frac{\partial n}{\partial x} q_{\sigma} (x) \label{eq:etaint}\,. 
\end{align}
We can resolve $q_{\sigma}(x)$ into symmetric and antisymmetric components: $q_{\sigma}(x) = q_{\sigma}^{(s)}(x) + q_{\sigma}^{(a)}(x)$.  Because of the even and odd symmetry of the two components of the integrand, we can write $\eta = \eta^{(s)} + \eta^{(a)}$ where:
\begin{subequations} \label{eq:etasa}
\begin{align} \label{eq:etas}
\eta^{(s)} &= - \frac{\pi}{4 h^{2}} \sum_{\sigma} \frac{(p_{F \sigma})^{4}}{m_{\sigma}^{*}} \int_{-\infty}^{\infty} dx \,  \frac{\partial n}{\partial x} q_{\sigma}^{(s)} (x)  \\ \label{eq:etaa}
\eta^{(a)} &= - \frac{\pi}{4 h^{2}} \sum_{\sigma} \frac{(p_{F \sigma})^{4}}{m_{\sigma}^{*}} \frac{1}{\beta \epsilon_{F \sigma}} \int_{-\infty}^{\infty} dx \, x   \frac{\partial n}{\partial x} q_{\sigma}^{(a)} (x) \,.
\end{align}
\end{subequations}
As in the previous section, the integrals over $q_{\sigma}$ can be evaluated exactly.~\cite{SykesBrooker1970} It is straightforward to show that $\eta^{(a)} /  \eta^{(s)} \sim O(T^{2})$, the same as in three dimensions, and thus we can ignore the contributions of the anti-symmetric part of $q_{\sigma}$ in the remainder of this discussion. 

In the absence of thermal or polarization gradients, the left hand side of the kinetic equation (\ref{eq:trans}) in leading order reduces to:~\cite{SykesBrooker1970} $ -  \left(\dfrac{\partial n_{\pvec \sigma}}{\partial \epsilon_{\pvec \sigma}} \right) \dfrac{1}{2} [ p_{x} v_{\pvec \sigma, y} +  p_{y} v_{\pvec \sigma, x}] \dfrac{\partial u_{x}}{\partial y}$.  Then with (\ref{eq:etadn}), the kinetic equation becomes:
\begin{equation} \label{eq:Iqkap}
\begin{split}
n_{1} \nbar_{1} &  v_{1y} p_{1x} =  \sum_{\pvec_{2}, \pvec_{3},\pvec_{4}}  \delta_{\pvec_{1} + \pvec_{2}, \pvec_{3}+\pvec_{4}} 
\delta(\epsilon_{1} + \epsilon_{2}  - \epsilon_{3} - \epsilon_{4}) n_{1} n_{2} \nbar_{3} \nbar_{4} \\
&\times \Big\{  W^{\sigone \sigone}_{f}(\theta)  
\left[  v_{1y} p_{1x}  q_{\sigone}(\epsilon_{1}) +  v_{2y} p_{2x}  q_{\sigone}(\epsilon_{2})  
     -   v_{3y} p_{3x}  q_{\sigone}(\epsilon_{3}) -   v_{4y} p_{4x}  q_{\sigone}(\epsilon_{4}) \right] \\
&+  W^{\sigone -\sigone}_{f}(\theta)  
\left[  v_{1y} p_{1x}  q_{\sigone}(\epsilon_{1}) +   v_{2y} p_{2x}  q_{-\sigone}(\epsilon_{2})  
     -   v_{3y} p_{3x}  q_{\sigone}(\epsilon_{3}) -  v_{4y} p_{4x} q_{-\sigone}(\epsilon_{4}) \right] \\
&+W^{\sigone -\sigone}_{b}(\theta) 
\left[  v_{1y} p_{1x}  q_{\sigone}(\epsilon_{1}) +  v_{2y} p_{2x}  q_{-\sigone}(\epsilon_{2})  
     -   v_{3y} p_{3x}  q_{\sigone}(\epsilon_{3}) -   v_{4y} p_{4x} q_{-\sigone}(\epsilon_{4}) \right] \Big\} \,,
\end{split}
\end{equation}
where we have canceled out a common factor of $\partial u_{x} / \partial y$.  This is similar to the kinetic equation for the thermal conductivity (\ref{eq:Iq}) except that in this system both the $x$ and $y$ directions play special roles. Thus we need to include information as to the directions of the momenta with respect to the $x$-direction, say. We introduce angle $\gamma$ which is the angle between $\pvec_{1}$ and the $x$-axis: $\pvec_{1} \cdot \bm{\hat{x}} \equiv p_{1} \cos{\gamma}$. 

At this point we need to emphasize a key difference between the thermal conductivity calculation, and that of the shear viscosity.   For the shear viscosity we need to permit the incoming and outgoing quasiparticle momenta to differ slightly from the zero-temperature $p_{F}$'s in order to obtain sensible results. Of course, energy and momentum still must be conserved in quasiparticle collisions.  This situation is also discussed by Novikov~\cite{Novikov06} in his treatment of the shear viscosity for a two-dimensional fermion system.

\begin{figure}[]
\includegraphics[trim = 0cm 24cm 14cm 0cm, clip]{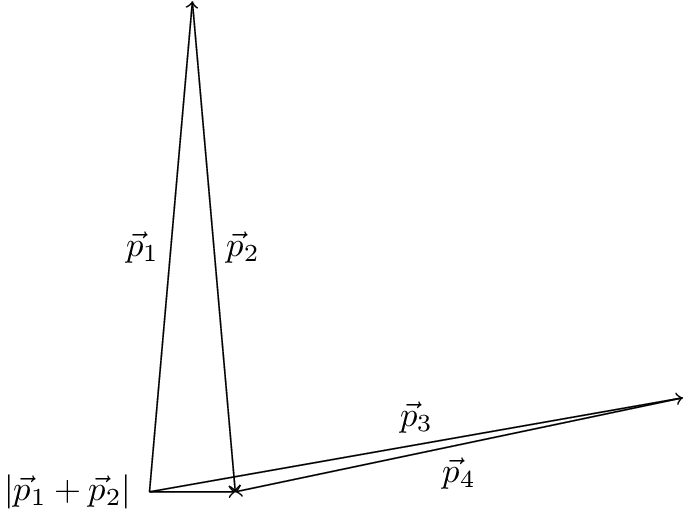}
\caption{\label{fig:head-on} Momentum space diagram illustrating the effects of letting the quasiparticle momenta move slightly off of the zero-temperature value $p_{F}$  for the backwards scattering case.  In this zero-polarization example we set $p_{1} = p_{2} = p_{F}$ as discussed in the text.  It is clear that $\Phi_{3} \neq \theta$, and also since $\pvec_{1}$ and $\pvec_{4}$ are not parallel $\Phi_{4} \neq 0$. }
\vspace{1truein}
\end{figure}

In principal all four momenta can be unequal to the zero-temperature $p_{F i}, i = 1,2,3,4$.  In our model however we shall fix $p_{1} = p_{F 1}$ and $ p_{2} = p_{F 2}$, and only permit $p_{3}$ and $p_{4}$ to differ from their zero-temperature values.  This simplifies the calculation while still maintaining the essential features of the the effect of finite temperature.  In addition, this maintains consistency with the MM treatment of the divergence in the integral over $\theta$ in the kinetic equation as discussed in Sec.~\ref{sec:tau}.  In Fig.~\ref{fig:head-on} we illustrate the sort of scattering process that can yield a nonzero value from the phase space integral that appears in the kinetic equation.
In terms of the angular variables, Eq.~(\ref{eq:Iqkap}) becomes:
\begin{equation} \label{eq:Igamkap}
\begin{split}
n_{1} \nbar_{1}  \sin{\gamma} \cos{\gamma} &=  \sum_{\pvec_{2}, \pvec_{3},\pvec_{4}}  \delta_{\pvec_{1} + \pvec_{2}, \pvec_{3}+\pvec_{4}} 
\delta(\epsilon_{1} + \epsilon_{2}  - \epsilon_{3} - \epsilon_{4}) n_{1} n_{2} \nbar_{3} \nbar_{4} \\
&\times \Bigg\{     W^{\sigone \, \sigone}_{f}(\theta) 
\Big[  \sin{\gamma} \cos{\gamma}  q_{\sigone}(\epsilon_{1}) +  \sin{(\gamma+\theta)} \cos{(\gamma+\theta)} q_{\sigone}(\epsilon_{2})   \\
         &\phantom{+ W^{\sigone \, \sigone}_{f}(\theta)  } -    \sin{(\gamma + \Phi_{3}^{f})} \cos{(\gamma + \Phi_{3}^{f})} q_{\sigone}(\epsilon_{3}) -   \sin{(\gamma + \Phi_{4}^{f})} \cos{(\gamma + \Phi_{4}^{f})} q_{\sigone}(\epsilon_{4}) \Big] \\
&+  W^{\sigone \, -\sigone}_{f}(\theta)  
\Big[ \sin{\gamma} \cos{\gamma}  q_{\sigone}(\epsilon_{1}) +  D_{\sigone} \sin{(\gamma+\theta)} \cos{(\gamma+\theta)}   q_{-\sigone}(\epsilon_{2})  \\
&\phantom{+  W^{\sigone \, -\sigone}_{f}(\theta) }     -   \sin{(\gamma + \Phi_{3}^{f})} \cos{(\gamma + \Phi_{3}^{f})}  q_{\sigone}(\epsilon_{3}) -  D_{\sigone} \sin{(\gamma + \Phi_{4}^{f})} \cos{(\gamma + \Phi_{4}^{f})} q_{-\sigone}(\epsilon_{4}) \Big] \\
&+W^{\sigone \, -\sigone}_{b}(\theta) 
\Big[  \sin{\gamma} \cos{\gamma}  q_{\sigone}(\epsilon_{1}) +  D_{\sigone} \sin{(\gamma+\theta)} \cos{(\gamma+\theta)}  q_{-\sigone}(\epsilon_{2})  \\
&\phantom{+  W^{\sigone \, \sigone}_{f}(\theta) }       -   \sin{(\gamma + \Phi_{3}^{b})} \cos{(\gamma  + \Phi_{3}^{b})}  q_{\sigone}(\epsilon_{3})  - D_{\sigone} 
\sin{(\gamma + \Phi_{4}^{b})} \cos{(\gamma  + \Phi_{4}^{b})} q_{-\sigone}(\epsilon_{4}) \Big] 
\Bigg\} \,,
\end{split}
\end{equation}
where for convenience  we have defined $D_{\sigma} \equiv (v_{-\sigma} p_{-\sigma}) / (v_{\sigma} p_{\sigma}) $.  

Expanding the trigonometric functions,  we obtain terms like $\sin\gamma \cos \gamma \cos(2\theta)$ and $\sin \theta \cos \theta \cos(2\gamma)$. The terms proportional to $\sin\theta$ give no contribution to the summation due to invariance under the transformation $\theta \rightarrow -\theta$. The right hand side of the equation can be  cast into a product of energy and angular integrals as was done in the previous section, and (\ref{eq:Igamkap}) becomes:
\begin{equation} \label{eq:KEetaqeps}
\begin{split}
n_{1} \nbar_{1}  &=   \left( k_{B} T  \right)^{2}  \frac{2}{h^{4}} \int_{-\infty}^{\infty} dx_{2} dx_{3} dx_{4}  \,
\delta(x_{1} + x_{2}  - x_{3} - x_{4})   n_{1} n_{2} \nbar_{3} \nbar_{4}   \\
&\times \Bigg\{\frac{(\mstar_{\sigone})^{3}}{p_{\sigone}^{2}}  \int_{0}^{\pi} d\theta \frac{W^{\sigone \, \sigone}_{f}(\theta)  }{\sqrt{\sin^{2}{\theta} - (\frac{\xi_{3}}{\epsilon_{F \sigone}})^{2}}} 
\Big[   q_{\sigone}(x_{1}) +   q_{\sigone}(x_{2}) \cos{(2 \theta)}  \\
&\phantom{\times \Bigg\{ \frac{(\mstar_{\sigone})^{3}}{p_{\sigone}^{2}}  \int_{0}^{\pi} d\theta \frac{W^{\sigone \, \sigone}_{f}(\theta)  }{\sqrt{\sin^{2}{\theta} - (\frac{\xi_{3}}{\epsilon_{F \sigone}})^{2}}} } -   q_{\sigone}(x_{3})\cos{(2 \Phi_{3}^{f})} -  q_{\sigone}(x_{4}) \cos{(2 \Phi_{4}^{f})} \Big]     \\
&+ \frac{\mstar _{\sigone}(\mstar_{-\sigone})^{2}}{p_{\sigone} p_{-\sigone}}  \int_{0}^{\pi} d\theta \frac{W^{\sigone \, -\sigone}_{f}(\theta)}{\sqrt{\sin^{2}{\theta} - (\frac{\xi_{3}}{\epsilon_{1 2}})}}  
\Big[   q_{\sigone}(x_{1}) +  D_{\sigone} q_{-\sigone}(x_{2}) \cos{(2 \theta)}  \\
&\phantom{+ \frac{\mstar _{\sigone}(\mstar_{-\sigone})^{2}}{p_{\sigone} p_{-\sigone}}  \int_{0}^{\pi} d\theta \frac{W^{\sigone \, -\sigone}_{f}(\theta)}{\sqrt{\sin^{2}{\theta} - (\frac{\xi_{3}}{\epsilon_{1 2}})}}}  -   q_{\sigone}(x_{3})\cos{(2 \Phi_{3}^{f})} - D_{\sigone} q_{-\sigone}(x_{4}) \cos{(2 \Phi_{4}^{f})} \Big]   \\   
&+ \frac{\mstar _{\sigone}(\mstar_{-\sigone})^{2}}{p_{\sigone} p_{-\sigone}}  \int_{0}^{\pi} d\theta  \frac{W^{\sigone \, -\sigone}_{b}(\theta)   }{\sqrt{\sin^{2}{\theta} - (\frac{\xi_{3}}{\epsilon_{1 2}})}}  
\Big[   q_{\sigone}(x_{1}) +  D_{\sigone} q_{-\sigone}(x_{2}) \cos{(2 \theta)}  \\
&\phantom{+ \frac{\mstar _{\sigone}(\mstar_{-\sigone})^{2}}{p_{\sigone} p_{-\sigone}}  \int_{0}^{\pi} d\theta  \frac{W^{\sigone \, -\sigone}_{b}(\theta)   }{\sqrt{\sin^{2}{\theta} - (\frac{\xi_{3}}{\epsilon_{1 2}})}} } -   q_{\sigone}(x_{3}) \cos{(2 \Phi_{3}^{b})} - D_{\sigone} q_{-\sigone}(x_{4}) \cos{(2 \Phi_{4}^{b})} \Big] \Bigg\}    \,.
\end{split}
\end{equation}

The second and third integrals in Eq. (\ref{eq:KEetaqeps}) describe forward and backward scattering between spin anti-parallel quasiparticles, and their exact forms depend on polarization. Appropriate expressions will be derived below where we will discuss three polarization ranges separately.  At this point we can evaluate the contribution from spin-parallel forward scattering.  The calculation  of $\cos{\Phi_{3}^{f}}$ and $\cos{\Phi_{4}^{f}}$  begins by determining $\cos{\theta_{3}}$ and $\cos{\theta_{4}}$. From Eq.~(\ref{eq:p4sq}):
\begin{align} \label{eq:costh3}
\cos{\theta_{3}} &= \frac{|\pvec_{1} + \pvec_{2}|^{2} + p_{3}^{2} - p_{4}^{2}}{2 p_{3} |\pvec_{1} + \pvec_{2}|} \,, \nonumber \\
&= \frac{|\pvec_{1} + \pvec_{2}|}{2 p_{3}} + \frac{ p_{3}^{2} - p_{4}^{2}}{2 p_{3} |\pvec_{1} + \pvec_{2}|} \,.
\end{align}
We note that the factor of $|\pvec_{1} + \pvec_{2}|$ in the denominator of (\ref{eq:costh3}) is the source of the enhancement of the role of head-on scattering as will be  discussed below.  

We shall begin with the spin-parallel channel.  Thus as per the above discussion we can set $p_{1} = p_{2} = p_{F}$.  From (\ref{eq:xi3}) we have the following definition:
\begin{equation} \label{eq:xi3T}
\xi_{3} = \dfrac{p_{3}^{2} - p_{F}^{2}}{2 m^{\ast}} \,.
\end{equation}
Using conservation of energy $2p_{F}^{2} = p_{3}^{2} + p_{4}^{2}$ we identify:
\begin{equation}
p_{3}^{2} - p_{4}^{2} = 4 m^{\ast} \xi_{3} \,.
\end{equation}
Thus since $ |\pvec_{1} + \pvec_{2}| = 2 p_{F} \cos{(\frac{\theta}{2})}$, we can write to lowest order in $T$:
\begin{equation} \label{eq:cos3}
\cos(\theta_{3}) = \cos{(\tfrac{\theta}{2})} + \dfrac{\Delta_{3}}{2 \cos{(\frac{\theta}{2}})}\,,
\end{equation}
where we have defined $\Delta_{3} \equiv \xi_{3}/\epsilon_{F}$. Then exchanging $p_{3}$ and $p_{4}$ in (\ref{eq:costh3}) we have:
\begin{equation} \label{eq:cos4}
\cos(\theta_{4}) = \cos{(\tfrac{\theta}{2})} - \dfrac{\Delta_{3}}{2 \cos{(\frac{\theta}{2}})}\,.
\end{equation}
Eqs.~(\ref{eq:cos3}) and (\ref{eq:cos4}) can now be used to determine $\Phi_{3}$ and $\Phi_{4}$.  In the case of forward scattering between parallel-spin quasiparticles, $\Phi_3^f$ and $\Phi_4^f$ can be significantly different from $0$ and $\theta$ respectively only at $\theta \approx \pi $, head-on scattering. Thus using (\ref{eq:cos3}) and (\ref{eq:cos4}), one has
\begin{align}
\sin \Phi_3^f & = \sin(\theta_3 - \theta_1) \notag \,, \\
                    & \approx \sin \theta_1 \cos \theta_1 - \sin \theta_1 \cos \theta_3 \notag \,, \\
                    & = \frac{\tan{(\tfrac{\theta}{2})}\Delta_3}{2}     \,,                
\end{align}
and similarly
\begin{equation}
\sin \Phi_4^f \approx \sin\theta - \frac{\tan{(\tfrac{\theta}{2})}\Delta_3}{2}    \,.
\end{equation}
Therefore
\begin{equation} \label{CosPhi3}
\cos (2\Phi_3^f) \approx 1 - \frac{\tan^2\left(\frac{\theta}{2}\right)\Delta_3^2}{2} \,,
\end{equation}
and
\begin{equation} \label{CosPhi4}
\cos (2\Phi_4^f) \approx \cos(2\theta) - \frac{\tan^2\left(\frac{\theta}{2}\right)\Delta_3^2}{2} \,,
\end{equation}
where we have used $\theta_{1} = 2 \pi - \tfrac{\theta}{2}$.  Substitute Eqs. (\ref{CosPhi3}) and (\ref{CosPhi4}) into the first integral in (\ref{eq:KEetaqeps}), and one obtains:
\begin{equation} \label{eq:parafw}
\frac{ (m^{*}_{\sigma_1})^3 }{p_{\sigma_1}^2} \int_{\Delta}^{\pi - \Delta} d\theta \frac{W_f^{\sigma_1 \sigma_1}(\theta)}{\sin\theta}\left\{q_{\sigma_1}(x_1) - \left[ 1 - \tan^2\left(\frac{\theta}{2}\right)\Delta_3^2\right]q_{\sigma_1}(x_2)\right\}
\end{equation}
Note we have assumed the parity of $q_{\sigma}(x)$ and the change of variables have already been applied.

\subsubsection{$\pol = 0$ Zero Polarization }

At zero polarization, we set $p_{F \uparrow} = p_{F \downarrow}, m_{\ua}^{*} = m_{\da}^{*} $, and $D_{\sigma} = 1$.  The contribution from the forward scattering of spin-antiparallel quasiparicles is identical to that of the forward scattering between spin-parallel quasiparticles as analyzed in (\ref{eq:parafw}). Thus, the second integral in (\ref{eq:KEetaqeps}) is identical to (\ref{eq:parafw}) with the substitution $W_f^{\sigma \, \sigma} \rightarrow W_f^{\sigma \,  -\sigma}$. 

For the backward scattering between spin-antiparallel quasiparticles, one notices that $\theta_3 \approx 2\pi - \theta_1$ and $\theta_4 \approx \theta_1$ as $\theta$ is not close to $\pi$, therefore

\begin{align}
\sin \Phi_3^b & = \sin(\theta_3 - \theta_1) \notag \,, \\
                    & \approx -\sin \theta_1 \cos \theta_1 - \sin \theta_1 \cos \theta_3 \notag \,, \\
                    & = \sin\theta + \frac{\tan \left(\frac{\theta}{2}\right)\Delta_3}{2}  \,,                   
\end{align}
and
\begin{align}
\sin \Phi_4^b & = \sin(\theta_4 - \theta_1) \notag \,, \\
                    & \approx \sin \theta_1 \cos \theta_1 - \sin \theta_1 \cos \theta_4 \notag \,, \\
                    & = - \frac{\tan \left(\frac{\theta}{2}\right)\Delta_3}{2}      \,.                
\end{align}
Then
\begin{align}
\cos(2\Phi_3^b) & = 1 - 2\sin^2\Phi_3^b \notag \,, \\
                          & \approx \cos(2\theta) - \frac{\tan^2 \left( \frac{\theta}{2} \right) \Delta_3^2}{2} \,,
\end{align}
and
\begin{align}
\cos(2\Phi_4^b) & = 1 - 2\sin^2(\Phi_4^b) \notag \,, \\
                          & \approx 1 - \frac{\tan^2\left( \frac{\theta}{2} \right) \Delta_3^2}{2} \,.
\end{align}                                           
Substituting these back into the backward scattering integral,  we obtain for (\ref{eq:KEetaqeps}):
\begin{align}
n_1\bar{n}_1 & = (k_BT)^2 \frac{2}{h^4} \frac{(m^{*})^3}{p_F^2}\int_{-\infty}^{+\infty} dx_2 dx_3 dx_4 \delta \left( x_1 + x_2 + x_3 + x_4 \right) n_1 n_2 n_3 n_4 \notag \\
& \int_{\Delta}^{\pi - \Delta} d\theta \frac{W_f^{\sigma_1 \, \sigma_1}(\theta) + W_f^{\sigma_1 \, -\sigma_1}(\theta) + W_b^{\sigma_1 \, -\sigma_1}(\theta)}{\sin \theta} \left\{ q_{\sigma_1}(x_1) - \left[ 1 - \tan^2\left( \frac{\theta}{2} \right) \Delta_3^2 \right] q_{\sigma_1}(x_2) \right\} \,.
\end{align}

As before the equation can be cast into the form:
\begin{equation} \label{eq:P0}
\frac{\pi^2}{4} \tau_{0} n_1 \bar{n}_1 = \int_{-\infty}^{+\infty} dx_2 dx_3 dx_4 \delta \left( x_1 + x_2 + x_3 + x_4 \right) n_1 n_2 n_3 n_4 \bigg[ q (x_1) - \Big( 1 - \nu_{0} \tau_{0} \Big) q(x_2) \bigg] \,,
\end{equation}
where $\tau_{0}$ is given by (\ref{eq:tauP0}), and we have defined a generalized frequency:
\begin{equation}
\nu_{0} = \frac{\pi^2}{2h^4}(k_BT)^2\frac{(m^*)^3}{p_F^2}\int_{\Delta}^{\pi - \Delta}
d\theta \frac{W_f^{\ua  \ua}(\theta) + W_f^{\ua \da}(\theta) + W_b^{\ua \da}(\theta)}{\sin \theta} \tan^2 \left(\frac{\theta}{2} \right) \Delta_3^2 \,.
\end{equation}
Following the same steps as for the thermal conductivity, Eq. (\ref{eq:P0}) can be cast into Sykes-Brooker form:
\begin{equation} \label{eq:etaKEP0}
\frac{\pi^2}{4}\tau_{0} = \int_{-\infty}^{+\infty} dx_2 K(x_1, x_2) \left[ q(x_1) - \lambda q(x_2) \right] \,.
\end{equation}
The solution is then
\begin{align}
Q & \triangleq -\int_{-\infty}^{+\infty} dx \frac{\partial n}{\partial x} q(x) \notag  = \frac{c(\lambda)}{2\nu_{0}} \,,
\end{align}
with the eigenvalue: 
\begin{equation} \label{eq:lamP0}
\lambda = 1 - \nu_{0} \tau_{0} \,,
\end{equation}
 and from Ref.~\onlinecite{SykesBrooker1970}:
\begin{align} \label{eq:SBcofl}
 \dfrac{ c(\lambda)}{(1-\lambda)} = \dfrac{1 }{4}  \sum_{n=0}^{\infty} \dfrac{(4n + 3)}{(n+1)(2n+1)[(n+1)(2n+1) - \lambda]} \,.
\end{align}
 If we keep only the zeroth order components of the transition rates, and simply set $\Delta_3 = \Delta$ to extract the correct order of temperature dependence in $\nu_{\sigma}$, we obtain
\begin{equation} \label{eq:nu0}
\nu_{0} \approx (k_BT)^2\frac{\pi^2}{h^4}\frac{(m^*)^3}{p_F^2} \left( W_{f, 0}^{\ua \ua}+ W_{f, 0}^{\ua \da}+ W_{b, 0}^{\ua \da} \right)
\end{equation}

The expression for the zero-polarization viscosity follows from (\ref{eq:etasa}):
\begin{equation} \label{eq:etaP0}
\eta(\pol = 0) = \frac{\pi}{4 h^{2}} \frac{p_{F}^{4}}{m^{\ast}} \tau_{0} \frac{c(\lambda)}{1 - \lambda} \,.
\end{equation}
Using $1 - \lambda = \nu_{0}  \tau \sim O(1/\ln{T})$, and $c(\lambda \approx 1) \approx \frac{3}{4}$, we find:
\begin{equation} \label{eq:etaP0l0}
\eta(\pol = 0) = \frac{3}{4} \frac{h^{2}}{\pi} \frac{v_{F}^{2} }{W_{f, 0}^{\ua \ua}+ W_{f, 0}^{\ua \da}+ W_{b, 0}^{\ua \da} } \left( \frac{\epsilon_{F}}{k_{B} T}  \right)^{2} \,.
\end{equation}
The zero-polarization temperature dependence $\eta^{-1} \sim T^{2}$ is the same as that found by Fu and Ebner.~\cite{FuEbner1974}  There is an important feature of this expression that needs to be pointed out.  Because $1 - \lambda$ in (\ref{eq:etaP0}) is proportional to $\tau_{0}$ the explicit dependence on the quasiparticle lifetime has canceled out. This is quite different from three dimensions where $\eta$ is proportional to $\tau$.  

At this juncture it is convenient to explain why the above unorthodox treatment of the kinetic equation is necessary for the fermion shear viscosity in two dimensions. If we had proceeded in the usual manner by fixing the Fermi momenta equal to their zero-temperature values then we would find no $\nu_{0} \tau_{0}$ term in(\ref{eq:lamP0}), and thus the eigenvalue $\lambda = 1$.  Unfortunately $\lambda = 1$ is an eigenvalue of the associated \textit{homogeneous} integral equation. Since the left hand side of (\ref{eq:etaKEP0}) is not zero, then there would be no solution. 

\subsubsection{$0 < \pol < 1$}

In this polarization range, we note that the second term in Eq. (\ref{eq:costh3}) is always negligible compared to the first term. Thus for the forward scattering between spin-antiparallel quasiparticles we set $\Phi_3^f = 0$ and $\Phi_4^f =  \theta$. Then the second term in the curly bracket of Eq.~(\ref{eq:KEetaqeps}) becomes
\begin{equation} \label{eq:antifw}
\frac{ m^{*}_{\sigma_1} (m^{*}_{-\sigma_1})^2 }{p_{\sigma_1} p_{-\sigma_1}} \int_0^{\pi}d\theta \frac{W_f^{\sigma_1\, -\sigma_1}(\theta)}{\sqrt{\sin^2\theta - \frac{\xi_3}{\epsilon_{12}} }} \bigg[
q_{\sigma_1}(x_1) - q_{\sigma_1}(x_2) \bigg] \,.
\end{equation}

The integral for the spin-antiparallel backward scattering has to be evaluated using Eqs. (3.18):
\begin{equation} \label{eq:antibw}
\frac{ m^{*}_{\sigma_1} (m^{*}_{-\sigma_1})^2 }{p_{\sigma_1} p_{-\sigma_1}} \int_0^{\pi}d\theta \frac{W_b^{\sigma_1, -\sigma_1}(\theta)}{\sqrt{\sin^2\theta - \frac{\xi_3}{\epsilon_{12}} }} \bigg[
q_{\sigma_1}(x_1) - \Big( 1 - 2\sin^2\Phi_3^b \Big) q_{\sigma_1}(x_2) + D_{\sigma_1} 2\Big(\cos^2\theta - \cos^2\Phi_4^b\Big) q_{-\sigma_1}(x_2)  \bigg] \,.
\end{equation}
The integral over the last term is much smaller than that for the first two terms. Indeed if one sets $W_b^{\sigma_1 \, -\sigma_1}(\theta) \approx W_{b, 0}^{\sigma_1 \, -\sigma_1}$, using Eqs. (3.18) one obtains:
\begin{equation} \label{eq:nu4}
\int_0^{\pi}d\theta \frac{\cos^2\theta - \cos^2\Phi_4^b}{\sqrt{\sin^2\theta - \frac{\xi_3}{\epsilon_{12}} }} =0 \,.
\end{equation}
Thus we can ignore the last term in the integral, and therefore the \textit{kinetic equation at finite polarization becomes spin-decoupled}. Combining Eqs. (\ref{eq:parafw}), (\ref{eq:antifw}), (\ref{eq:antibw}) and (\ref{eq:nu4}), the kinetic equation becomes:
\begin{align}
n_1\bar{n}_1 =  & (k_BT)^2\frac{2}{h^4} \int_{-\infty}^{+\infty}dx_2dx_3dx_4\delta(x_1 + x_2 + x_3 + x_4)n_1n_2 n_3 n_4 \notag \\
& \Bigg\{ \frac{ (m^{*}_{\sigma_1})^3 }{p_{\sigma_1}^2} 
\int_{\Delta}^{\pi - \Delta} d\theta \frac{W_f^{\sigma_1 \, \sigma_1}(\theta)}{\sin\theta}\left(q_{\sigma_1}(x_1) - \left[ 1 - \tan^2\left(\frac{\theta}{2}\right)\Delta_3^2\right]q_{\sigma_1}(x_2)\right) \notag \\
& + \frac{ m^{*}_{\sigma_1} (m^{*}_{-\sigma_1})^2 }{p_{\sigma_1} p_{-\sigma_1}} \int_0^{\pi}d\theta \frac{W_f^{\sigma_1 \, -\sigma_1}(\theta)}{\sqrt{\sin^2\theta - \frac{\xi_3}{\epsilon_{12}} }} \bigg[
q_{\sigma_1}(x_1) - q_{\sigma_1}(x_2) \bigg] \notag \\
& + \frac{ m^{*}_{\sigma_1} (m^{*}_{-\sigma_1})^2 }{p_{\sigma_1} p_{-\sigma_1}} \int_0^{\pi}d\theta \frac{W_b^{\sigma_1 \, -\sigma_1}(\theta)}{\sqrt{\sin^2\theta - \frac{\xi_3}{\epsilon_{12}} }} \bigg[
q_{\sigma_1}(x_1) - \Big( 1 - 2\sin^2\Phi_3^b \Big) q_{\sigma_1}(x_2) \bigg] \Bigg\} \,.
\end{align}
With the aid of the definition of the quasiparticle lifetime (\ref{eq:tauP}) the kinetic equation becomes:
\begin{align}
\frac{\pi^2}{4} \tau_{\sigma_1} n_1 \bar{n}_1 = & \int_{-\infty}^{+\infty} dx_2 dx_3 dx_4 \delta \left( x_1 + x_2 + x_3 + x_4 \right) n_1 n_2 n_3 n_4 \notag \\
& \left\{ q_{\sigma_1} (x_1) - \Big[ 1 - \left(\nu_{\sigma_1}^f + \nu_{\sigma_1}^{(3)} \right) \tau_{\sigma_1} \Big] q_{\sigma_1}(x_2) \right\} \,,
\end{align}
with the definitions
\begin{align} 
\nu_{\sigma}^f &= \frac{\pi^2}{2h^4}(k_BT)^2\frac{(m_{\sigma}^*)^3}{p_{\sigma}^2}\int_{\Delta}^{\pi - \Delta}
d\theta \frac{W_f^{\sigma \, \sigma}(\theta)}{\sin \theta} \tan^2 \left(\frac{\theta}{2} \right) \Delta_3^2 \,, \label{nuf} \\
\nu^{(3)}_{\sigma} &= \frac{\pi^{2}}{h^{4}}  \left( k_{B} T  \right)^{2}   \frac{\mstar _{\sigma}(\mstar_{-\sigma})^{2}}{p_{\sigma} p_{-\sigma}}  \int_{\Delta}^{\pi -\Delta} d\theta  \frac{W^{\sigma \, -\sigma}_{b}(\theta)  }{\sin{\theta}}  \sin^{2}{\Phi_{3}} \,.
\end{align}
The kinetic equation can now be cast into Sykes-Brooker form:
\begin{equation} 
\frac{\pi^2}{4}\tau_{\sigma} = \int_{-\infty}^{+\infty} dx_2 K(x_1, x_2) \left[ q_{\sigma}(x_1) - \lambda_{\sigma} q_{\sigma}(x_2) \right] \,,
\end{equation}
with
\begin{equation} \label{eq:etaLamP}
\lambda_{\sigma} = 1 - \left(\nu_{\sigma}^f + \nu_{\sigma}^{(3)} \right) \tau_{\sigma} \,.
\end{equation}
The solution is 
\begin{equation} \label{eq:etaQP}
Q_{\sigma}  = \frac{c(\lambda_{\sigma})}{2\left(\nu_{\sigma}^f + \nu_{\sigma}^{(3)} \right)} \,,
\end{equation}
where $c(\lambda)$ is given in Eq.~(\ref{eq:SBcofl}).  In lowest order of $W$, we can obtain approximate analytic expressions for the generalized frequencies:
\begin{align} 
\nu_{\sigma}^f  &\approx (k_BT)^2\frac{\pi^2}{h^4}\frac{(m_{\sigma}^*)^3}{p_{\sigma}^2} W_{f, 0}^{\sigma \, \sigma} \,, \label{nuf1} \\
\nu_{\sigma}^{(3)} &\approx \frac{2\pi^2}{h^4} (k_B T)^2 \frac{(m^{*}_{-\sigma})^2}{\epsilon_{F \, \sigma}} W_{b,0}^{\sigma \, -\sigma} \frac{p_{-\sigma}}{p_{\sigma}}\left( 1 + \frac{p_{\sigma}^2 - p_{-\sigma}^2}{2 p_{\sigma} p_{-\sigma}} \ln \left | \frac{p_{\sigma} + p_{-\sigma}}{p_{\sigma} - p_{-\sigma}} \right |  \right) \,.
\end{align}
We note they are both on the order of $T^2$.

For the polarization range $0 < \pol < 1$ the shear viscosity becomes:
\begin{equation}  \label{eq:etaP}
\eta =  \frac{3 \pi}{32 h^{2}} \sum_{\sigma} \frac{(p_{F \sigma})^{4}}{m_{\sigma}^{*}} \frac{1}{\left(\nu_{\sigma}^f + \nu_{\sigma}^{(3)} \right)} \,,
\end{equation}
where $c(\lambda) \approx 3/4$ since $\nu_{\sigma} \tau_{\sigma} \sim O(1/\ln{T})$.  In lowest order we find:
\begin{equation} \label{eq:etaPell0}
\eta = \frac{3}{8} \frac{h^{2}}{\pi} \sum_{\sigma} \frac{v_{F  \sigma}^{2}}{W^{\sigma \, \sigma}_{f,0} + 4 \left(\frac{v_{F -\sigma}}{v_{F \sigma}}\right)  
                        \left(\frac{m^{*}_{-\sigma}}{ m^{*}_{\sigma}}\right)^{3} 
			W_{b,0}^{\sigma \, -\sigma} \left( 1 + \frac{p_{\sigma}^2 - p_{-\sigma}^2}{2 p_{\sigma} p_{-\sigma}} \ln \left | \frac{p_{\sigma} + p_{-\sigma}}{p_{\sigma} - p_{-\sigma}} \right |  \right)  } \left( \frac{\epsilon_{F \sigma}}{k_{B} T}  \right)^{2} \,.
\end{equation}
Thus, at finite polarization we find $\eta^{-1} \sim T^{2}$, the same as at zero polarization. Further, the quasiparticle lifetime has canceled out in the same manner as at zero polarization.  

\subsubsection{Full polarization $\pol = 1$}
At full polarization we ignore those terms that involve scattering between anti-parallel spin quasiparticles since there are no particles in the minority Fermi sea. Thus we ignore the $\nu^{(3)}$ term in (\ref{eq:etaLamP}), and therefore $\lambda_{\ua} =  1 - \nu_{\ua}^{f} \tau_{1}$. The solution becomes $Q_{\ua} = c(\lambda_{\ua}) / (2 \nu_{\ua}^{f})  \approx 3/(8 \nu_{\ua}^{f})$.  From Eqs.~(\ref{eq:etaP}) and (\ref{eq:etaPell0}) the shear viscosity at full polarization becomes:
\begin{equation}  \label{eq:etaP1}
\eta(\pol = 1) =  \frac{3 \pi}{32 h^{2}}  \frac{(p_{F \ua})^{4}}{m_{\ua}^{*}} \frac{1}{\nu_{\ua}^f } \approx 
\frac{3}{8} \frac{h^{2}}{\pi}  \frac{v_{F  \ua}^{2}}{W^{\ua \, \ua}_{f,0}  } \left( \frac{\epsilon_{F \ua}}{k_{B} T}  \right)^{2} \,.
\end{equation} 

\subsubsection*{Summary}
We find that the temperature dependence for the shear viscosity at arbitrary polarization $0 \leq \Pee \leq 1$ is $\eta^{-1} \sim T^{2}$.  This is in agreement with the zero-polarization results of Fu and Ebner.~\cite{FuEbner1974}

\subsection{\label{sec:diff} Spin diffusion}

As noted in the Introduction, the longitudinal spin diffusion coefficient $D$ for a two dimensional Fermi liquid at arbitrary polarization was calculated by Miyake and Mullin.~\cite{MiyakeMullin1983,*MiyakeMullin1984} We shall include a very brief discussion of the calculation of $D$ in order to  compare this kinetic equation solution to the closely related solutions for the thermal conductivity and the shear viscosity as discussed in the previous sections. Further, using our results for the Landau parameters for \he3 films we can calculate predicted values for $D$ as a function of density and polarization.  These numerical results will be presented in the following section. 

In this case the driving field is a chemical potential gradient.  The left hand side of the kinetic equation (\ref{eq:trans}) can be written: $n_{1}\nbar_{1} \vvec_{1} \cdot (\bm{\nabla} \mu_{1})$, and the linearized ansatz for the non-equilibrium distribution function becomes:
\begin{equation} \label{eq:zetaD}
\zeta_{i} = \vvec_{i} \cdot (\bm{\nabla} \mu_{i}) q_{\sigma}(\epsilon_{i}) \,.
\end{equation}

Choosing $\vvec_{1} \parallel (\bm{\nabla} \mu_{1})$ with no loss of generality, the kinetic equation becomes:
\begin{equation} \label{eq:DKE}
\begin{split}
n_{1} \nbar_{1} 
&=  \sum_{\pvec_{2}, \pvec_{3},\pvec_{4}}  \delta_{\pvec_{1} + \pvec_{2}, \pvec_{3}+\pvec_{4}} 
\delta(\epsilon_{1} + \epsilon_{2}  - \epsilon_{3} - \epsilon_{4}) n_{1} n_{2} \nbar_{3} \nbar_{4} \\
\Big\{  &W^{\sigone \sigone}_{f}(\theta)  
\Big[   q_{\sigone}(\epsilon_{1}) +   q_{\sigone}(\epsilon_{2}) \cos{\theta}  
      -   q_{\sigone}(\epsilon_{3}) -  q_{\sigone}(\epsilon_{4}) \cos{\theta} \Big] \\
+  &W^{\sigone -\sigone}_{f}(\theta)  
\Big[  q_{\sigone}(\epsilon_{1})  +  A_{\sigone}{v_{\sigone}} q_{-\sigone}(\epsilon_{2})  \cos{\theta} 
      -   q_{\sigone}(\epsilon_{3}) -   A_{\sigone} q_{-\sigone}(\epsilon_{4}) \cos{\theta} \Big] \\
+ &W^{\sigone -\sigone}_{b}(\theta) 
\Big[  q_{\sigone}(\epsilon_{1}) +    A_{\sigone} q_{-\sigone}(\epsilon_{2})   \cos{\theta}
      - q_{\sigone}(\epsilon_{3}) \cos{\Phi_{3}} -  A_{\sigone} q_{-\sigone}(\epsilon_{4}) \cos{\Phi_{4}} \Big] \Big\} \,,
\end{split}
\end{equation}
where following the notation of MM we introduce:
\begin{equation}
A_{\sigone} \equiv \frac{v_{-\sigone}}{v_{\sigone}} \frac{R_{-\sigone} / N_{0}^{-\sigone}}{R_{\sigone} / N_{0}^{\sigone}} \,.
\end{equation}
The $R_{\sigma}$ parameters  are the proportionality constants that connect the chemical potential gradients with magnetization gradients: $\bm{\nabla} \mu_{\sigma} = (R_{\sigma}/N_{\sigma}(0) ) \bm{\nabla} \left(\nbar \, \Pee \right)$ where $\nbar = \nbar_{\ua} + \nbar_{\da} = N/A$ is the total number density, and the $N_{0}^{\sigma} = m_{\sigma}^{*}/(2 \pi \hbar^{2}) $ are the single spin-state density of states at the Fermi surface. For completeness, we shall write down the explicit expression for the $R_{\sigma}$ since this quantity finds its way into the final expression for the diffusion constant:
\begin{equation} \label{eq:Rsig}
R_{\sigma} = \frac{\sigma \nbar_{-\sigma}}{N_{0}^{-\sigma} }  \frac{(1 + F_{0}^{\sigma \sigma})(1 + F_{0}^{-\sigma -\sigma}) - F_{0}^{\sigma -\sigma}F_{0}^{-\sigma \sigma} }
{\nbar_{\sigma}(1 + F_{0}^{\sigma \sigma} + F_{0}^{\sigma -\sigma})/N_{0}^{\sigma} + \nbar_{-\sigma}(1 + F_{0}^{-\sigma -\sigma} + F_{0}^{-\sigma \sigma})/N_{0}^{-\sigma} }\,,
\end{equation}
and we associate $\sigma = \{+1,-1\}$ with $\sigma =\, \{\ua, \da\}$, respectively.  We note that from the requirement of stable equilibrium~\cite{LAM_PRB2012} the numerator of (\ref{eq:Rsig}) must be positive. The Landau parameters are defined by:
\begin{equation} \label{eq:F0}
F^{\sigma \sigma'}_{0} = N_{0}^{\sigma} \int_{0}^{2 \pi} \,\frac{d \theta}{2 \pi} f^{\sigma \sigma'}_{\pvec \pvep} \,
\end{equation}

The reduction of (\ref{eq:DKE}) to Sykes-Brooker form follows the same steps as for the thermal conductivity.  
One obtains a kinetic equation in non-diagonal Sykes-Brooker form:
\begin{equation}  \label{eq:MMB5}
\begin{split}
\left( \frac{\pi^{2}}{4} \right) \tau_{\sigone} 
=  \int_{-\infty}^{\infty} dx_{2} \, K(x_{1}, x_{2}) \Big[ q_{\sigone}(x_{1}) -  \left( 1 - \nu_{\sigone} \tau_{\sigone} \right) q_{\sigone}(x_{2})  + \nu_{-\sigone} \tau_{\sigone}  q_{-\sigone}(x_{2}) \Big]    \,,
\end{split}
\end{equation}
where we have introduced the the spin-parallel and spin anti-parallel scattering times (\ref{eq:taup}) and (\ref{eq:tauanti}),  the generalized frequency $\nu_{\sigma}$:
\begin{equation} \label{eq:nusig}
\begin{aligned}
\nu_{\sigma} \equiv   \frac{\pi^{2}}{h^{4}}  (k_{B}T)^{2} m^{\ast}_{\sigma} ( m^{\ast}_{-\sigma})^{2} \frac{p_{-\sigma}}{p_{\sigma}}\int_{\Delta}^{\pi - \Delta}  d\theta \, \frac{W^{\sigma -\sigma}_{b}(\theta) \sin{\theta}}{\ell^{2}} \,,
\end{aligned}
\end{equation}
and in addition we have also used:
\begin{equation}
\nu_{-\sigma} = - \frac{p_{\sigma}}{p_{-\sigma}} A_{\sigma} \nu_{\sigma} \,.
\end{equation}
Eq.~(\ref{eq:MMB5}) agrees with (B5) in MM which marks the start of their analysis.  The only differences are in the definition of some parameters:  $(\tau_{\sigma})^{\text{MM}}=(\pi^{2}/4) \tau_{\sigma}$ and $(\nu_{\sigma}/f_{\sigma})^{\text{MM}} = \nu_{\sigma} \tau_{\sigma}$. 

We rewrite (\ref{eq:MMB5}) with a matrix representation of the coefficients of  $q_{\sigma}(x_{2})$:
\begin{equation}  
\begin{split}
\left( \frac{\pi^{2}}{4} \right) \tau_{\sigma} 
=  \int_{-\infty}^{\infty} dx_{2} \, K(x_{1}, x_{2}) \Big[ q_{\sigma}(x_{1}) -  \sum_{\sigma'} \lambda_{\sigma \sigma'} q_{\sigma'}(x_{2})   \Big]    \,,
\end{split}
\end{equation}
where for simplicity we have switched from $\sigone$ to $\sigma$.  The coefficient matrix is given by: 
\begin{equation}
\lambda = \begin{pmatrix}
1 - \nu_{\ua}\tau_{\ua} & -\nu_{\da}\tau_{\ua} \\
-\nu_{\ua}\tau_{\da}     & 1 - \nu_{\da}\tau_{\da}
\end{pmatrix} \,.
\end{equation}
The eigenvalues of $\lambda$ are:
\begin{equation} \label{eq:lambdas}
\lambda_{+} = 1\,, \qquad \lambda_{-} = 1 - \left(  \nu_{\ua}\tau_{\ua} +  \nu_{\da}\tau_{\da} \right) \,,
\end{equation}
where as in Sec.~\ref{sec:therm} the $\pm$ subscripts on the $\lambda$'s denote the rotated up and down spin states. 
The diagonalization is accomplished by transforming to variables $\tilde{\tau} = S \tau$ and $\tilde{q} = Sq$ where:
\begin{equation}
\begin{aligned}
S = \frac{1}{\nu_{\ua}\tau_{\ua} +  \nu_{\da}\tau_{\da}} \begin{pmatrix} -\tau_{\da} &\tau_{\ua} \\ \nu_{\ua} &\nu_{\da} \end{pmatrix} \,.
\end{aligned}
\end{equation}
In terms of the transformed variables the diagonalized pair of kinetic equations are:
\begin{align} \label{eq:KEdiag}
0 &=  \int_{-\infty}^{\infty} dx_{2} \, K(x_{1}, x_{2}) \Big[ \tilde{q}_{+}(x_{1}) - \tilde{q}_{+}(x_{2})   \Big]  \,, \\
\frac{\pi^{2}}{4} &= \int_{-\infty}^{\infty} dx_{2} \, K(x_{1}, x_{2}) \Big[ \tilde{q}_{-}(x_{1}) - \lambda_{-}\tilde{q}_{-}(x_{2})   \Big] \,.
\end{align}
We have written the pair of equations separately in order to emphasize that the equation with the unit eigenvalue is indeed homogeneous. 

As with the other transport coefficients, the important quantity is not $\tilde{q}_{\sigma}$ itself but rather the integrated quantities:
\begin{equation}
\tilde{Q}_{\sigma} \equiv \int_{-\infty}^{\infty} dx \frac{\partial n}{\partial x} \tilde{q}_{\sigma} \,.
\end{equation}
This definition differs from that of MM by a minus sign. From Sykes and Brooker the solutions can be written:
\begin{align}
&\tilde{Q}_{+} = C \,, \\ \label{eq:Qtilm}
&\tilde{Q}_{-} = - \frac{c(\lambda_{-})}{2 (1 - \lambda_{-})} \,,
\end{align}
where $C$ is an arbitrary constant to be determined below, and $c(\lambda_{-})$ is a series involving the eigenvalue $\lambda_{-}$ defined in  Eq.~(\ref{eq:SBcofl}). The diffusion coefficient is given in terms of $Q$, obtained from the inverse transform: $Q = S^{-1} \tilde{Q}$.

The spin current for the $\sigma$th component is $\mathbf{j}_{\sigma} = \hlf v_{\sigma}^{2} R_{\sigma}Q_{\sigma}(\mathbf{\nabla} m)$.   MM fixed the value of $C$ by assuming that the system is in the frame of reference where the spin current due to bulk motion vanishes, thus $\sum_{\sigma}\mathbf{j}_{\sigma}=0$.

The diffusion current is given by $\mathbf{j} = \sum_{\sigma} \sigma \mathbf{j}_{\sigma} \equiv -D \mathbf{\nabla}m$.  Thus, $D = \hlf \sum_{\sigma} v_{\sigma}^{2} R_{\sigma} Q_{\sigma}$, and  we have:
\begin{equation}
D = \hlf v_{\ua}^{2}v_{\da}^{2}R_{\ua}R_{\da} \left(\frac{\nu_{\ua}\tau_{\ua} + \nu_{\da}\tau_{\da}}{\nu_{\ua}v_{\da}^{2} R_{\da} -  \nu_{\da}v_{\ua}^{2}R_{\ua}} \right)  \frac{c(\lambda_{-})}{(1 - \lambda_{-})} \,.
\end{equation}
This is in agreement with MM's fundamental result.  Using  (\ref{eq:lambdas}) we can also write:
\begin{equation} \label{eq:Dfinal}
D = \hlf \frac{v_{\ua}^{2}v_{\da}^{2}R_{\ua} R_{\da} }{\nu_{\ua}v_{\da}^{2} R_{\da} -  \nu_{\da}v_{\ua}^{2}R_{\ua}}   \frac{(1 - \lambda_{-})}{4} 
\Big[\sum_{n=0}^{\infty} \frac{4n+3}{(n+1)(2n+1) [(n+1)(2n+1) - \lambda_{-}]}\Big] \,.
\end{equation}
The temperature dependence of $D$ depends on whether one is at zero polarization or non-zero polarization. These two cases will be discussed in turn. 

\subsubsection {Zero polarization}
At zero polarization we set $v_{\sigma} = v_{F}, R_{\ua} = -R_{\da} = R$ and $\nu_{\sigma} = \nu $ in (\ref{eq:Dfinal}) where:
\begin{align}
R &= \frac{1 + F_{0}^{a}}{2} \,, \\
\nu &= (k_{B}T)^{2} \frac{\pi^{2}}{h^{4}} (m^{\ast})^{3}  \int_{0}^{\pi - \Delta} d\theta \,  \frac{W^{\sigma -\sigma}_{b}(\theta) \sin{\theta}}{\ell^{2}} \,,\end{align}
and $\Delta = k_{B}T/\epsilon_{F}$.  The only pole is at the upper limit of the $\theta$-integral, thus we have extended the lower limit to zero. The spin diffusion coefficient reduces to:
\begin{align}
D(\Pee = 0) &= \frac{v_{F}^{2} (1 + F_{0}^{a})}{8 \nu} c(\lambda_{-}) \,.
\end{align}
In the $\ell = 0$ approximation the frequency $\nu$ is given by:
\begin{align}
\nu &= \hlf  \frac{\pi^{2}}{h^{4}} (m^{\ast})^{2} W_{b, 0}^{\sigma \, -\sigma} \frac{(k_{B}T)^{2}}{\epsilon_{F}} \ln{\left(\frac{2 \epsilon_{F}}{k_{B}T} \right)} \,, \label{eq:nuDP0}  
\end{align}
and the quasiparticle lifetime $\tau_{0}$ is defined in (\ref{eq:tauP0}).
$D$ depends on both the spin parallel and the spin anti-parallel transition probabilities through the the eigenvalue $\lambda_{-}$:
\begin{equation} \label{eq:lambdam}
\lambda_{-} = 1 - 2 \nu \tau_{0} = 1 - 2 \left[ \frac{W_{b, 0}^{\sigma \, -\sigma}}{W_{f, 0}^{\sigma \, \sigma} +  W_{f, 0}^{\sigma \, -\sigma} +  W_{b, 0}^{\sigma \, -\sigma}}  \right]  \,.
\end{equation}
Written out in detail, the diffusion coefficient at zero polarization is:
\begin{equation} \label{eq:DP0}
\begin{split}
D = &\hlf \left(\frac{h^{4}}{\pi^{2} } \right) \frac{(1 + F_{0}^{a})}{(m^{\ast})^{3} W^{\sigma \, -\sigma}_{b, 0}} \frac{1}{\left(\frac{k_{B}T}{\epsilon_{F}} \right)^{2} \ln{\left(\frac{2 \epsilon_{F}}{k_{B}T}\right)} } \\
&\times \frac{(1 - \lambda_{-})}{4} \Big[\sum_{n=0}^{\infty} \frac{4n+3}{(n+1)(2n+1) [(n+1)(2n+1) - \lambda_{-}]}\Big] \,.
\end{split}
\end{equation}
This result is in agreement with MM (41).  The dependence of $D$ on the spin-anti-parallel transition probabilities through the eigenvalue mimics the solution for bulk Fermi liquids.~\cite{SykesBrooker1970}   In a more general form than the $\ell = 0$ approximation, the eigenvalue $\lambda_{-}$ depends on the transition probabilities $W^{\sigma \sigma}_{f}, W^{\sigma -\sigma}_{f}, \text{ and }W^{\sigma -\sigma}_{b}$ through the angular averages that appear in $\nu_{\sigma}$ and $\tau_{\sigma}$.

\subsubsection{Nonzero polarization}
The low-temperature physics changes qualitatively at any non-zero polarization. The important contribution is from the frequency $\nu_{\sigma}$ (\ref{eq:nusig}):
\begin{equation}
\begin{aligned}
\nu_{\sigma} &=           \frac{\pi^{2}}{h^{4}}  (k_{B}T)^{2} m^{\ast}_{\sigma} ( m^{\ast}_{-\sigma})^{2} \Big(\frac{p_{-\sigma}}{p_{\sigma}} \Big) 
\int_{0}^{\pi}  d\theta \, \frac{W^{\sigma \, -\sigma}_{b}(\theta) \sin{\theta}}{\ell^{2}} \,, \\
\nu_{\sigma} &\approx  \frac{\pi^{2}}{h^{4}}  (k_{B}T)^{2} m^{\ast}_{\sigma} ( m^{\ast}_{-\sigma})^{2} \Big(\frac{p_{-\sigma}}{p_{\sigma}} \Big) \frac{W^{\sigma \, -\sigma}_{b, 0}}{p_{\sigma} p_{-\sigma}} \ln{\Big|\frac{p_{\sigma} + p_{-\sigma}}{p_{\sigma} - p_{-\sigma}}\Big|} \,,
\end{aligned}
\end{equation}
where the integration limits are not cut-off since at finite polarization there are no poles. As pointed out by MM there is a logarithmic singularity in the limit of zero polarization with no singular behavior as a function of temperature.   The lifetimes at finite polarization are given by (\ref{eq:tauP}). Thus we find $\tau_{\sigma} \nu_{\sigma} \sim O(T^{-2} \ln^{-1}(T))$. This should be compared with (\ref{eq:lambdam}) above. At low temperature then we have $\lambda_{-} \approx 1 + O(T^{-2} \ln^{-1}(T)) $.  As discussed in Refs.~\onlinecite{SykesBrooker1970} and \onlinecite{APQ1987}, in the limit $\lambda \rightarrow 1^{+}$, we can replace the sum by its first term yielding $c(\lambda) \approx 3/4$.

Thus, the spin diffusion coefficient at $\Pee \ne 0$ is given by:
\begin{equation} \label{eq:DP}
D =  \frac{3}{2} \left( \frac{h^{4}}{\pi^{2}} \right)  
\frac{\epsilon_{F  \ua} \epsilon_{F \da} R_{\ua} |R_{\da}|}
{m^{\ast}_{\ua}m^{\ast}_{\da}
\left[  m^{\ast}_{\ua}  \frac{p_{\da}^{2}}{p_{\ua}^{2}} |R_{\da}| \right. 
+  \left.  m^{\ast}_{\da} \frac{p_{\ua}^{2}}{p_{\da}^{2}} R_{\ua}  \right] \left(W^{\ua \da}_{b, 0} \right) \ln{\Big|\frac{p_{\ua} + p_{\da}}{p_{\ua} - p_{\da}} \Big|}}  
(k_{B}T)^{-2} \,.
\end{equation}
The absence of the $\ln{(T)}$ divergence in $D$  at finite polarization was first noticed by MM who pointed that the change in temperature dependence from zero polarization was due to the inability of the system to conserve momentum in spin anti-parallel collisions at non-zero polarization and low enough temperature. In the limit of full polarization the spin diffusion coefficient vanishes.  From (\ref{eq:DP}) it can be seen that $D(\Pee \rightarrow 1) \sim O(p_{\da}^{3})$.

The finite-polarization spin diffusion coefficient given by (\ref{eq:DP}) does not go smoothly into the zero-polarization result (\ref{eq:DP0}).  We can easily calculate the value of the polarization $\Pee_{c}$ at which the diffusion coefficients from Eqs. (\ref{eq:DP}) and (\ref{eq:DP0}) are equal.  In the limit of very small polarization $|\Pee| \ll 1$ (\ref{eq:DP}) becomes:
\begin{equation} \label{eq:DPsmall}
D(|\Pee| \ll 1) \approx  \frac{3}{8} \left(\frac{h^{4}}{\pi^{2}} \right)  \frac{(1 + F_{0}^{a})}{(m^{\ast})^{3} W^{\ua \da}_{b, 0} }   
 \frac{1}{ \ln{\left( \frac{1}{2 |\Pee|} \right) }}  \left(\frac{\epsilon_{F}}{k_{B}T} \right)^{2} \,,
\end{equation}
where all terms have been set to their $\Pee = 0$ values except for the term that is singular in that limit.  We can set $c(\lambda_{-}) = 3/4$ in (\ref{eq:DP0}) with little error since its range is $3/4 \leq c(\lambda_{-}) \leq 1$.~\cite{SykesBrooker1970}  Then the only difference between (\ref{eq:DP0}) and (\ref{eq:DPsmall}) are the logarithmic terms.   By inspection:
\begin{equation}
\Pee_{c}  = (k_{B}T/ 4 \epsilon_{F}) \,.
\end{equation}
For polarizations less than $\Pee_{c}$ the diffusion coefficient may be measurably larger than the zero-polarization diffusion coefficient. Of course this would be within a very small regime since this analysis is valid only in the limit $k_{B}T \ll \epsilon_{F}$.

\subsubsection*{Summary}
The temperature dependence for the spin diffusion coefficient at zero polarization is $D^{-1} \sim T^{2} \ln{T}$.  This is in agreement with the zero-polarization results of Fu and Ebner.~\cite{FuEbner1974}  At finite polarization $0 < \pol < 1$ the temperature dependence for the spin diffusion coefficient is $D^{-1} \sim T^{2}$. These results were obtained by MM.~\cite{MiyakeMullin1983,*MiyakeMullin1984}

\section{Application to thin $^{3}$H\lowercase{e} films}\label{sec:he3}
In this section we calculate transport coefficients for thin-film \he3 systems. The system specific information is provided by the angular integrals of the transition rates $W^{\sigma \sigma'}(\theta)$ that appear in the quasiparticle lifetimes $\tau$ and the generalized frequencies $\nu$.   The  transition rates can be written in terms of the scattering amplitudes: $W^{\sigma \sigma'}_{f,b}(\theta) = \frac{2 \pi}{\hbar}|a^{\sigma \sigma'}_{f,b}(\theta)|^{2}$.  Dimensionless scattering amplitudes can be defined by: 
\begin{equation} \label{eq:Atilde}
\tilde{A}^{\sigma \sigma'}_{f,b}(\theta) = \tilde{N}_{0} a^{\sigma \sigma'}_{f,b}(\theta) \,,
\end{equation}
where $\tilde{N}_{0} = m/(2 \pi \hbar^{2}).$   The transition rates then become
\begin{equation} \label{eq:WAtilde}
W^{\sigma \sigma'}_{f,b}(\theta) = \frac{h^{3}}{m^{2}}|\tilde{A}^{\sigma \sigma'}_{f,b}(\theta)|^{2} \,.
\end{equation}
In turn, for forward scattering, the Fourier components of the scattering amplitudes can be written in terms of the Landau parameters:~\cite{AM2011}
\begin{subequations}  \label{eq:scattA}
\begin{align}
a_{f, \ell}^{\ua \ua} &= \frac{ f_{\ell}^{\ua \ua}(1 + N_{0}^{\da}  f_{\ell}^{\da \da}) - N_{0}^{\da} ( f_{\ell}^{\ua \da})^{2} }{(1 + N_{0}^{\ua}  f_{\ell}^{\ua \ua})(1 + N_{0}^{\da}  f_{\ell}^{\da \da}) - N_{0}^{\ua}N_{0}^{\da}  (f_{\ell}^{\ua \da})^{2}} \,, \\
a_{f, \ell}^{\ua \da} &= \frac{ f_{\ell}^{\ua \da} }{(1 + N_{0}^{\ua}  f_{\ell}^{\ua \ua})(1 + N_{0}^{\da}  f_{\ell}^{\da \da}) - N_{0}^{\ua}N_{0}^{\da}  (f_{\ell}^{\ua \da})^{2}} \,.
\end{align}
\end{subequations}
We note in passing that in this notation the forward scattering sum rules~\cite{Hone1962} become $W_{f}^{\sigma \sigma}(0) = 0$. 

At zero polarization one can also write the backward scattering transition probability $W_{b}^{\sigma \, -\sigma}$ in terms of the forward scattering amplitudes:~\cite{MiyakeMullin1983}
\begin{equation} \label{eq:aback}
a_{b}^{\sigma \, -\sigma}(\theta) = a_{f}^{\sigma \, -\sigma}(\theta) - a^{\sigma \sigma}_{f}(\theta) \,.
\end{equation}
It is not known whether a comparable exact result can be obtained for nonzero polarization in two dimensions. The important point is that at zero polarization in two dimensions $D$  can be written solely in terms of the Landau parameters.  This point was made by Miyake and Mullin. In fact this is valid for all of the transport coefficients.  In order to compute backward scattering contributions at nonzero polarization, and also head-on transition rates used for the shear viscosity, we shall proceed by making some reasonable assumptions. For backward scattering at nonzero polarization:
\begin{equation}
a_{b}^{\sigma \, -\sigma} \approx a_{f}^{\sigma \, -\sigma} \,,
\end{equation}
and for the head-on scattering needed for the shear viscosity:
\begin{equation}
a_{head-on}^{\sigma \, \sigma'}(\theta = \pi, \phi=0 \text{ or } \pi) \approx a_{f}^{\sigma \, \sigma'}(\theta = \pi) \,.
\end{equation}

In three dimensions one faces a similar problem because one also needs information concerning the $\phi$-dependence of the scattering amplitudes where $\phi$ is the angle between the planes formed by the momenta of the incoming and outgoing quasiparticles.  An approximate solution at zero polarization was obtained by Dy and Pethick.~\cite{DyPethick1969} Unfortunately the \textit{s-p} approximation does not generalize to nonzero polarization.  In two dimensions as noted previously  $\phi$ can only take on the values $0$ and $\pi$.

\begin{figure}[t]
\includegraphics[]{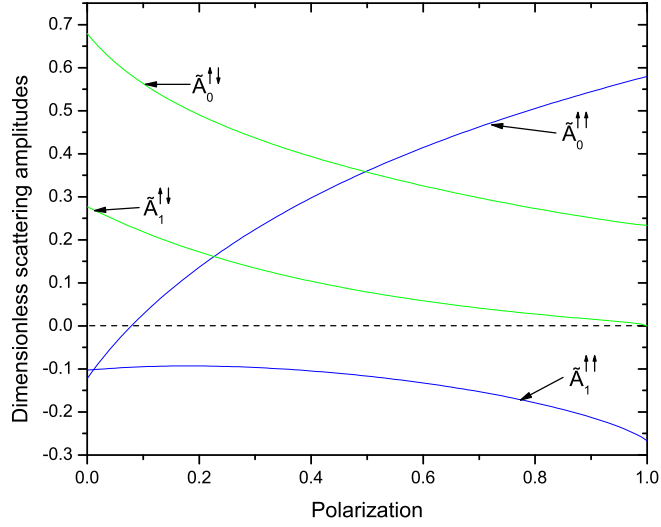}
\caption{\label{fig:amp01} Dimensionless scattering amplitudes (\ref{eq:Atilde}) versus polarization at $\nbar = 0.0132$~\AA$^{-2}$  on a graphite substrate.  Comparison of $\tilde{A}^{\ua \ua}_{0}$ with $\tilde{A}^{\ua \ua}_{1}$, and $\tilde{A}^{\ua \da}_{0}$ with $\tilde{A}^{\ua \da}_{1}$.  We note that at $\pol = 0$ we find $\tilde{A}^{\ua \ua}_{0} \approx \tilde{A}^{\ua \ua}_{1}$.  For all polarizations $\tilde{A}^{\ua \da}_{0} >\tilde{A}^{\ua \da}_{1}$.}
\vspace{1truein}
\end{figure}

For \he3 in two dimensions we can calculate Landau parameters to high orders~\cite{LAM_PRB2012} by utilizing effective \textit{s}-wave and \textit{p}-wave \textit{T}-matrix elements determined by experimental data.    Thus in principle we can also calculate the Fourier sum for $W_{f}^{\sigma \sigma'}(\theta)$ to high orders.  For the numerical work to be discussed below, however, we shall use the lowest-order $\ell = 0$ approximation for the transition rates.  In Fig.~\ref{fig:amp01} we compare $\tilde{A}^{\ua \ua}_{\ell}$ and $\tilde{A}^{\ua \da}_{\ell}$ for $\ell = 0, 1$ at $\nbar = 0.0132$~\AA$^{-2}$  on a graphite substrate. We see, at zero polarization, $\tilde{A}^{\ua \da}_{0}$ dominates the other three components:  as one expects, the singlet channel dominates the \textit{s}-wave scattering. As the polarization increases however, $\tilde{A}^{\ua \ua}_{0}$ increases rapidly and eventually becomes the dominant component. Therefore we can approximate the transition rates by simply keeping the $\ell = 0$ components over the whole polarization range.

We begin by examining the transport coefficients in the forms $\kappa T \ln{(2 T_{F}/T)}$,  $\eta T^{2} $, and  $DT^{2} \ln{(2 T_{F}/T)}$ to analyze their density dependence for $\pol = 0$ and $\pol = 1$.   The results are presented in this manner in order to take advantage of the fact that at zero polarization and full polarization the explicit temperature dependence factors, see Table~\ref{tab:TransP0P1}.  In Table~\ref{tab:GraphiteP0} we show the values for the system of second layer \he3 on graphite.  We include in the table the Fermi energies, and the effective masses. The values of the effective masses at full polarization come from Ref.~\onlinecite{LAM_PRB2012}.   We can compare the qualitative behavior of the transport coefficients with their bulk \he3 analogs. Figure~3 in Bedell and Pines~\cite{BedellPines1980} shows the pressure dependence at zero polarization of $\kappa T$, $\eta T^{2}$, and $D T^{2}$. In three dimensions each of these quantities appears to be a monotonically decreasing function of pressure. In two dimensions this is not necessarily the case. One can easily extract the explicit density and effective mass dependence of the transport coefficients by examining Eqs.~(\ref{eq:tauP0}), (\ref{eq:kappaP0}), (\ref{eq:etaP0}), and (\ref{eq:DP0}).  By inspection of Table~\ref{tab:TransP0P1}, we find  $\kappa \sim \nbar^{2}/ ({\mstar})^{4}$,  $\eta \sim \nbar^{3}/ ({\mstar})^{4}$, and $D \sim \nbar^{2}/({\mstar})^{5}$.   Thus, the explicit density dependence tries to increase the transport coefficients with increasing density whereas the explicit effective mass dependence tries to decrease the transport coefficients with increasing effective mass.  
\begin{table}[]
\caption{\label{tab:TransP0P1} Expressions for the transport coefficients at $\pol =0$ and $\pol = 1$ rewritten in forms that are useful for analyzing the density dependence.}
\begin{ruledtabular}
\begin{tabular}{ccl}
Transport coefficient  & Polarization & Expression \\
\hline
$\kappa$  & $\pol =0$   &   $\kappa T \ln{\left(\frac{2T_{F}}{T}\right)} =  \left(\dfrac{h^{3}}{8 \pi m^{2}} \right) H(1) \dfrac{\nbar^{2}}{(m^{\ast}/m)^{4}} 
                                         \dfrac{1}{|\tilde{A}^{\ua \ua}_{f,0}|^{2} + |\tilde{A}^{\ua \da}_{f,0}|^{2} + |\tilde{A}^{\ua \da}_{b,0}|^{2}} $ \\ 
$\kappa$  & $\pol = 1$  &   $\kappa T \ln{\left(\frac{2T_{F}}{T}\right)} =  \left(\dfrac{h^{3}}{4 \pi m^{2}} \right) H(1) \dfrac{\nbar^{2}}{(m^{\ast}_{\ua}/m)^{4}} 
                                         \dfrac{1}{|\tilde{A}^{\ua \ua}_{f,0}|^{2} } $ \\
$\eta$     & $\pol = 0$  &   $\eta T^{2} =  \left(\dfrac{3 \pi \hbar^{5}}{4 k_{B}^{2} m^{2}} \right)  \dfrac{\nbar^{3}}{(m^{\ast}/m)^{4}} 
                                         \dfrac{1}{|\tilde{A}^{\ua \ua}_{f,0}|^{2} + |\tilde{A}^{\ua \da}_{f,0}|^{2} + |\tilde{A}^{\ua \da}_{b,0}|^{2}} $ \\ 
$\eta$     & $\pol = 1$  &   $\eta T^{2} =  \left(\dfrac{3 \pi \hbar^{5}}{k_{B}^{2} m^{2}} \right)  \dfrac{\nbar^{3}}{(m^{\ast}_{\ua}/m)^{4}} 
                                         \dfrac{1}{|\tilde{A}^{\ua \ua}_{f,0}|^{2} } $ \\ 
$D$        &  $\pol = 0$  &   $D T^{2} \ln{\left(\frac{2T_{F}}{T}\right)} = \left(\dfrac{\pi \hbar^{5}}{k_{B}^{2} m^{3}} \right)  c(1) \dfrac{\nbar^{2}}{(m^{\ast}/m)^{5}} 
                                         \dfrac{1 + F^{a}_{0}}{|\tilde{A}^{\ua \da}_{b,0}|^{2} } $ \\ 
\end{tabular}
\end{ruledtabular}
\end{table}

At  $\pol = 0$,   $\kappa T \ln{(2T_{F} /T )}$ decreases monotonically with increasing density. This behavior is dominated by the increase in the effective mass.   There is additional density dependence carried by the scattering amplitudes.   In Table~\ref{tab:GraphiteA} we include the most important scattering amplitudes for the \he3 system of Table~\ref{tab:GraphiteP0}.   The contribution of the scattering amplitudes to $\kappa$ is shown in the second column of Table~\ref{tab:GraphiteA}.  The quantity $|\tilde{A}^{\ua \ua}_{f,0}|^{2} + |\tilde{A}^{\ua \da}_{f,0}|^{2} + |\tilde{A}^{\ua \da}_{b,0}|^{2}$ is non-monotonic, however,  the extent of variation is small compared to that of the effective mass.  This is also the case for the spin diffusion coefficient.  The important scattering amplitude in this case is $|\tilde{A}^{\ua \da}_{b,0}|^{2}$  which is fairly constant. Thus,   the density dependence of the spin diffusion coefficient  is  dominated by that of the effective mass.  The case of $\eta T^{2}$ is more intriguing, as it seems first to increase, and then to decrease with density. At low density, the cubic density dependence  dominates the viscosity.  Thus at low densities the  viscosity increases with increasing $\nbar$.  At higher densities the increase in the effective mass eventually takes over, and the viscosity starts to decrease. This ``bump" feature is also present in the density dependent behavior of $T_{F}$.  

In contrast to zero polarization, at full polarization the transport coefficients exhibit an \textit{increase} with increasing density. This behavior can be understood by referring to the effective masses at $\pol = 1$ shown in Table~\ref{tab:GraphiteP0}.  In the limit of full polarization the  effective mass  shows only a slow increase with increasing density.  Thus, at full polarization the explicit increases with density dominate the small increases in the effective masses.

\begin{table}[]
\caption{\label{tab:GraphiteP0} \he3 on a graphite substrate. The zero-polarization $\pol = 0$  and full polarization $\pol = 1$   thermal conductivity $\kappa$, shear viscosity $\eta$, and spin diffusion coefficient $D$ as functions of areal density $\nbar$ with the explicit temperature dependencies factored out. The units of $\kappa T $ are (10$^{-5}$ erg  s$^{-1}$) ; the units of $\eta T^{2}$ are (10$^{-9}$ g s$^{-1}$ mK$^{2}$); the units of $D T^{2}$ are  (cm$^{2}$ s$^{-1}$ mK$^{2}$).  We also include the  effective masses, the  Fermi energies, and we note that $D(\pol = 1) = 0$ from its definition. The values of $\kappa$ are obtained from (\ref{eq:kappaP0}) and (\ref{eq:kappaP1}), for $\eta$ from (\ref{eq:etaP0l0}) and (\ref{eq:etaP1}), and for $D$ from (\ref{eq:DP0}), for zero and full polarization respectively.  }
\begin{ruledtabular}
\begin{tabular}{cccccccccc}
Density (\AA$^{-2}$) & \multicolumn{2}{c}{$m^{\ast}/m$} & \multicolumn{2}{c}{$\epsilon_{F}$ (K)} & \multicolumn{2}{c}{$\kappa T \ln{(2 T_{F}/T)}$}  & \multicolumn{2}{c}{$\eta T^{2} $} & \multicolumn{1}{c}{$D T^{2} \ln{(2 T_{F}/T)}$} \\
&$\pol = 0$&$\pol = 1$&$\pol = 0$&$\pol = 1$&$\pol = 0$&$\pol = 1$&$\pol = 0$&$\pol = 1$&$\pol = 0$\\
\hline
0.013  & 1.29 & 0.82    & 0.52  & 1.64  &  0.138     &   5.4     & 4.88    & $0.38  \times 10^{3}$  & 5.42   \\
0.025  & 1.72 & 0.81    & 0.75  & 3.16  &  0.123     &  10.4    & 8.34    & $1.41  \times 10^{3}$  & 2.67   \\
0.037  & 2.64 & 0.86    & 0.72  & 4.36  &  0.047     &  15.2    & 4.70    & $3.01  \times 10^{3}$  & 0.47   \\
0.046  & 3.66 & 0.92    & 0.64  & 5.10  &  0.040     &  18.7    & 4.91    & $4.61  \times 10^{3}$  & 0.29   \\
0.054  & 4.88 & 0.95    & 0.57  & 5.82  &  0.014     &  23.3    & 2.06    & $6.79  \times 10^{3}$  & 0.06   
\end{tabular}
\end{ruledtabular}
\end{table}

\begin{table}[]
\caption{\label{tab:GraphiteA} \he3 on a graphite substrate. The dimensionless scattering amplitudes $\tilde{A}_{f/b, \, 0}^{\sigma \sigma'}$, and Landau parameter $F^{a}_{0}$ that are the input into calculating the transport coefficients shown in Table~\ref{tab:GraphiteP0}.  }
\begin{ruledtabular}
\begin{tabular}{ccccc}
Density (\AA$^{-2}$) & $|\tilde{A}^{\ua \ua}_{f,0}|^{2} + |\tilde{A}^{\ua \da}_{f,0}|^{2} + |\tilde{A}^{\ua \da}_{b,0}|^{2}$ & $F^{a}_{0}$ & $|\tilde{A}^{\ua \da}_{b,0}|^{2}$  & $|\tilde{A}^{\ua \ua}_{f,0}|^{2}$ \\
&$\pol = 0$&$\pol = 0$&$\pol = 0$&$\pol = 1$\\
\hline
0.013  & 1.12    & -0.51  & 0.64 & 0.35    \\
0.025  & 1.44    & -0.62  & 0.88 & 0.68    \\
0.037  & 1.46    & -0.72  & 0.93 & 0.79    \\
0.046  & 0.72    & -0.71  & 0.46 & 0.78    \\
0.054  & 0.90    & -0.79  & 0.59 & 0.76    
\end{tabular}
\end{ruledtabular}
\end{table}

In Tables~\ref{tab:He4314P0}, \ref{tab:He4314A}, and \ref{tab:He4433P0}, \ref{tab:He4433A} we show analogous results for \he3 adsorbed to {3.14~\AA} and {4.33~\AA} superfluid \he4 films, respectively. It is important to note that these results are restricted to a much smaller density range than for \he3 on the second layer of graphite. The reason for this difference is that in the superfluid \he4 environment the \he3 undergoes a transition to a transverse excited state at an areal density $\nbar = 0.036$~\AA$^{-2}$.~\cite{AMH1999} The data in Tables~\ref{tab:He4314P0}-\ref{tab:He4433A} cover a density range less than the first three data points in Tables~\ref{tab:GraphiteP0}, \ref{tab:GraphiteA}.  Using Tables~\ref{tab:GraphiteA}, \ref{tab:He4314A}, and \ref{tab:He4433A}, we can compare the Landau parameter $F_{0}^{a}$, and some of the scattering amplitudes for the two substrates.  Over the same density range $F_{0}^{a}$ is markedly smaller in magnitude in the mixture film than on graphite.  The denominators for $\kappa$ and $\eta$ $|\tilde{A}^{\ua \ua}_{f,0}|^{2} + |\tilde{A}^{\ua \da}_{f,0}|^{2} + |\tilde{A}^{\ua \da}_{b,0}|^{2}$ are considerably  smaller for the mixture films than for graphite. We also note that the effective masses only increase moderately with increasing density. However, they are fairly constant over the density range of interest.  As a consequence, we can identify the density $\nbar$ as the major component driving the increases in  $\kappa T \ln{(2 T_{F}/T)}$ and $\eta T^{2} $ for $\pol = 0$ and $\pol = 1$.

On the other hand, for the mixture film in Table~\ref{tab:He4314P0} $D T^{2} \ln{(2 T_{F} / T)}$ follows an irregular pattern with increasing density. This is primarily due to the drastic variation of $|\tilde{A}^{\ua \da}_{b,0}|^{2}$ as can be seen in column 4 Table~\ref{tab:He4314A}.  This behavior is due to the fact that $\tilde{A}^{\ua \da}_{b,0}$ is calculated from Eq.~(\ref{eq:aback}): $\tilde{A}^{\ua \da}_{b,0} = \tilde{A}^{\ua \da}_{f,0} - \tilde{A}^{\ua \ua}_{f,0} $.  Thus a small change in the difference between the two forward scattering amplitudes can result in a significant change in the backward scattering amplitude.  We note that in comparing the denominators of the mixture film  transport coefficients $|\tilde{A}^{\ua \da}_{b,0}|^{2} \ll  |\tilde{A}^{\ua \ua}_{f,0}|^{2} + |\tilde{A}^{\ua \da}_{f,0}|^{2} + |\tilde{A}^{\ua \da}_{b,0}|^{2}$, and thus we expect for example  that the spin diffusion coefficient for the mixture films may be the most sensitive quantity with regard to our use of the lowest order ``$\ell = 0$'' approximation for numerical calculations.  
\begin{table}[]
\caption{\label{tab:He4314P0} \he3 in a 3.14~layer film of \he4. The zero-polarization $\pol = 0$  and full polarization $\pol = 1$  thermal conductivity $\kappa$, shear viscosity $\eta$, and spin diffusion coefficient $D$ as functions of areal density $\nbar$ with the explicit temperature dependencies factored out. The units of $\kappa T $ are (10$^{-5}$ erg  s$^{-1}$) ; the units of $\eta T^{2}$ are (10$^{-7}$ g s$^{-1}$ mK$^{2}$); the units of $D T^{2}$ are  (10$^{2}$ cm$^{2}$ s$^{-1}$ mK$^{2}$). For this mixture film $m_{H} = 1.56 m$ is the hydrodynamic effective mass.~\cite{PhysRevB.73.012507} The values of $\kappa$ are obtained from (\ref{eq:kappaP0}) and (\ref{eq:kappaP1}), for $\eta$ from (\ref{eq:etaP0l0}) and (\ref{eq:etaP1}), and for $D$ from (\ref{eq:DP0}), for zero and full polarization respectively.   }
\begin{ruledtabular}
\begin{tabular}{cccccccccc}
Density (\AA$^{-2}$) & \multicolumn{2}{c}{$m^{\ast}/m_{H}$} & \multicolumn{2}{c}{$\epsilon_{F}$ (K)} & \multicolumn{2}{c}{$\kappa T \ln{(2 T_{F}/T)}$}  & \multicolumn{2}{c}{$\eta T^{2} $} & \multicolumn{1}{c}{$D T^{2} \ln{(2 T_{F}/T)}$} \\
&$\pol = 0$&$\pol = 1$&$\pol = 0$&$\pol = 1$&$\pol = 0$&$\pol = 1$&$\pol = 0$&$\pol = 1$&$\pol = 0$\\
\hline
0.013  & 1.31 & 0.84    & 0.32  & 0.99  &  0.285     &  0.92     & 0.098    & 0.63    & 1.34   \\
0.016  & 1.47 & 0.87    & 0.36  & 1.21  &  0.341     &  1.24     & 0.147    & 1.07    & 3.33   \\
0.019  & 1.60 & 0.88    & 0.39  & 1.43  &  0.398     &  1.63     & 0.206    & 1.69    & 5.21   \\
0.024  & 1.70 & 0.88    & 0.46  & 1.78  &  0.510     &  2.52     & 0.330    & 3.26    & 1.99   \\
0.029  & 1.77 & 0.88    & 0.53  & 2.14  &  0.615     &  3.66     & 0.476    & 5.67    & 1.07   
\end{tabular}
\end{ruledtabular}
\end{table}
\begin{table}[]
\caption{\label{tab:He4314A} \he3 in a 3.14~layer film of \he4. The dimensionless scattering amplitudes $\tilde{A}_{f/b, \, 0}^{\sigma \sigma'}$, and Landau parameter $F^{a}_{0}$ that are the input into calculating the transport coefficients shown in Table~\ref{tab:He4314P0}.  }
\begin{ruledtabular}
\begin{tabular}{ccccc}
Density (\AA$^{-2}$) & $|\tilde{A}^{\ua \ua}_{f,0}|^{2} + |\tilde{A}^{\ua \da}_{f,0}|^{2} + |\tilde{A}^{\ua \da}_{b,0}|^{2}$ & $F^{a}_{0}$ & $|\tilde{A}^{\ua \da}_{b,0}|^{2}$  & $|\tilde{A}^{\ua \ua}_{f,0}|^{2}$ \\
&$\pol = 0$&$\pol = 0$&$\pol = 0$&$\pol = 1$\\
\hline
0.013  & 0.20 & -0.11    & 0.0098  & 0.71    \\
0.016  & 0.16 & -0.08    & 0.0036  & 0.75    \\
0.019  & 0.14 & -0.07    & 0.0022  & 0.76    \\
0.024  & 0.14 & -0.12    & 0.0063  & 0.76    \\
0.029  & 0.14 & -0.17    & 0.0132  & 0.77    
\end{tabular}
\end{ruledtabular}
\end{table}

\begin{table}[]
\caption{\label{tab:He4433P0} \he3 in a 4.33~layer film of \he4. The zero-polarization $\pol = 0$  and full polarization $\pol = 1$  thermal conductivity $\kappa$, shear viscosity $\eta$, and spin diffusion coefficient $D$ as functions of areal density $\nbar$ with the explicit temperature dependencies factored out. The units of $\kappa T $ are (10$^{-5}$ erg  s$^{-1}$) ; the units of $\eta T^{2}$ are (10$^{-7}$ g s$^{-1}$ mK$^{2}$); the units of $D T^{2}$ are  (10$^{2}$ cm$^{2}$ s$^{-1}$ mK$^{2}$). For this mixture film $m_{H} = 1.29 m$ is the hydrodynamic effective mass.~\cite{HoHallock2001} The values of $\kappa$ are obtained from (\ref{eq:kappaP0}) and (\ref{eq:kappaP1}), for $\eta$ from (\ref{eq:etaP0l0}) and (\ref{eq:etaP1}), and for $D$ from (\ref{eq:DP0}), for zero and full polarization respectively.  }
\begin{ruledtabular}
\begin{tabular}{cccccccccc}
Density (\AA$^{-2}$) & \multicolumn{2}{c}{$m^{\ast}/m_{H}$} & \multicolumn{2}{c}{$\epsilon_{F}$ (K)} & \multicolumn{2}{c}{$\kappa T \ln{(2 T_{F}/T)}$}  & \multicolumn{2}{c}{$\eta T^{2} $} & \multicolumn{1}{c}{$D T^{2} \ln{(2 T_{F}/T)}$} \\
&$\pol = 0$&$\pol = 1$&$\pol = 0$&$\pol = 1$&$\pol = 0$&$\pol = 1$&$\pol = 0$&$\pol = 1$&$\pol = 0$\\
\hline
0.015  & 1.22 & 0.83    & 0.50  & 1.46  &  0.68     &  2.29     & 0.28    & 1.88    & 1.74   \\
0.019  & 1.32 & 0.84    & 0.56  & 1.77  &  0.80     &  3.03     & 0.40    & 3.03    & 2.05   \\
0.022  & 1.37 & 0.84    & 0.62  & 2.03  &  0.93     &  3.87     & 0.53    & 4.46    & 1.93   \\
0.025  & 1.40 & 0.84    & 0.70  & 2.34  &  1.08     &  5.18     & 0.72    & 6.89    & 1.61   \\
0.028  & 1.45 & 0.84    & 0.76  & 2.63  &  1.21     &  6.40     & 0.91    & 9.60    & 1.59   \\
0.031  & 1.50 & 0.84    & 0.82  & 2.92  &  2.17     &  7.73     & 1.08    & 12.9    & 1.42 
\end{tabular}
\end{ruledtabular}
\end{table}
\begin{table}[]
\caption{\label{tab:He4433A} \he3 in a 4.33~layer film of \he4. The dimensionless scattering amplitudes $\tilde{A}_{f/b, \, 0}^{\sigma \sigma'}$, and Landau parameter $F^{a}_{0}$ that are the input into calculating the transport coefficients shown in Table~\ref{tab:He4433P0}.  }
\begin{ruledtabular}
\begin{tabular}{ccccc}
Density (\AA$^{-2}$) & $|\tilde{A}^{\ua \ua}_{f,0}|^{2} + |\tilde{A}^{\ua \da}_{f,0}|^{2} + |\tilde{A}^{\ua \da}_{b,0}|^{2}$ & $F^{a}_{0}$ & $|\tilde{A}^{\ua \da}_{b,0}|^{2}$  & $|\tilde{A}^{\ua \ua}_{f,0}|^{2}$ \\
&$\pol = 0$&$\pol = 0$&$\pol = 0$&$\pol = 1$\\
\hline
0.015  & 0.23 & -0.16    & 0.03  & 0.64    \\
0.019  & 0.21 & -0.16    & 0.02  & 0.68    \\
0.022  & 0.21 & -0.18    & 0.03  & 0.71    \\
0.025  & 0.22 & -0.21    & 0.04  & 0.71    \\
0.028  & 0.22 & -0.22    & 0.04  & 0.72    \\
0.031  & 0.22 & -0.24    & 0.04  & 0.73  
\end{tabular}
\end{ruledtabular}
\end{table}

\begin{figure}
\includegraphics[]{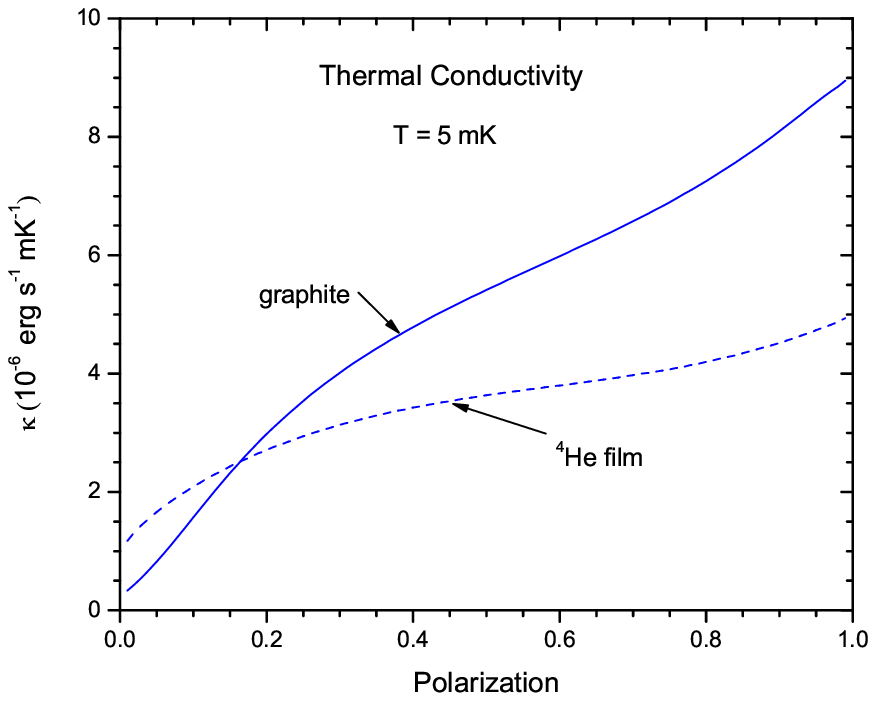}
\caption{\label{fig:KappaT5mK} The thermal conductivity $\kappa$  (\ref{eq:kappafinal}) as a function of polarization for \he3 on substrates of  graphite (solid line), and a $4.33$~\AA \,  superfluid \he4 film (dashed line).  Both results are at $T = 5$~mK, and the \he3 areal densities are $0.0252$~\AA$^{-2}$ and $0.0248$~\AA$^{-2}$ on graphite and \he4, respectively. }
\vspace{1truein}
\end{figure}

\begin{figure}
\includegraphics[]{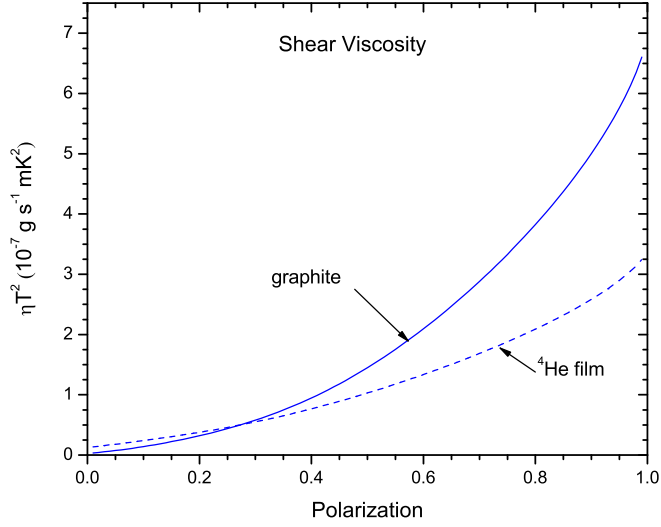}
\caption{\label{fig:ViscosityT5mK} The shear viscosity $\eta$ (\ref{eq:etaPell0})  times temperature squared  as a function of polarization for \he3 on substrates of  graphite (solid line), and a $4.33$~\AA \,  superfluid \he4 film (dashed line).  The results are shown for \he3 areal densities of $0.0252$~\AA$^{-2}$ and $0.0248$~\AA$^{-2}$ on graphite and \he4, respectively. }
\vspace{1truein}
\end{figure}

\begin{figure}
\includegraphics[]{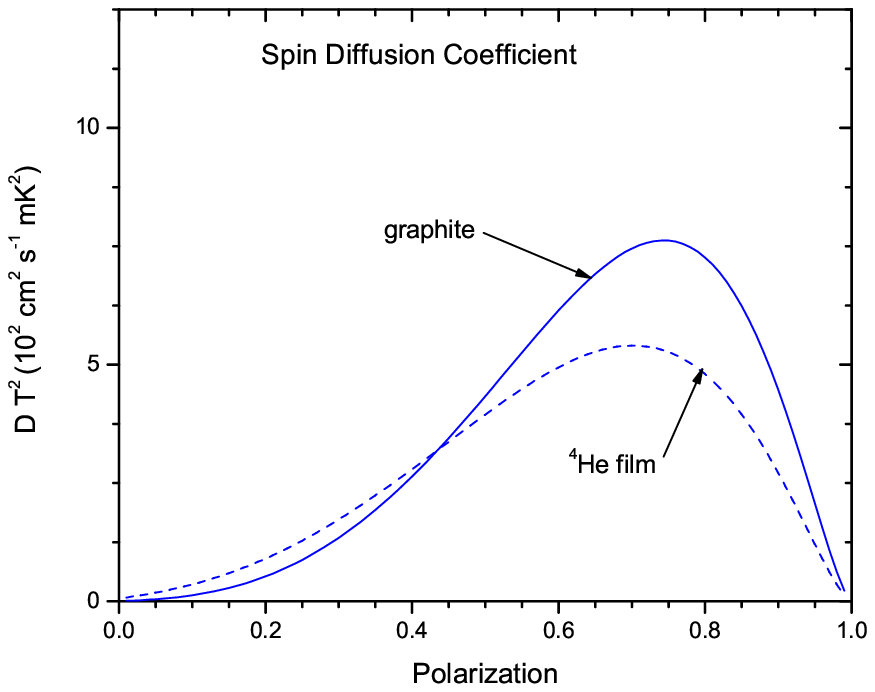}
\caption{\label{fig:DT5mK} The spin diffusion coefficient $D$ (\ref{eq:Dfinal}) times temperature squared  as a function of polarization for \he3 on substrates of  graphite (solid line), and a $4.33$~\AA \,  superfluid \he4 film (dashed line).  The results are shown for \he3 areal densities of $0.0252$~\AA$^{-2}$ and $0.0248$~\AA$^{-2}$ on graphite and \he4, respectively.  }
\vspace{1truein}
\end{figure}

In Figs.~\ref{fig:KappaT5mK}, \ref{fig:ViscosityT5mK}, and \ref{fig:DT5mK} we show the \textit{polarization} dependence of the thermal conductivity, shear viscosity times temperature squared, and the spin diffusion coefficient times temperature squared, respectively, for \he3 on graphite at $\nbar = 0.0252$~\AA$^{-2}$, and \he3 on a $4.33$~\AA \, \he4 film at $\nbar = 0.0248$~\AA$^{-2}$.  The data for $\kappa$ were calculated at a temperature $T = 5$~mK which was chosen to ensure that the inequality $T < T_{F \, \da}$ is obeyed at all polarizations.  The units for $\kappa$  are different in the figure than in the table  because for this quantity the temperature dependence is not factorable for $0 < \pol < 1$. The major prediction for this section then is that $\kappa$ and $\eta$ increase dramatically, by an order of magnitude, as $\pol$ increases from 0 to 1 for both substrates.  The spin diffusion coefficient goes through a similar large increase from its zero-polarization value to its maximum value in the region $\pol \approx 0.74$ for both graphite and \he4, and then vanishes in the full polarization limit.  $D$ vanishes like $(1 - \pol)^{3/2}$ in the limit of full polarization (see Sec.~\ref{sec:diff}).  Thus we predict an increase in $D$ from zero polarization to its maximum value of $1 \sim 2$ orders of magnitude.

In a recent interesting development, Kovtun, Son, and Starinets~\cite{KSS2005} have conjectured that there is a \textit{universal} lower bound to the ratio of the the shear viscosity to entropy density:
\begin{equation} \label{eq:Lowerbound}
4 \pi \frac{\eta/\hbar}{s/k_{B}} \geq 1 \,,
\end{equation}
where $s = S/A$, and $S$ is the entropy.  The authors describe the distance from the lower bound as a way to characterize how close a fluid is to being perfect. They argue that possible systems that may satisfy the lower bound ought to be strongly interacting systems that are normally characterized by a small viscosity (i.e. a small mean free path). They suggested that quark-gluon plasmas, and ultra-cold gases at the unitarity limit are candidates. There is now evidence that an ultra-cold Fermi gas nearly satisfies the lower bound.~\cite{Thomas2009}  Using the results from Sec.~\ref{sec:visco} we can estimate the value of this expression for a \he3 film. From Fig.~\ref{fig:ViscosityT5mK} we see  that the viscosity is a monotonically increasing function of polarization. The polarization dependent entropy density is given by $s/k_{B} = (\pi/6 \hbar^{2}) \left(m^{*}_{\ua} + m^{*}_{\da}   \right) k_{B}T$.~\cite{LAM_PRB2012} This entropy is a monotonically decreasing function of polarization. Thus, we need only to concern ourselves with the zero-polarization limit. Using (\ref{eq:etaP0l0}) we find for the left hand side:
\begin{align}
4 \pi \frac{\eta/\hbar}{s/k_{B}} &= \frac{9}{\pi^{2}} \frac{(m/m^{*})^{2}}{\left[ |\tilde{A}^{\ua \ua}_{f, 0}|^{2} +  |\tilde{A}^{\ua \da}_{f, 0}|^{2} + |\tilde{A}^{\ua \ua}_{f, 0} - \tilde{A}^{\ua \da}_{f, 0}|^{2} \right]} \left( \frac{T_{F}}{T}  \right)^{3} \nonumber \,, \\
4 \pi \frac{\eta/\hbar}{s/k_{B}} &\approx 0.28 \left( \frac{T_{F}}{T}  \right)^{3} \,,
\end{align}
where we have used (\ref{eq:WAtilde}) to write the transition rates in terms of the dimensionless scattering amplitudes. The numbers come from Table~\ref{tab:GraphiteP0} and Fig.~\ref{fig:amp01}, and so they refer to the second layer of \he3 on graphite at $\nbar = 0.025~\AA^{-2}$ ($T_{F} = 0.74$~K). It is clear from the inverse cubic temperature dependence that deep in the Fermi-liquid regime the system satisfies the lower bound.  At higher temperatures this expression passes through one when $T \approx 0.5$~K. This is not that high, and suggests that at temperatures on the order of 100's~mK the ratio may not be very far from one for this \he3 thin film system. 

\section{Conclusion} \label{sec:Conclusion}

We have derived exact expressions for the transport coefficients $\kappa$ and $\eta$ utilizing methods developed by numerous groups~\cite{BP1991} for application to bulk \he3.  We calculated predicted values for the polarization dependence of $\kappa$, $\eta$, and $D$ for thin, degenerate \he3 films using previously determined Landau parameters.  The key to performing the principal angular integration in phase space is the procedure suggested by Miyake and Mullin~\cite{MiyakeMullin1983,*MiyakeMullin1984} for avoiding a finite temperature singularity. The Miyake-Mullin approach is discussed in detail in Sec.~\ref{sec:tau}.  In that section we derive the polarization dependent expression for the quasiparticle lifetime due to quasiparticle-quasiparticle collisions in the relaxation time approximation. We compare that result with that of a previous derivation of the quasiparticle lifetime using completely different techniques, and note that they are identical up to factor of order one. 

The derivation of the transport coefficients in Sec.~\ref{sec:transport} follows the methods developed by Abrikosov and Khalatnikov,~\cite{AK1958} and Sykes and Brooker.~\cite{SykesBrooker1970}  The calculation of $\kappa$ is very similar to that of the spin diffusion coefficient $D$ as described by Miyake and Mullin.  The collision integral is reduced to an integral eigenvalue problem whose integrand depends on both spin-up and spin-down fluctuations. The system is diagonalized by standard methods, and is reduced to an independent  pair of equations in Sykes-Brooker form. The temperature dependencies for the transport coefficients are in agreement with older work at zero polarization by Fu and Ebner.~\cite{FuEbner1974} Further, we find that, unlike spin diffusion, these dependencies ($T \ln{T}$ for $\kappa$ and $T^{2}$ for $\eta$) are not changed by polarization.  The solution for the shear viscosity is unlike that of any other fermion transport coefficient. The key physics lies in including the contributions of scattering from quasiparticles whose momenta differ slightly from their zero-temperature values but are still allowed by energy and momentum conservation at non-zero temperature.  We introduced a simplified model in which we fix the incoming quasiparticle momenta at the zero-temperature values, and allow the outgoing momenta to vary (see Fig.~\ref{fig:head-on}). We find that in lowest order the viscosity is formally independent of the quasiparticle lifetime (see Eq.~(\ref{eq:etaP0}, for example). We note however that $1/\nu_{0}$ (\ref{eq:nu0}) is very similar to $\tau_{0}$. In Ref.~\onlinecite{Novikov06} Novikov, in the zero-polarization limit, allows all four quasiparticle momenta to drift from their zero-temperature values.  We find at zero polarization, in agreement with Novikov, that the head-on collisions between quasiparticles with momenta in opposite directions dominate the scattering process, and we also find that they are the dominant process in the scattering between spin-parallel quasiparticles at finite polarization.  Our final result for the shear viscosity temperature dependence differs from that of Novikov because Novikov assumes the Landau parameters have a divergence at $\theta = \pi$,  and this gives an extra factor of $\ln^{2}{(T_{F} /T)}$ in the final result for the viscosity.

In Sec.~\ref{sec:he3} we apply these results to a system of thin \he3 films both in the second layer on a graphite substrate, and also in a thin \he3-\he4 film mixture.   In Table~\ref{tab:Tsummary} we gather together the main results from this paper concerning the temperature dependence of the thermal conductivity and the shear viscosity, and we have also included the spin diffusion coefficient results from Miyake and Mullin.~\cite{MiyakeMullin1983,*MiyakeMullin1984}  
\begin{table}[]
\caption{\label{tab:Tsummary} The temperature dependencies of the inverse transport coefficients as a function of polarization. The thermal conductivity $\kappa$ and the shear viscosity $\eta$ are calculated in Sec.~\ref{sec:transport}.  The spin diffusion coefficient $D$  is from Ref.~\onlinecite{MiyakeMullin1983,*MiyakeMullin1984}.  We note that $D^{-1}$ is undefined at $\pol = 1$. }
\begin{ruledtabular}
\begin{tabular}{ccc}
Coefficient & $\pol = 0$ & $0 < \pol \leq 1$ \\
\hline
$\kappa^{-1}$  & $T \ln{T}$        & $T \ln{T}$   \\
$\eta^{-1}$      & $T^{2}$           & $T^{2}$      \\
$D^{-1}$          & $T^{2} \ln{T}$ & $T^{2}$      \\
\end{tabular}
\end{ruledtabular}
\end{table}

The predicted polarization dependence of the transport coefficients for \he3 on the second layer of graphite and also for the 4.33~\AA-thick  \he4 film is shown in  Figs.~\ref{fig:KappaT5mK}, \ref{fig:ViscosityT5mK}, and \ref{fig:DT5mK}.  These results show a dramatic increase in the magnitudes of the coefficients as the polarization increases from zero. We showed in Sec.~\ref{sec:therm} that for the thermal conductivity in two dimensions $\kappa$ is proportional to the quasiparticle lifetime. Further, we showed in previous work, see Fig.~7 in Ref.~\onlinecite{LAM_PRB2012}, that the magnitude of the contribution to the quasi-particle lifetime  from the majority spin component \textit{decreases} dramatically as a function of increasing polarization.   Thus, for the thermal conductivity a fairly simple qualitative picture emerges of the role of polarization: increasing $\pol$ induces an increase in the quasiparticle lifetime, and thus the transport coefficient. For very dilute systems this mechanism is basically understood as the quenching of $s$-wave scattering with increased $\pol$.  However for the shear viscosity such a simple picture does not seem to be relevant if for no other reason than because the quasi-particle lifetime does not contribute directly to the transport coefficient. In this case we must consider instead the complicated dynamical question of the relative importance of the spin anti-parallel head-on scattering to the spin parallel head-on scattering as per the discussion in Sec.~\ref{sec:visco}, which itself is related to the balance of $s$-wave and $p$-wave scattering.

In lowest order of temperature the derivation of the expressions for the transport coefficients is essentially exact. The calculation of explicit results for \he3 films suffers from the use of the $\ell = 0$ approximation for the scattering amplitudes. An improvement in the present results would be the inclusion of additional Fourier components in the expressions for the transition probabilities in terms of the scattering amplitudes, see Eq.~(\ref{eq:WFourier}). The approximations used in the determination of the \he3 film Landau parameters from experimental measurements of the specific heat effective mass, and the spin susceptibility have been discussed in Ref.~\onlinecite{LAM_PRB2012}.

At this time to the best of our knowledge there have been no measurements of any transport coefficient in a thin \he3 film. In addition there have been no measurements at all in a polarized thin \he3 film. These experiments would be very difficult. In fact the first measurement of zero sound in a thin, unpolarized \he 3 film was only reported in 2010 by Godfrin,  Meschke,  Lauter,  B\"{o}hm,  Krotscheck,  and Panholzer.~\cite{GodfrinEtAl2010}  Our Landau parameters do yield excellent agreement with this zero sound measurement. For bulk \he3 there has been some work on the polarization dependence of transport coefficients. A recent review~\cite{BPWChapter2005} summarizes the state of the field.


\newpage

\begin{thebibliography}{37}%
\makeatletter
\providecommand \@ifxundefined [1]{%
 \@ifx{#1\undefined}
}%
\providecommand \@ifnum [1]{%
 \ifnum #1\expandafter \@firstoftwo
 \else \expandafter \@secondoftwo
 \fi
}%
\providecommand \@ifx [1]{%
 \ifx #1\expandafter \@firstoftwo
 \else \expandafter \@secondoftwo
 \fi
}%
\providecommand \natexlab [1]{#1}%
\providecommand \enquote  [1]{``#1''}%
\providecommand \bibnamefont  [1]{#1}%
\providecommand \bibfnamefont [1]{#1}%
\providecommand \citenamefont [1]{#1}%
\providecommand \href@noop [0]{\@secondoftwo}%
\providecommand \href [0]{\begingroup \@sanitize@url \@href}%
\providecommand \@href[1]{\@@startlink{#1}\@@href}%
\providecommand \@@href[1]{\endgroup#1\@@endlink}%
\providecommand \@sanitize@url [0]{\catcode `\\12\catcode `\$12\catcode
  `\&12\catcode `\#12\catcode `\^12\catcode `\_12\catcode `\%12\relax}%
\providecommand \@@startlink[1]{}%
\providecommand \@@endlink[0]{}%
\providecommand \url  [0]{\begingroup\@sanitize@url \@url }%
\providecommand \@url [1]{\endgroup\@href {#1}{\urlprefix }}%
\providecommand \urlprefix  [0]{URL }%
\providecommand \Eprint [0]{\href }%
\providecommand \doibase [0]{http://dx.doi.org/}%
\providecommand \selectlanguage [0]{\@gobble}%
\providecommand \bibinfo  [0]{\@secondoftwo}%
\providecommand \bibfield  [0]{\@secondoftwo}%
\providecommand \translation [1]{[#1]}%
\providecommand \BibitemOpen [0]{}%
\providecommand \bibitemStop [0]{}%
\providecommand \bibitemNoStop [0]{.\EOS\space}%
\providecommand \EOS [0]{\spacefactor3000\relax}%
\providecommand \BibitemShut  [1]{\csname bibitem#1\endcsname}%
\let\auto@bib@innerbib\@empty
\bibitem [{\citenamefont {Landau}(1956)}]{Landau56}%
  \BibitemOpen
  \bibfield  {author} {\bibinfo {author} {\bibfnamefont {L.~D.}\ \bibnamefont
  {Landau}},\ }\href@noop {} {\bibfield  {journal} {\bibinfo  {journal}
  {Zh.~Eskp.~Teor.~Fiz.}\ }\textbf {\bibinfo {volume} {30}},\ \bibinfo {pages}
  {1058} (\bibinfo {year} {1956})},\ \bibinfo {note} {[Sov. Phys. JETP
  \textbf{3}, 920-925 (1957)]}\BibitemShut {NoStop}%
\bibitem [{\citenamefont {Landau}(1957)}]{Landau57}%
  \BibitemOpen
  \bibfield  {author} {\bibinfo {author} {\bibfnamefont {L.~D.}\ \bibnamefont
  {Landau}},\ }\href@noop {} {\bibfield  {journal} {\bibinfo  {journal}
  {Zh.~Eskp.~Teor.~Fiz.}\ }\textbf {\bibinfo {volume} {32}},\ \bibinfo {pages}
  {59} (\bibinfo {year} {1957})},\ \bibinfo {note} {[Sov. Phys. JETP
  \textbf{5}, 101-108 (1957)]}\BibitemShut {NoStop}%
\bibitem [{\citenamefont {Landau}(1958)}]{Landau58}%
  \BibitemOpen
  \bibfield  {author} {\bibinfo {author} {\bibfnamefont {L.~D.}\ \bibnamefont
  {Landau}},\ }\href@noop {} {\bibfield  {journal} {\bibinfo  {journal}
  {Zh.~Eskp.~Teor.~Fiz.}\ }\textbf {\bibinfo {volume} {35}},\ \bibinfo {pages}
  {97} (\bibinfo {year} {1958})},\ \bibinfo {note} {[Sov. Phys. JETP
  \textbf{8}, 70-74 (1959)]}\BibitemShut {NoStop}%
\bibitem [{\citenamefont {Abrikosov}\ and\ \citenamefont
  {Khalatnikov}(1957)}]{AK1958}%
  \BibitemOpen
  \bibfield  {author} {\bibinfo {author} {\bibfnamefont {A.~A.}\ \bibnamefont
  {Abrikosov}}\ and\ \bibinfo {author} {\bibfnamefont {I.~M.}\ \bibnamefont
  {Khalatnikov}},\ }\href@noop {} {\bibfield  {journal} {\bibinfo  {journal}
  {Zh.~Eskp.~Teor.~Fiz.}\ }\textbf {\bibinfo {volume} {33}},\ \bibinfo {pages}
  {1154} (\bibinfo {year} {1957})},\ \bibinfo {note} {[Sov. Phys. JETP
  \textbf{6}, 888-892 (1958)]}\BibitemShut {NoStop}%
\bibitem [{\citenamefont {Pines}\ and\ \citenamefont
  {Nozi{\`e}res}(1966)}]{PinesNoz1966}%
  \BibitemOpen
  \bibfield  {author} {\bibinfo {author} {\bibfnamefont {D.}~\bibnamefont
  {Pines}}\ and\ \bibinfo {author} {\bibfnamefont {P.}~\bibnamefont
  {Nozi{\`e}res}},\ }\href@noop {} {\emph {\bibinfo {title} {The Theory of
  Quantum Liquids}}}\ (\bibinfo  {publisher} {W. A. Benjamin, Inc},\ \bibinfo
  {address} {New York},\ \bibinfo {year} {1966})\BibitemShut {NoStop}%
\bibitem [{\citenamefont {Baym}\ and\ \citenamefont {Pethick}(1991)}]{BP1991}%
  \BibitemOpen
  \bibfield  {author} {\bibinfo {author} {\bibfnamefont {G.}~\bibnamefont
  {Baym}}\ and\ \bibinfo {author} {\bibfnamefont {C.}~\bibnamefont {Pethick}},\
  }\href@noop {} {\emph {\bibinfo {title} {Landau Fermi-Liquid Theory}}}\
  (\bibinfo  {publisher} {Wiley},\ \bibinfo {address} {New York},\ \bibinfo
  {year} {1991})\BibitemShut {NoStop}%
\bibitem [{\citenamefont {Anderson}\ \emph {et~al.}(1987)\citenamefont
  {Anderson}, \citenamefont {Pethick},\ and\ \citenamefont {Quader}}]{APQ1987}%
  \BibitemOpen
  \bibfield  {author} {\bibinfo {author} {\bibfnamefont {R.~H.}\ \bibnamefont
  {Anderson}}, \bibinfo {author} {\bibfnamefont {C.~J.}\ \bibnamefont
  {Pethick}}, \ and\ \bibinfo {author} {\bibfnamefont {K.~F.}\ \bibnamefont
  {Quader}},\ }\href {\doibase 10.1103/PhysRevB.35.1620} {\bibfield  {journal}
  {\bibinfo  {journal} {Phys. Rev. B}\ }\textbf {\bibinfo {volume} {35}},\
  \bibinfo {pages} {1620} (\bibinfo {year} {1987})}\BibitemShut {NoStop}%
\bibitem [{\citenamefont {Meyerovich}(1983)}]{Meyerovich1983}%
  \BibitemOpen
  \bibfield  {author} {\bibinfo {author} {\bibfnamefont {A.}~\bibnamefont
  {Meyerovich}},\ }\href {\doibase 10.1007/BF00682490} {\bibfield  {journal}
  {\bibinfo  {journal} {J. Low Temp. Phys.}\ }\textbf {\bibinfo {volume}
  {53}},\ \bibinfo {pages} {487} (\bibinfo {year} {1983})}\BibitemShut
  {NoStop}%
\bibitem [{\citenamefont {Buu}\ \emph {et~al.}(1999)\citenamefont {Buu},
  \citenamefont {Forbes}, \citenamefont {Puech},\ and\ \citenamefont
  {Wolf}}]{Buu...1999}%
  \BibitemOpen
  \bibfield  {author} {\bibinfo {author} {\bibfnamefont {O.}~\bibnamefont
  {Buu}}, \bibinfo {author} {\bibfnamefont {A.~C.}\ \bibnamefont {Forbes}},
  \bibinfo {author} {\bibfnamefont {L.}~\bibnamefont {Puech}}, \ and\ \bibinfo
  {author} {\bibfnamefont {P.~E.}\ \bibnamefont {Wolf}},\ }\href {\doibase
  10.1103/PhysRevLett.83.3466} {\bibfield  {journal} {\bibinfo  {journal}
  {Phys. Rev. Lett.}\ }\textbf {\bibinfo {volume} {83}},\ \bibinfo {pages}
  {3466} (\bibinfo {year} {1999})}\BibitemShut {NoStop}%
\bibitem [{\citenamefont {Akimoto}\ \emph {et~al.}(2002)\citenamefont
  {Akimoto}, \citenamefont {Xia}, \citenamefont {Adams}, \citenamefont
  {Candela}, \citenamefont {Mullin},\ and\ \citenamefont
  {Sullivan}}]{Akimoto...2002}%
  \BibitemOpen
  \bibfield  {author} {\bibinfo {author} {\bibfnamefont {H.}~\bibnamefont
  {Akimoto}}, \bibinfo {author} {\bibfnamefont {J.~S.}\ \bibnamefont {Xia}},
  \bibinfo {author} {\bibfnamefont {E.~D.}\ \bibnamefont {Adams}}, \bibinfo
  {author} {\bibfnamefont {D.}~\bibnamefont {Candela}}, \bibinfo {author}
  {\bibfnamefont {W.~J.}\ \bibnamefont {Mullin}}, \ and\ \bibinfo {author}
  {\bibfnamefont {N.~S.}\ \bibnamefont {Sullivan}},\ }\href {\doibase
  10.1142/S0217979202013705} {\bibfield  {journal} {\bibinfo  {journal} {Int.
  J. Mod. Phys. B}\ }\textbf {\bibinfo {volume} {16}},\ \bibinfo {pages} {3117}
  (\bibinfo {year} {2002})}\BibitemShut {NoStop}%
\bibitem [{\citenamefont {Sawkey}\ \emph {et~al.}(2006)\citenamefont {Sawkey},
  \citenamefont {Puech},\ and\ \citenamefont {Wolf}}]{Sawkey...2006}%
  \BibitemOpen
  \bibfield  {author} {\bibinfo {author} {\bibfnamefont {D.}~\bibnamefont
  {Sawkey}}, \bibinfo {author} {\bibfnamefont {L.}~\bibnamefont {Puech}}, \
  and\ \bibinfo {author} {\bibfnamefont {P.~E.}\ \bibnamefont {Wolf}},\ }\href
  {\doibase 10.1103/PhysRevLett.96.215301} {\bibfield  {journal} {\bibinfo
  {journal} {Phys. Rev. Lett.}\ }\textbf {\bibinfo {volume} {96}},\ \bibinfo
  {pages} {215301} (\bibinfo {year} {2006})}\BibitemShut {NoStop}%
\bibitem [{\citenamefont {Sykes}\ and\ \citenamefont
  {Brooker}(1970)}]{SykesBrooker1970}%
  \BibitemOpen
  \bibfield  {author} {\bibinfo {author} {\bibfnamefont {J.}~\bibnamefont
  {Sykes}}\ and\ \bibinfo {author} {\bibfnamefont {G.~A.}\ \bibnamefont
  {Brooker}},\ }\href@noop {} {\bibfield  {journal} {\bibinfo  {journal} {Ann.
  Phys. (NY)}\ }\textbf {\bibinfo {volume} {56}},\ \bibinfo {pages} {1}
  (\bibinfo {year} {1970})}\BibitemShut {NoStop}%
\bibitem [{\citenamefont {Jensen}\ \emph {et~al.}(1969)\citenamefont {Jensen},
  \citenamefont {Smith},\ and\ \citenamefont {Wilkins}}]{JSW1969}%
  \BibitemOpen
  \bibfield  {author} {\bibinfo {author} {\bibfnamefont {H.~H.}\ \bibnamefont
  {Jensen}}, \bibinfo {author} {\bibfnamefont {H.}~\bibnamefont {Smith}}, \
  and\ \bibinfo {author} {\bibfnamefont {J.~W.}\ \bibnamefont {Wilkins}},\
  }\href {\doibase 10.1103/PhysRev.185.323} {\bibfield  {journal} {\bibinfo
  {journal} {Phys. Rev.}\ }\textbf {\bibinfo {volume} {185}},\ \bibinfo {pages}
  {323} (\bibinfo {year} {1969})}\BibitemShut {NoStop}%
\bibitem [{\citenamefont {Li}\ \emph {et~al.}(2013)\citenamefont {Li},
  \citenamefont {Anderson},\ and\ \citenamefont {Miller}}]{LAM_PRB2013}%
  \BibitemOpen
  \bibfield  {author} {\bibinfo {author} {\bibfnamefont {D.~Z.}\ \bibnamefont
  {Li}}, \bibinfo {author} {\bibfnamefont {R.~H.}\ \bibnamefont {Anderson}}, \
  and\ \bibinfo {author} {\bibfnamefont {M.~D.}\ \bibnamefont {Miller}},\
  }\href@noop {} {\bibfield  {journal} {\bibinfo  {journal} {Phys. Rev. B}\
  }\textbf {\bibinfo {volume} {87}},\ \bibinfo {pages} {104519} (\bibinfo
  {year} {2013})}\BibitemShut {NoStop}%
\bibitem [{\citenamefont {Li}\ \emph {et~al.}(2014)\citenamefont {Li},
  \citenamefont {Anderson}, \citenamefont {Miller},\ and\ \citenamefont
  {Crowell}}]{LAMC2014}%
  \BibitemOpen
  \bibfield  {author} {\bibinfo {author} {\bibfnamefont {D.~Z.}\ \bibnamefont
  {Li}}, \bibinfo {author} {\bibfnamefont {R.~H.}\ \bibnamefont {Anderson}},
  \bibinfo {author} {\bibfnamefont {M.~D.}\ \bibnamefont {Miller}}, \ and\
  \bibinfo {author} {\bibfnamefont {E.}~\bibnamefont {Crowell}},\ }\href
  {\doibase 10.1088/1742-5468/2014/07/P07021} {\bibfield  {journal} {\bibinfo
  {journal} {J. Stat. Mech.}\ }\textbf {\bibinfo {volume} {2014}},\ \bibinfo
  {pages} {P07021} (\bibinfo {year} {2014})}\BibitemShut {NoStop}%
\bibitem [{\citenamefont {Li}\ \emph {et~al.}(2012)\citenamefont {Li},
  \citenamefont {Anderson},\ and\ \citenamefont {Miller}}]{LAM_PRB2012}%
  \BibitemOpen
  \bibfield  {author} {\bibinfo {author} {\bibfnamefont {D.~Z.}\ \bibnamefont
  {Li}}, \bibinfo {author} {\bibfnamefont {R.~H.}\ \bibnamefont {Anderson}}, \
  and\ \bibinfo {author} {\bibfnamefont {M.~D.}\ \bibnamefont {Miller}},\
  }\href {\doibase 10.1103/PhysRevB.85.224511} {\bibfield  {journal} {\bibinfo
  {journal} {Phys. Rev. B}\ }\textbf {\bibinfo {volume} {85}},\ \bibinfo
  {pages} {224511} (\bibinfo {year} {2012})}\BibitemShut {NoStop}%
\bibitem [{\citenamefont {Fu}\ and\ \citenamefont {Ebner}(1974)}]{FuEbner1974}%
  \BibitemOpen
  \bibfield  {author} {\bibinfo {author} {\bibfnamefont {H.-H.}\ \bibnamefont
  {Fu}}\ and\ \bibinfo {author} {\bibfnamefont {C.}~\bibnamefont {Ebner}},\
  }\href {\doibase 10.1103/PhysRevA.10.338} {\bibfield  {journal} {\bibinfo
  {journal} {Phys. Rev. A}\ }\textbf {\bibinfo {volume} {10}},\ \bibinfo
  {pages} {338} (\bibinfo {year} {1974})}\BibitemShut {NoStop}%
\bibitem [{\citenamefont {Miyake}\ and\ \citenamefont
  {Mullin}(1983)}]{MiyakeMullin1983}%
  \BibitemOpen
  \bibfield  {author} {\bibinfo {author} {\bibfnamefont {K.}~\bibnamefont
  {Miyake}}\ and\ \bibinfo {author} {\bibfnamefont {W.~J.}\ \bibnamefont
  {Mullin}},\ }\href {\doibase 10.1103/PhysRevLett.50.197} {\bibfield
  {journal} {\bibinfo  {journal} {Phys. Rev. Lett.}\ }\textbf {\bibinfo
  {volume} {50}},\ \bibinfo {pages} {197} (\bibinfo {year} {1983})}\BibitemShut
  {NoStop}%
\bibitem [{\citenamefont {Miyake}\ and\ \citenamefont
  {Mullin}(1984)}]{MiyakeMullin1984}%
  \BibitemOpen
  \bibfield  {author} {\bibinfo {author} {\bibfnamefont {K.}~\bibnamefont
  {Miyake}}\ and\ \bibinfo {author} {\bibfnamefont {W.~J.}\ \bibnamefont
  {Mullin}},\ }\href {\doibase 10.1007/BF00681808} {\bibfield  {journal}
  {\bibinfo  {journal} {J. Low Temp. Phys.}\ }\textbf {\bibinfo {volume}
  {56}},\ \bibinfo {pages} {499} (\bibinfo {year} {1984})}\BibitemShut
  {NoStop}%
\bibitem [{\citenamefont {Baym}\ and\ \citenamefont
  {Ebner}(1968)}]{BaymEbner1970}%
  \BibitemOpen
  \bibfield  {author} {\bibinfo {author} {\bibfnamefont {G.}~\bibnamefont
  {Baym}}\ and\ \bibinfo {author} {\bibfnamefont {C.}~\bibnamefont {Ebner}},\
  }\href {\doibase 10.1103/PhysRev.170.346} {\bibfield  {journal} {\bibinfo
  {journal} {Phys. Rev.}\ }\textbf {\bibinfo {volume} {170}},\ \bibinfo {pages}
  {346} (\bibinfo {year} {1968})}\BibitemShut {NoStop}%
\bibitem [{\citenamefont {Stern}(1967)}]{SternPRL1967}%
  \BibitemOpen
  \bibfield  {author} {\bibinfo {author} {\bibfnamefont {F.}~\bibnamefont
  {Stern}},\ }\href {\doibase 10.1103/PhysRevLett.18.546} {\bibfield  {journal}
  {\bibinfo  {journal} {Phys. Rev. Lett.}\ }\textbf {\bibinfo {volume} {18}},\
  \bibinfo {pages} {546} (\bibinfo {year} {1967})}\BibitemShut {NoStop}%
\bibitem [{\citenamefont {Khalatnikov}\ and\ \citenamefont
  {Abrikosov}(1957)}]{Khalatnikov1958}%
  \BibitemOpen
  \bibfield  {author} {\bibinfo {author} {\bibfnamefont {I.~M.}\ \bibnamefont
  {Khalatnikov}}\ and\ \bibinfo {author} {\bibfnamefont {A.~A.}\ \bibnamefont
  {Abrikosov}},\ }\href@noop {} {\bibfield  {journal} {\bibinfo  {journal}
  {Zh.~Eskp.~Teor.~Fiz.}\ }\textbf {\bibinfo {volume} {33}},\ \bibinfo {pages}
  {110} (\bibinfo {year} {1957})},\ \bibinfo {note} {[Sov. Phys. JETP
  \textbf{6}, 84 (1958)]}\BibitemShut {NoStop}%
\bibitem [{\citenamefont {Abrikosov}\ and\ \citenamefont
  {Khalatnikov}(1959)}]{AK1959}%
  \BibitemOpen
  \bibfield  {author} {\bibinfo {author} {\bibfnamefont {A.~A.}\ \bibnamefont
  {Abrikosov}}\ and\ \bibinfo {author} {\bibfnamefont {I.~M.}\ \bibnamefont
  {Khalatnikov}},\ }\href {http://stacks.iop.org/0034-4885/22/i=1/a=310}
  {\bibfield  {journal} {\bibinfo  {journal} {Rep. Prog. Phys.}\ }\textbf
  {\bibinfo {volume} {22}},\ \bibinfo {pages} {329} (\bibinfo {year}
  {1959})}\BibitemShut {NoStop}%
\bibitem [{\citenamefont {Morel}\ and\ \citenamefont
  {Nozi\`eres}(1962)}]{MorelNozieres1962}%
  \BibitemOpen
  \bibfield  {author} {\bibinfo {author} {\bibfnamefont {P.}~\bibnamefont
  {Morel}}\ and\ \bibinfo {author} {\bibfnamefont {P.}~\bibnamefont
  {Nozi\`eres}},\ }\href {\doibase 10.1103/PhysRev.126.1909} {\bibfield
  {journal} {\bibinfo  {journal} {Phys. Rev.}\ }\textbf {\bibinfo {volume}
  {126}},\ \bibinfo {pages} {1909} (\bibinfo {year} {1962})}\BibitemShut
  {NoStop}%
\bibitem [{\citenamefont {Abramowitz}\ and\ \citenamefont
  {Stegun}(1965)}]{A&S}%
  \BibitemOpen
  \bibinfo {editor} {\bibfnamefont {M.}~\bibnamefont {Abramowitz}}\ and\
  \bibinfo {editor} {\bibfnamefont {I.~A.}\ \bibnamefont {Stegun}},\ eds.,\
  \href@noop {} {\emph {\bibinfo {title} {Handbook of Mathematical Functions
  with Formulas, Graphs, and Mathematical Tables}}}\ (\bibinfo  {publisher}
  {Dover Publications},\ \bibinfo {address} {New York},\ \bibinfo {year}
  {1965})\BibitemShut {NoStop}%
\bibitem [{\citenamefont {Novikov}()}]{Novikov06}%
  \BibitemOpen
  \bibfield  {author} {\bibinfo {author} {\bibfnamefont {D.~S.}\ \bibnamefont
  {Novikov}},\ }\href@noop {} {\bibinfo  {journal} {arXiv:cond-mat/0603184
  [cond-mat.mes-hall]}\ }\BibitemShut {NoStop}%
\bibitem [{\citenamefont {Anderson}\ and\ \citenamefont
  {Miller}(2011)}]{AM2011}%
  \BibitemOpen
\bibfield  {journal} {  }\bibfield  {author} {\bibinfo {author} {\bibfnamefont
  {R.~H.}\ \bibnamefont {Anderson}}\ and\ \bibinfo {author} {\bibfnamefont
  {M.~D.}\ \bibnamefont {Miller}},\ }\href {\doibase
  10.1103/PhysRevB.84.024504} {\bibfield  {journal} {\bibinfo  {journal} {Phys.
  Rev. B}\ }\textbf {\bibinfo {volume} {84}},\ \bibinfo {pages} {024504}
  (\bibinfo {year} {2011})}\BibitemShut {NoStop}%
\bibitem [{\citenamefont {Hone}(1962)}]{Hone1962}%
  \BibitemOpen
  \bibfield  {author} {\bibinfo {author} {\bibfnamefont {D.}~\bibnamefont
  {Hone}},\ }\href {\doibase 10.1103/PhysRev.125.1494} {\bibfield  {journal}
  {\bibinfo  {journal} {Phys. Rev.}\ }\textbf {\bibinfo {volume} {125}},\
  \bibinfo {pages} {1494} (\bibinfo {year} {1962})}\BibitemShut {NoStop}%
\bibitem [{\citenamefont {Dy}\ and\ \citenamefont
  {Pethick}(1969)}]{DyPethick1969}%
  \BibitemOpen
  \bibfield  {author} {\bibinfo {author} {\bibfnamefont {K.~S.}\ \bibnamefont
  {Dy}}\ and\ \bibinfo {author} {\bibfnamefont {C.~J.}\ \bibnamefont
  {Pethick}},\ }\href {\doibase 10.1103/PhysRev.185.373} {\bibfield  {journal}
  {\bibinfo  {journal} {Phys. Rev.}\ }\textbf {\bibinfo {volume} {185}},\
  \bibinfo {pages} {373} (\bibinfo {year} {1969})}\BibitemShut {NoStop}%
\bibitem [{\citenamefont {Bedell}\ and\ \citenamefont
  {Pines}(1980)}]{BedellPines1980}%
  \BibitemOpen
  \bibfield  {author} {\bibinfo {author} {\bibfnamefont {K.}~\bibnamefont
  {Bedell}}\ and\ \bibinfo {author} {\bibfnamefont {D.}~\bibnamefont {Pines}},\
  }\href {\doibase 10.1103/PhysRevLett.45.39} {\bibfield  {journal} {\bibinfo
  {journal} {Phys. Rev. Lett.}\ }\textbf {\bibinfo {volume} {45}},\ \bibinfo
  {pages} {39} (\bibinfo {year} {1980})}\BibitemShut {NoStop}%
\bibitem [{\citenamefont {Anderson}\ \emph {et~al.}(1999)\citenamefont
  {Anderson}, \citenamefont {Miller},\ and\ \citenamefont {Hallock}}]{AMH1999}%
  \BibitemOpen
  \bibfield  {author} {\bibinfo {author} {\bibfnamefont {R.~H.}\ \bibnamefont
  {Anderson}}, \bibinfo {author} {\bibfnamefont {M.~D.}\ \bibnamefont
  {Miller}}, \ and\ \bibinfo {author} {\bibfnamefont {R.~B.}\ \bibnamefont
  {Hallock}},\ }\href {\doibase 10.1103/PhysRevB.59.3345} {\bibfield  {journal}
  {\bibinfo  {journal} {Phys. Rev. B}\ }\textbf {\bibinfo {volume} {59}},\
  \bibinfo {pages} {3345} (\bibinfo {year} {1999})}\BibitemShut {NoStop}%
\bibitem [{\citenamefont {Akimoto}\ \emph {et~al.}(2006)\citenamefont
  {Akimoto}, \citenamefont {Cummings},\ and\ \citenamefont
  {Hallock}}]{PhysRevB.73.012507}%
  \BibitemOpen
  \bibfield  {author} {\bibinfo {author} {\bibfnamefont {H.}~\bibnamefont
  {Akimoto}}, \bibinfo {author} {\bibfnamefont {J.~D.}\ \bibnamefont
  {Cummings}}, \ and\ \bibinfo {author} {\bibfnamefont {R.~B.}\ \bibnamefont
  {Hallock}},\ }\href {\doibase 10.1103/PhysRevB.73.012507} {\bibfield
  {journal} {\bibinfo  {journal} {Phys. Rev. B}\ }\textbf {\bibinfo {volume}
  {73}},\ \bibinfo {pages} {012507} (\bibinfo {year} {2006})}\BibitemShut
  {NoStop}%
\bibitem [{\citenamefont {Ho}\ and\ \citenamefont
  {Hallock}(2001)}]{HoHallock2001}%
  \BibitemOpen
  \bibfield  {author} {\bibinfo {author} {\bibfnamefont {P.-C.}\ \bibnamefont
  {Ho}}\ and\ \bibinfo {author} {\bibfnamefont {R.~B.}\ \bibnamefont
  {Hallock}},\ }\href {\doibase 10.1103/PhysRevLett.87.135301} {\bibfield
  {journal} {\bibinfo  {journal} {Phys. Rev. Lett.}\ }\textbf {\bibinfo
  {volume} {87}},\ \bibinfo {pages} {135301} (\bibinfo {year}
  {2001})}\BibitemShut {NoStop}%
\bibitem [{\citenamefont {Kovtun}\ \emph {et~al.}(2005)\citenamefont {Kovtun},
  \citenamefont {Son},\ and\ \citenamefont {Starinets}}]{KSS2005}%
  \BibitemOpen
  \bibfield  {author} {\bibinfo {author} {\bibfnamefont {P.~K.}\ \bibnamefont
  {Kovtun}}, \bibinfo {author} {\bibfnamefont {D.~T.}\ \bibnamefont {Son}}, \
  and\ \bibinfo {author} {\bibfnamefont {A.~O.}\ \bibnamefont {Starinets}},\
  }\href {\doibase 10.1103/PhysRevLett.94.111601} {\bibfield  {journal}
  {\bibinfo  {journal} {Phys. Rev. Lett.}\ }\textbf {\bibinfo {volume} {94}},\
  \bibinfo {pages} {111601} (\bibinfo {year} {2005})}\BibitemShut {NoStop}%
\bibitem [{\citenamefont {Thomas}(2009)}]{Thomas2009}%
  \BibitemOpen
  \bibfield  {author} {\bibinfo {author} {\bibfnamefont {J.}~\bibnamefont
  {Thomas}},\ }\href {\doibase
  http://dx.doi.org/10.1016/j.nuclphysa.2009.09.055} {\bibfield  {journal}
  {\bibinfo  {journal} {Nuclear Physics A}\ }\textbf {\bibinfo {volume}
  {830}},\ \bibinfo {pages} {665c } (\bibinfo {year} {2009})}\BibitemShut
  {NoStop}%
\bibitem [{\citenamefont {Godfrin}\ \emph {et~al.}(2010)\citenamefont
  {Godfrin}, \citenamefont {Meschke}, \citenamefont {Lauter}, \citenamefont
  {B\"{o}hm}, \citenamefont {Krotscheck},\ and\ \citenamefont
  {Panholzer}}]{GodfrinEtAl2010}%
  \BibitemOpen
  \bibfield  {author} {\bibinfo {author} {\bibfnamefont {H.}~\bibnamefont
  {Godfrin}}, \bibinfo {author} {\bibfnamefont {M.}~\bibnamefont {Meschke}},
  \bibinfo {author} {\bibfnamefont {H.-J.}\ \bibnamefont {Lauter}}, \bibinfo
  {author} {\bibfnamefont {H.}~\bibnamefont {B\"{o}hm}}, \bibinfo {author}
  {\bibfnamefont {E.}~\bibnamefont {Krotscheck}}, \ and\ \bibinfo {author}
  {\bibfnamefont {M.}~\bibnamefont {Panholzer}},\ }\href@noop {} {\bibfield
  {journal} {\bibinfo  {journal} {J. Low Temp. Phys.}\ }\textbf {\bibinfo
  {volume} {158}},\ \bibinfo {pages} {147} (\bibinfo {year}
  {2010})}\BibitemShut {NoStop}%
\bibitem [{\citenamefont {Buu}\ \emph {et~al.}(2005)\citenamefont {Buu},
  \citenamefont {Puech},\ and\ \citenamefont {Wolf}}]{BPWChapter2005}%
  \BibitemOpen
  \bibfield  {author} {\bibinfo {author} {\bibfnamefont {O.}~\bibnamefont
  {Buu}}, \bibinfo {author} {\bibfnamefont {L.}~\bibnamefont {Puech}}, \ and\
  \bibinfo {author} {\bibfnamefont {P.~E.}\ \bibnamefont {Wolf}},\ }in\
  \href@noop {} {\emph {\bibinfo {booktitle} {Progress in Low Temperature
  Physics}}},\ Vol.~\bibinfo {volume} {XV},\ \bibinfo {editor} {edited by\
  \bibinfo {editor} {\bibfnamefont {W.~P.}\ \bibnamefont {Halperin}}}\
  (\bibinfo  {publisher} {Elsevier},\ \bibinfo {address} {Amsterdam},\ \bibinfo
  {year} {2005})\ Chap.~\bibinfo {chapter} {3}\BibitemShut {NoStop}%
\end{thebibliography}
%

\end{document}